%% file: main.tex
\documentclass[twocolumn]{autart}

\newcommand{\rmd}{\mathrm{d}}
\newcommand{\expm}{\text{expm}}

\usepackage{graphicx}
\usepackage{subcaption}

\usepackage{hyperref}
\usepackage{xcolor}
\usepackage{mathtools}
\usepackage{algorithm}
\usepackage{algpseudocode}
\usepackage{amssymb}

\input{macros}

\begin{document}

\begin{frontmatter}
  \title{Likelihood-based generalization of Markov parameter estimation and multiple shooting objectives in system identification}

  \thanks[cor]{Corresponding author.}  

  \author[cor]{Nicholas Galioto}\ead{ngalioto@umich.edu},
  \author{Alex Arkady Gorodetsky}\ead{goroda@umich.edu}

  \address{Department of Aerospace Engineering, University of Michigan, Ann Arbor, MI 48109, USA}


\begin{abstract}
  This paper considers the problem of system identification (ID) of linear and nonlinear non-autonomous systems from noisy and sparse data. We propose and analyze an objective function derived from a Bayesian formulation for learning a hidden Markov model with stochastic dynamics. We then analyze this objective function in the context of several state-of-the-art approaches for both linear and nonlinear system ID. In the former, we analyze least squares approaches for Markov parameter estimation, and in the latter, we analyze the multiple shooting approach. We demonstrate the limitations of the optimization problems posed by these existing methods by showing that they can be seen as special cases of the proposed optimization objective under certain simplifying assumptions: conditional independence of data and zero model error. Furthermore, we observe that our proposed approach has improved smoothness and inherent regularization that make it well-suited for system ID and provide mathematical explanations for these characteristics' origins. Finally, numerical simulations demonstrate a mean squared error over 8.7 times lower compared to multiple shooting when data are noisy and/or sparse. Moreover, the proposed approach can identify accurate and generalizable models even when there are more parameters than data or when the underlying system exhibits chaotic behavior.
\end{abstract}
\end{frontmatter}

\section{Introduction}\label{sec:intro}
Learning accurate and generalizable models of dynamical systems is of interest in a wide variety of research areas. Examples include, but are not limited to, predicting blood glucose levels in diabetes patients~\cite{plis2014,fox2018,li2019}, weather forecasting~\cite{gneiting2005,abhishek2012,karevan2020}, model predictive control~\cite{baros2022,zhang2020}, and load forecasting of smart grids~\cite{khan2016,fallah2018,hossain2019}. In a previous work~\cite{galioto2020}, we analyzed a Bayesian system identification (ID) algorithm that uses a hidden Markov model (HMM) with stochastic dynamics to model the underlying system. Oftentimes, the process noise in a stochastic model represents external disturbances to the system's nominal dynamics, but we showed that it can also be used to account for the uncertainty in an estimated model to yield improved performance.

All of the examples and analysis considered in~\cite{galioto2020}, however, were of autonomous systems. In this paper, we consider non-autonomous systems and provide new theory and comparisons to prevalent system ID approaches within this different class of systems. Identification of such systems is useful in a variety of controls-based applications, but non-autonomous systems also appear in a much broader range of applications beyond control. Examples include modeling the dynamics of lake temperatures with time-varying atmospheric conditions~\cite{jia2019} and modeling the transmission dynamics of COVID-19 that depend on time-varying government interventions~\cite{fang2020}.

There are a number of challenges within system ID~\cite{ljung2010}, but this work focuses only on a subset that pertains to the formulation of the learning objective function. Specifically, we address (1) dealing with error accumulation in the objective that leads to optimization difficulties, (2) finding generalizable models with few data, and (3) deriving a measure of model fit that is robust to model, measurement, and parameter uncertainty.

This work considers the objective functions related to two existing system ID approaches that are widely used on autonomous and non-autonomous systems alike. The two approaches are the eigensystem realization algorithm (ERA)~\cite{juang1985}, originally known as the Ho-Kalman algorithm~\cite{ho1966}, and the multiple shooting (MS) objective~\cite{bock1984}. The ERA is used within various algorithms for linear system ID, and the MS objective is commonly used to enable nonlinear system ID. One challenge with using the ERA is that it requires the Markov parameters of the system, and these are typically unknown and must be estimated. Since the performance of the ERA strongly depends on the quality of the Markov parameter estimates, we narrow our focus to consider algorithms used for estimating these Markov parameters. We then analyze the objective functions of the Markov parameter estimation and MS approaches and show that their objectives are special cases of the posterior distribution of the HMM formulation. This analysis leads to our three main contributions:
\begin{enumerate}
\item establish that many Markov parameter estimation methods implicitly assume that the data are conditionally independent given the initial condition, inputs, and Markov parameters (Proposition~\ref{prop:equiv}),
  \item illustrate how the variance parameters in the Bayesian posterior affect the smoothness of the optimization surface similarly to the time horizon parameter in MS (Section~\ref{sec:smooth}),
  \item empirically demonstrate that significant performance gains in system ID can be achieved over these two approaches by using the Bayesian posterior (Section~\ref{sec:results}).
\end{enumerate}
In addition to these main contributions, we also highlight a number of supporting contributions:
\begin{itemize}
\item extend the Bayesian algorithm used in~\cite{galioto2020} to nonlinear, non-autonomous systems, including chaotic systems and partial differential equations,
  \item identify how regularization arises in the marginal likelihood (Eq.~\ref{eq:linear_like}) and empirically demonstrate that the Bayesian posterior consequently resists overfitting,
  \item establish that least squares-based metrics can assign unfavorable rankings to strong model estimates identified by the Bayesian posterior (Section~\ref{sec:duffing}).
\end{itemize}

\subsection{Related work}
This section provides a brief literature review on works that have sought to address the challenges of error accumulation and generalizability, with a focus on approaches similar to those of the Bayesian algorithm. The third challenge of handling model, measurement, and parameter uncertainty was discussed in depth in a previous work~\cite{galioto2020}.

First, we review the existing work that attempts to address error accumulation in the objective function. The formation of local extrema due to error accumulation is a well-known problem in system ID~\cite{piroddi2003} and there are various heuristic approaches to solving it that revolve around discounting the data in one form or another. One example is  simulated annealing~\cite{kirkpatrick1983}, which  smooths the posterior by scaling down the influence of the likelihood by a discount factor. As optimization progresses, the likelihood is gradually scaled up to its full weight by increasing the discount factor according to a ``cooling schedule.'' This algorithm can be difficult to use in practice since its effectiveness strongly depends on the cooling schedule, which must be chosen by the user, and, although there are certain schedules that are commonly used, choosing a good schedule is largely problem-dependent. Furthermore, the algorithm does not address the issue of error accumulation, so the posterior is still filled with local minima that make finding a good schedule especially difficult. A variation of simulated annealing known as data annealing~\cite{green-sa2015} starts with a small subset of training data and gradually introduces more data. This has a similar ``cooling'' effect of incrementally increasing the influence of the likelihood on the posterior. In this algorithm, using data from a shorter training period at the early iterations can prevent the large error accumulation associated with long simulation times, but it still lacks a mechanism for handling error accumulation once the full dataset has been introduced into the posterior. As will be shown later in the paper, objectives that do not account for model error/uncertainty can sometimes assign low rankings to model estimates whose dynamical characteristics closely match those of the underlying system.

Other algorithms~\cite{zhong2019} directly address error accumulation by simulating the system output at many different initial times and using only the data within a specified time horizon of the initial time to evaluate the fit of each trajectory. An example of an extreme case of this is standard and exact DMD~\cite{tu2014}. In these algorithms, the Koopman operator is estimated by finding the best linear mapping of the data forward only one step in time, leading to a convex optimization problem. Algorithms that use a time horizon greater than one fall generally under the category of MS. In such algorithms, the selected time horizon usually depends on the approximate time scale of the system. The main issue with these algorithms is that they do not allow for flexibility in the case that the state components have different time scales. In contrast, the Bayesian algorithm studied in this paper can account for error accumulated at different rates within the state by estimating a process noise covariance matrix. The covariance matrix has the additional advantages of being continuous to allow for greater precision compared to discrete-time horizons and being able to be tuned automatically during optimization.

Second, we review observations in the literature on how randomized/probabilistic approaches affect estimation. Specifically, we note that algorithms that account for uncertainty within the model estimation process tend to inherently yield regularizing effects without requiring user-specified regularizing terms/assumptions. For example, in Bayesian model selection, the form of the model is treated as uncertain. As a result, a term that penalizes extracting more information from the data than is needed to reasonably fit the model arises in the log marginal likelihood~\cite{green-model2015,mackay2003}. The effect of this term is that simpler models that provide good fitting of the data are favored over extremely expressive models that could fit a wider range of output behavior. Thus, the preference for low complexity models is encoded automatically into the method. Other algorithms introduce randomness/uncertainty into the model estimation process by training with random subsets of the training dataset. Examples include stochastic gradient descent and sketch-and-solve algorithms, and both these methods have also observed implicit regularization benefits~\cite{zhang2021,smith2018,drineas2016}. Analagous to these other probabilistic algorithms, the proposed Bayesian algorithm will be shown to naturally introduce regularization into the realm of system ID.

\subsection{Paper outline}
The rest of the paper is organized as follows. Section~\ref{sec:method} details the probabilistic formulation of the dynamics model and provides the algorithm for computing the unnormalized log posterior. This section also provides an example of the proposed marginal likelihood in closed form for a linear-Gaussian system. In Section~\ref{sec:related}, background is provided on existing system ID approaches for evaluating model fit. Then, Section~\ref{sec:results} presents the results of numerical experiments with comparisons to the existing approaches described in the previous section. Lastly, conclusions are given in Section~\ref{sec:conclusion}.

\section{Background}\label{sec:method}
In this section, we provide the HMM framework, the recursive algorithm for efficiently solving for the posterior, and a representative example from a linear system.

The notation used throughout this paper is as follows: matrices are represented with uppercase and bold font $\mA$, and vectors are represented with lowercase and bold font $\vx$. Matrices and vectors are indexed with square brackets, e.g., the $(i,j)$th element of a matrix $\mA$ is denoted $\mA[i,j]$. The norm of a vector $\vx$ weighted by a positive definite matrix $\mW$ is defined as $\lVert\vx\rVert_{\mW}^2\coloneqq\vx^*\mW^{-1}\vx$, where $^*$ denotes the transpose. The $L_2$ norm of a vector and the induced $L_2$ matrix norm, also known as the spectral norm, are denoted as $\lVert\cdot\rVert_2$. The norm $\lvert\cdot\rvert$ represents the element-wise absolute value. The notation $\N(\vm,\mP)$ denotes a normal distribution with mean $\vm$ and covariance $\mP$. If $\vy=\lvert\vx\rvert$ and $\vx$ follows the normal distribution $\N(\vzero,\mP)$, then $\vy$ is said to follow a half-normal distribution denoted as $\text{half--}\N(\vzero,\mP)$.

\subsection{Hidden Markov model}\label{sec:formulation}
Consider a dynamical system modeled as an HMM~\cite{elliott2008}
\begin{subequations}
  \label{eq:hmm}
  \begin{align}
    \vx_{k+1} &= \Psi(\vx_k,\vu_k,\thetdyn) + \vxi_k, \hspace{-5mm}&\vxi_k\sim\N(\vzero,\mSigma(\thetsig)), \label{eq:dynamics}\\
    \vy_k &= h(\vx_k,\vu_k,\thetobs) + \veta_k,  &\veta_k\sim\N(\vzero,\mGamma(\thetgam)), \label{eq:observation}
  \end{align}
\end{subequations}
with uncertain initial condition $\vx_0(\thetic)$. The states are denoted by $\vx_k\in\reals^{\dimx}$, the measurements are $\vy_k\in\reals^{\dimy}$, and the inputs are $\vu_k\in\reals^{\dimu}$. The subscript $k\in\mathbb{Z}_+\cup\{0\}$ is an index corresponding to time $t_k\in[0,\infty)$. The function $\Psi:\reals^{\dimx}\times\reals^{\dimu}\times\reals^{\dimpar}\mapsto\reals^{\dimx}$ models the dynamics of the hidden state, and $h:\reals^{\dimx}\times\reals^{\dimu}\times\reals^{\dimpar}\mapsto\reals^{\dimy}$ is an observation function that maps the hidden space to the observable space. The additive, zero-mean Gaussian noise in the dynamics $\vxi_k\in\reals^{\dimx}$ and in the observations $\veta_k\in\reals^{\dimy}$ represent the process and measurement uncertainty, respectively. The uncertain parameters $\vtheta=\begin{bmatrix}\thetic^*&\thetdyn^*&\thetobs^*&\thetsig^*&\thetgam^*\end{bmatrix}^*\in\reals^{\dimpar}$ correspond to the initial condition, dynamics, observation function, process noise covariance $\mSigma\in\reals^{\dimx\times\dimx}$, and measurement noise covariance $\mGamma\in\reals^{\dimy\times\dimy}$, respectively.

\subsection{Bayesian learning}
The goal of Bayesian system ID is to characterize the posterior probability distributions of the parameters $\vtheta$ after collecting $\numObs+1$ data points $\vy_0,\ldots,\vy_{\numObs}$. This posterior distribution is denoted by $\probd(\vtheta|\yn)$, where $\probd$ is a probability density function, and $\yn\coloneqq(\vy_0,\ldots,\vy_{\numObs})$ is the collection of data points. Bayes' rule represents this posterior distribution in a computable form
\begin{equation}\label{eq:bayes}
  \probd(\vtheta|\yn) = \frac{\like(\vtheta;\yn)\probd(\vtheta)}{\probd(\yn)},
\end{equation}
where $\like(\vtheta;\yn)\coloneqq\probd(\yn|\vtheta)$ is the likelihood, $\probd(\vtheta)$ is the prior, and $\probd(\yn)$ is the evidence. The main computational challenge in evaluating the posterior is computing the likelihood.  In this case, the uncertainty in the states further increases the computational difficulty by inducing the joint likelihood $\probd(\vtheta,\xn|\yn)$, where $\xn\coloneqq(\vx_0,\ldots,\vx_{\numObs})$, rather than the target marginal likelihood $\like(\vtheta;\yn).$ This marginal likelihood is obtained through high-dimensional integration over all the uncertain states $\like(\vtheta;\yn) = \int\like(\vtheta,\xn;\yn)\rmd\xn$. Specifically, this is an integral over the $\dimx$-dimensional state at each of the $\numObs+1$ time instances. Typically high-dimensional integration is intractable, but this integral can be computed efficiently though recursion. The recursive procedure begins by observing that the marginal likelihood can be factored according to:
\begin{equation}\label{eq:marg_like}
  \like(\vtheta;\yn) = \probd(\vy_0\mid\vtheta)\prod_{k=1}^{\numObs}\probd(\vy_k\mid\mathcal{Y}_{k-1},\vtheta).
\end{equation}
Then, Algorithm~\ref{alg:marg_like} provides a recursive approach~\cite{sarkka2013} to evaluate each of the terms in the product. In linear-Gaussian systems, the marginal likelihood $\like(\vtheta;\yk)$, prediction distribution $\probd(\vx_{k+1}|\yk,\vtheta)$, and update distribution $\probd(\vx_{k}|\mathcal{Y}_{k},\vtheta)$ are all Gaussian, and, as a result, can be found in closed-form using the Kalman filter. If the system is not linear-Gaussian, other Bayesian filters must be used. For example, in this and past works~\cite{galioto2020}, the unscented Kalman filter~\cite{julier1997} is used to find an approximation of the marginal likelihood when the system is nonlinear.
\begin{algorithm}
  \caption{{\scriptsize Recursive marginal likelihood evaluation~\cite{sarkka2013}}}
  \label{alg:marg_like}
  \begin{algorithmic}[1]{\scriptsize
    \Require $\probd(\vx_0|\vtheta)$, $\yn$
    \Ensure $\like(\vtheta;\yn)$
    \State Initialize $\probd(\vx_0|\mathcal{Y}_{-1},\vtheta)\coloneqq\probd(\vx_0|\vtheta)$ and $\like(\vtheta;\mathcal{Y}_{-1})\coloneqq1$    
    \For{$k=0,\ldots\numObs$}
    \State Marginalize:
    \vspace{-4mm}
    \begin{align*}  \probd(\vy_k | \mathcal{Y}_{k-1},\vtheta) &\gets \int\probd(\vy_k|\vx_k,\vtheta)\probd(\vx_k|\mathcal{Y}_{k-1},\vtheta)\rmd\vx_k \\
    \like(\vtheta;\mathcal{Y}_k) &\gets \like(\vtheta;\mathcal{Y}_{k-1})\probd(\vy_k | \mathcal{Y}_{k-1},\vtheta)\end{align*}\vspace{-5mm}
    \If{$k < \numObs$}
    \State {\raggedright Update: $\displaystyle\probd(\vx_{k}|\mathcal{Y}_{k},\vtheta) \gets \frac{\probd(\vy_k|\vx_k,\vtheta)}{\probd(\vy_k | \mathcal{Y}_{k-1},\vtheta)}\probd(\vx_k|\mathcal{Y}_{k-1},\vtheta)$}
    \State {\raggedright Predict: $\displaystyle\probd(\vx_{k+1}|\yk,\vtheta) \gets \int\probd(\vx_{k+1}|\vx_{k},\vtheta)$}
      \Statex \begin{flushright}$\displaystyle\times\probd(\vx_{k}|\mathcal{Y}_{k},\vtheta)\rmd\vx_{k}$\end{flushright}
    \EndIf
    \EndFor}
  \end{algorithmic}
\end{algorithm}

Once the marginal likelihood is evaulated with Algorithm~\ref{alg:marg_like}, it can be plugged into Bayes' rule~\eqref{eq:bayes}. The resulting posterior is generally non-Gaussian and not analytically tractable. As a result, we use a Markov chain Monte Carlo (MCMC) algorithm to generate samples from the posterior. These samples can then be used for subsequent predictions. For improved convergence, we use a DRAM within Gibbs procedure to sequentially sample the parameter groups $\{\thetic\}$, $\{\thetdyn\}$, and $\{\thetsig,\thetgam\}$. The observation parameters $\thetobs$ are fixed in order to restrict the coordinate frame and thereby mitigate the sampling challenges that arise due to non-uniqueness of the parameters. A detailed description of the implementation of this MCMC algorithm on our examples is given in Section~\ref{sec:results}.

\subsection{Linear time-invariant systems}\label{sec:lti}
For certain systems, the marginal likelihood is analytically tractable. Here we show the approach in the context of linear time-invariant (LTI) models defined as
\begin{equation}
  \label{eq:linear_dynamics}
  \begin{aligned}
    \vx_{k+1} &= \mA(\vtheta)\vx_k +\mB(\vtheta)\vu_k + \vxi_k, &\vxi_k\sim\N(\vzero,\mSigma(\vtheta)), \\
    \vy_k &= \mH(\vtheta)\vx_k + \mD(\vtheta)\vu_k + \veta_k, &\veta_k\sim\N(\vzero,\mGamma(\vtheta)).
  \end{aligned}
\end{equation}
In this system, if $\vx_0$ is either given or Gaussian-distributed, then the equations in Algorithm~\ref{alg:marg_like} have closed-form solutions. Next, we consider two different, but technically equivalent, approaches for evaluating the closed-form marginal likelihood.

\subsubsection{State-space approach}

The first approach uses the state-space models with a Kalman filter to evaluate the marginal likelihood. Following~\cite{sarkka2013}, let $\vm_k(\vtheta)$ and $\mP_k(\vtheta)$ denote the mean and covariance of the Gaussian distribution $\probd(\vx_k\mid\ykk,\vtheta)$, i.e., $\probd(\vx_k\mid\ykk,\vtheta)=\N(\vm_k(\vtheta),\mP_k(\vtheta))$, at time $t_k$. The value of the mean $\vm_k$ and covariance $\mP_k$ can be found via a Kalman filter. Then, each term in Eq.~\eqref{eq:marg_like}  becomes $\probd(\vy_k\mid\ykk,\vtheta)=\N(\vmu_k(\vtheta),\mS_k(\vtheta))$, where $\vmu_{k}(\vtheta)=\mH(\vtheta)\vm_{k}(\vtheta)+\mD(\vtheta)\vu_{k}$ and $\mS_k(\vtheta) = \mH(\vtheta)\mP_{k}(\vtheta)\mH(\vtheta)^*+\mGamma(\vtheta)$. Finally, the log marginal likelihood from line 3 of Algorithm~\ref{alg:marg_like} becomes
\begin{equation}\label{eq:linear_like}
  \begin{aligned}
    \log\like(\vtheta;\yn) = &-\frac{1}{2}\sum_{k=0}^{\numObs}\Big(\lVert \vy_k - \vmu_{k}(\vtheta) \rVert^2_{\mS_k(\vtheta)} \\
    &\quad+\log\left(\text{det}\left( \mS_k(\vtheta)\right)\right) + \dimy\log(2\pi)\Big),
  \end{aligned}
\end{equation}
where $\text{det}(\cdot)$ is the determinant~\cite[Th. 12.3]{sarkka2013}.

The form of the marginal likelihood~\eqref{eq:linear_like} resembles a least squares metric plus a regularization term $\log\left(\text{det}\left(\mS_k\right)\right)$. The inclusion of this term differs from the typical approach in which regularization is introduced through a prior distribution or some sort of heuristic penalty placed on the parameters (e.g., the $L_2$ norm used in ridge regression). Here, the regularization term has arisen in the likelihood directly from the probabilistic model of the dynamical system and does not require any assumptions on the parameters. The effect of this additional term is a penalty on systems where the estimated output has a large covariance, which is oftentimes a sign of overfitting. For example, consider two sets of parameters that produce the exact same output on the training data, but one set of parameters is much more sensitive to the model inputs such that the variance of its output is much greater. Intuitively, one would prefer the less sensitive model because it is more likely to be generalizable to unseen data. The regularization term $\log\left(\text{det}\left(\mS_k\right)\right)$ encodes this preference automatically in the likelihood. 

\subsubsection{Input-output Markov parameter approach}\label{sec:iomarkov}
The other common approach to LTI system ID is to first estimate the Markov parameters and then use the eigensystem realization algorithm (ERA)~\cite{juang1985} to extract a realization of the system matrices $(\mA, \mB, \mH, \mD)$. The ERA is one of the most popular algorithms for identification of LTI state-space models due to its speed, scalability, approximation error guarantees, and ease of use. The algorithm essentially consists of a singular value decomposition (SVD) followed by a linear least squares solve. The only requirements for its use are the system Markov parameters and the state-space dimension $\dimx$. In general, the ERA is considered to be a foundational system ID method~\cite{van2012,viberg1995} with a number of variations~\cite{kramer2018,lale2020}, and its use can be found in system ID applications ranging from subspace identification~\cite{pappalardo2018} to model reduction~\cite{ma2011,almunif2020}. Details on the ERA procedure itself are provided in Appendix~\ref{app:era}. Here we discuss how the Markov parameters can be obtained from data.

The Markov parameters are directly obtained by rewriting the linear system~\eqref{eq:linear_dynamics} in a form that removes the states through recursive substitution into the observation equations to yield
{\small
\begin{subequations}
  \begin{align}
    \begin{split}
    \vy_k &= \mH\mA^{k}\vx_0 + \sum_{i=1}^{k}\mH\mA^{i-1}\left(\mB\vu_{k-i} + \vxi_{k-i}\right)+\mD\vu_k +\veta_k,
    \end{split}\\
      &= \mH\mA^{k}\vx_0 + \sum_{i=0}^{k}\mG_k \vu_{k-i} + \sum_{i=1}^{k}\mH\mA^{i-1}\vxi_{k-i} +\veta_k,
      \label{eq:io}
  \end{align}
\end{subequations}}
where the Markov parameters are $\mG_0=\mD$ and $\mG_k=\mH\mA^{k-1}\mB$ for $k=1,2,\ldots$. Defining a new random variable $\vnu_k = \sum_{i=1}^{k}\mH\mA^{i-1}\vxi_{k-i}$, we obtain the compact form
\begin{equation}
    \vy_k= \mH\mA^{k}\vx_0+\sum_{i=0}^{k}\mG_i\vu_{k-i}+\vnu_k, \hfill \vnu_k\sim\N(\vzero,\mLambda_k), \label{eq:io_compact}  
\end{equation}
where  $\mLambda_k = \sum_{i=1}^{k}\mH\mA^{i-1}\mSigma(\mH\mA^{i-1})^* + \mGamma$ if $k>0$, and $\mLambda_k=\mGamma$ if $k=0$. Note that the $\vnu_k$ are not independent due to their sharing of the process noise $\vxi_k$. The covariance $\mLambda_{j,k}\coloneqq\mathbb{C}\text{ov}[\vnu_j, \vnu_k]$ is
\begin{equation}
  \mLambda_{j,k} = \sum_{i=1}^{j}\mH\mA^{i-1}\mSigma(\mH\mA^{k-j+i-1})^*,
\end{equation}
for $0<j<k$. If $j=0$ and $j\neq k$, then $\mLambda_{j,k}=\vzero$. Lastly, if $k < j$, then $\mLambda_{j,k}=\mLambda_{k,j}^*$.

The task then is to learn the Markov parameters for use within the ERA.  One can use Bayesian inference again to learn a posterior over the Markov parameters. Assuming $\vx_0=\vzero$, the likelihood model implied by Eq.~\eqref{eq:io_compact} is
\begin{equation} \label{eq:likelihood_io}
  p(\yn \mid \mG_{0:\numObs}, \mU_{0:\numObs}) = \mathcal{N}\left(\mG_{0:\numObs}\mU_{0:\numObs}, \mLambda \right),
\end{equation}
where $\mG_{0:\numObs} = \begin{bmatrix}\mG_0 & \mG_1 & \cdots & \mG_{\numObs}\end{bmatrix}$ and
\begin{equation*}
  \mLambda = \begin{bmatrix}
    \mLambda_{0} & \mLambda_{0,1} & \cdots & \mLambda_{0,\numObs} \\
                & \mLambda_{1}   & \cdots & \vdots \\
                &               & \ddots &  \vdots  \\
    \text{Sym}  &               &        & \mLambda_{\numObs}
  \end{bmatrix},
  \mU_{0:\numObs} = \begin{bmatrix}
    \vu_0  & \vu_1  &  \cdots & \vu_{\numObs} \\
    \vzero & \vu_0  &  \cdots & \vu_{\numObs-1} \\
    \vdots & \vdots & \vdots & \vdots \\
    \vzero & \vzero & \cdots & \vu_0
  \end{bmatrix}.
\end{equation*}
The log of this likelihood~\eqref{eq:likelihood_io} is equivalent to the state-space log likelihood~\eqref{eq:linear_like} in the following sense: if we evaluate~\eqref{eq:likelihood_io} with matrices $\mG_{0:\numObs}$ and $\mLambda$ determined by a set of state-space matrices, the result equals the evaluation of~\eqref{eq:linear_like} using that same set of state-space matrices.
  
For comparison with current state-of-the-art methods in Section~\ref{sec:era}, it will be useful to also consider a maximum likelihood estimate (MLE). The likelihood~\eqref{eq:likelihood_io} is maximized by solving
\begin{equation}
  \hat{\mG}_{0:\numObs} = \argmin_{\mG_{0:\numObs}}\left\lVert\text{vec}\left(\mY_{0:\numObs} - \mG_{0:\numObs}\mU_{0:\numObs}\right)\right\rVert_{\mLambda}^2,
  \label{eq:mle_io}
\end{equation}
where $\text{vec}(\cdot)$ denotes the vectorization of a matrix, and $\mY_{0:\numObs} = \begin{bmatrix}\vy_0 & \vy_1 & \cdots & \vy_{\numObs}\end{bmatrix}$. The solution for this generalized least squares problem can be computed as $\text{vec}(\hat{\mG}_{0:\numObs}) = (\mV^*\mLambda^{-1}\mV)^{\dagger}\mV^*\mLambda^{-1}\text{vec}(\mY_{0:\numObs})$, where $^\dagger$ denotes the pseudo-inverse, and $\mV^*\coloneqq\mU_{0:\numObs}\otimes\mI_{\dimy}$, where $\otimes$ is the Kronecker product.

There are two main issues with the MLE approach. The first is that the covariance matrix $\mLambda$ depends on the unknown state-space matrices and is therefore itself unknown. To address this issue, the usual solutions are to remove the weighting entirely and simply use the $L_2$ norm~\cite{chen2012} or to estimate a realization of the state-space matrices directly~\cite{beck2010}. 

The second main issue is that the Markov parameters are dramatically overparameterized since they are direct functions of the system matrices $(\mA, \mB, \mH, \mD).$ Specifically, overparameterization happens when $\numObs\dimy\dimu > \dimx^2 + \dimx\dimu + \dimx\dimy + \dimy\dimu$, as is usually the case.  As a result, the number of data points $(\numObs\dimy)$ is typically smaller than the number of unknowns $(\numObs \dimy \dimu)$, --- except in the case where $\dimu=1$ --- and the optimum is not unique. Moreover, any optimum found with this approach, including when $\dimu=1$, will necessarily overfit the data when there is noise. This issue motivates multiple approaches in the literature: some approaches use multiple trajectories/rollouts with differing inputs\footnote{The inputs must differ by more than a scalar multiplier to avoid underdetermination.} to increase the number of data points~\cite{zheng2020,sun2020}, and other approaches break a single trajectory into multiple ones to decrease the effective number of Markov parameters~\cite{oymak2019,sarkar2021}. We will show that these existing works use implicit simplifying assumptions and are still at risk of underdetermination for certain input signals. The state-space approach used by the proposed Bayesian algorithm, on the other hand, makes all assumptions explicit and is viable regardless of the type of control inputs. These approaches and relationships are futher described in Section~\ref{sec:era}.

\section{Theoretical foundations and analysis}\label{sec:related}
In this section, an analysis of how optimization objectives for Markov parameter estimation in LTI systems and multiple shooting in nonlinear systems arise from simplifications of Algorithm~\ref{alg:marg_like}.

\subsection{Estimation of stochastic LTI systems with conditionally independent data}\label{sec:era}

To begin our analysis, we consider learning state-space realizations of stochastic LTI systems. First, we analyze how existing algorithms approach the problem of unknown covariance $\mLambda$. Second, we describe how single and multiple rollout resolve the issue of overparameterized Markov parameters that leads to the aforementioned underdetermined system. Finally, we demonstrate how learning the system matrices rather than the Markov parameters leads to faster convergence.

\subsubsection{Markov parameter estimation}\label{sec:markov_analysis}
All the approaches previously referenced avoid knowledge of $\mLambda$ by minimizing some sort of least squares objective with an unweighted $L_2$ norm. Here, we describe the implicit assumptions that these approaches make, in light of the setup provided in Section~\ref{sec:iomarkov}.

The main assumption these approaches make is that observations are conditionally independent given the system parameters, inputs, and initial condition. This assumption fixes the off-diagonal blocks of the covariance matrix according to $\mLambda_{j,k} = \vzero$. Under this assumption, the optimal estimator for such a system is given in Proposition~\ref{prop:blue}.
\begin{prop}\label{prop:blue}
  Assume that $\vx_0=\vzero$, the inputs $\vu_k$ are known, and the outputs $\vy_k$ are conditionally independent given $\mG_{0:\numObs}$ and $\mU_{0:\numObs}$, $\forall k=0,1,\ldots,\numObs$. Then, the MLE of an LTI system's Markov parameters is
\begin{equation}\label{eq:fulltraj}
  \hat{\mG}_{0:\numObs} = \argmin_{\{\mG_i\}_{i=0}^{\numObs}}\sum_{k=0}^{\numObs}\left\lVert \vy_k - \sum_{i=0}^{k}\mG_i\vu_{k-i} \right\rVert_{\mLambda_k}^2.
\end{equation}
\end{prop}
\begin{pf}
  The likelihood distribution~\eqref{eq:likelihood_io} can be factored as follows:
  \begin{equation}
    \begin{split}
      \probd(\yn|\mG_{0:\numObs},\mU_{0:\numObs}) &= \probd(\vy_{0}|\mG_{0:\numObs},\mU_{0:\numObs})\\
      &\times\prod_{k=1}^{\numObs}\probd(\vy_{k}|\mathcal{Y}_{0:k-1},\mG_{0:\numObs},\mU_{0:\numObs}).
    \end{split}
\end{equation}
The conditional independence assumption yields $\probd(\yn|\mG_{0:\numObs},\mU_{0:\numObs}) = \prod_{k=0}^{\numObs}\probd(\vy_{k}|\mG_{0:\numObs},\mU_{0:\numObs})$. With $\vx_0=\vzero$, the marginal distributions follow straightforwardly from the input-output relation in Eq.~\eqref{eq:io_compact} as $\probd(\vy_{k}|\mG_{0:\numObs},\mU_{0:\numObs})=\N(\sum_{i=0}^{k}\mG_i\vu_{k-i}, \mLambda_k)$. Then, taking the negative log yields the MLE of the Markov parameters~\eqref{eq:fulltraj}.\qed
\end{pf}

The independence assumption, however, is not sufficient to convert Eq.~\eqref{eq:mle_io} to the least squares objectives commonly used in the literature. The additional assumption that $\mLambda_k\propto\mI$ is needed to convert the weighted norm to a scalar multiple of the $L_2$ norm. This assumption holds trivially when $\dimy=1$. However, if the outputs are multi-dimensional, there is no reasonable assumption to enable $\mLambda_k$ to be proportional to the identity. Nevertheless, if one considers an approximate objective where this is assumed to be so, one obtains
\begin{equation}\label{eq:l2obj}
  \hat{\mG}_{0:\numObs} = \argmin_{\{\mG_i\}_{i=0}^{\numObs}}\sum_{k=0}^{\numObs}\left\lVert \vy_k - \sum_{i=0}^{k}\mG_i\vu_{k-i} \right\rVert_{2}^2.
\end{equation}

This approximate objective no longer requires knowledge of the system matrices and is used as the basis for a large number of approaches~\cite{tsiamis2019,sarkar2021,zheng2020}. These approaches differ based on how they resolve the underdetermination issue depending on if the data are collected from a multiple or single rollout procedure. This work considers learning from single trajectories, so methods based on multiple rollout are not applicable.

For the single rollout procedure, the data are divided into $K$ overlapping subtrajectories of length $\bar{n}$ such that $\numObs=\bar{n}+K-2$. To address underdetermination, one must also require $\bar{n}<\frac{\numObs+1}{\dimu}$.

After dividing the single trajectory into multiple trajectories, the final output of each trajectory follows the same form as Eq.~\eqref{eq:io_compact},
\begin{equation}
  \vy_{k} = \mH\mA^{\bar{n}-1}\vx_{k-\bar{n}+1} + \sum_{i=0}^{\bar{n}-1}\mG_i\vu_{k-i} + \vnu_{k},
\end{equation}
for $k=\bar{n},\ldots,\numObs$. Assuming that the inputs are zero-mean, then the expected value of each $\vx_{k-\bar{n}+1}$ is zero with respect to the inputs and noise variables. This is sometimes used as justification to eliminate $\vx_{k-\bar{n}+1}$ from the estimation problem~\cite{oymak2019}, and we therefore also adopt this ansatz. Adding the approximation that the noise covariance be proportional to the identity yields the following optimization problem
\begin{equation}\label{eq:subtraj}
  \hat{\mG}_{0:\bar{n}-1} = \argmin_{\{\mG_i\}_{i=0}^{\bar{n}-1}}\sum_{k=\bar{n}}^{\numObs}\left\lVert \vy_{k} - \sum_{i=0}^{\bar{n}-1}\mG_i\vu_{\bar{n}-i}\right\rVert^2_{2}.
\end{equation}
This optimization problem now only has $\bar{n}$ unknown Markov parameters rather than $\numObs+1$, mitigating the problem of undetermination. The least squares solution is given by $\hat{\mG}_{0:\bar{n}-1} = \mY_{\bar{n}:\numObs}\bar{\mU}_{\bar{n}:\numObs}^{\dagger}$, where
\begin{equation*}
  \bar{\mU}_{\bar{n}:\numObs} = \begin{bmatrix}
    \vu_{\bar{n}} & \vu_{\bar{n}+1} & \cdots & \vu_{\numObs} \\
    \vu_{\bar{n}-1}   & \vu_{\bar{n}} & \cdots & \vu_{\numObs-1} \\
    \vdots        & \vdots         & \cdots & \vdots \\
    \vu_{1}   & \vu_{2} & \cdots & \vu_{K}
  \end{bmatrix}.
\end{equation*}
This equation is equivalent to the slightly different form provided by~\cite{oymak2019}\footnote{The unlabeled equation following Eq. 5 in~\cite{oymak2019}.}. 

Although the system of equations now has more equations than unknowns for proper choice of $\bar{n}$, the system can still suffer from undetermination for certain input signals. For example, sinusoidal inputs generate a $\bar{\mU}_{\bar{n}:\numObs}$ with rank of only 2. Additionally, each estimated data point requires exactly $\bar{n}$ inputs. Consequently, the outer sum skips the first $\bar{n}$ data points since the inputs $\vu_{0}, \vu_{-1},\ldots$ are typically assumed unknown. 

This $L_2$ optimization problem~\eqref{eq:subtraj} is equivalent to the weighted $L_2$ optimization problem~\eqref{eq:fulltraj} under additional assumptions stated in Proposition~\ref{prop:equiv}. 
\begin{prop}\label{prop:equiv}
  Assume the assumptions of Proposition~\ref{prop:blue} are met and additionally that $\sum_{i=\bar{n}}^k\mG_i\vu_{k-i}=\vzero,$ $\mA^{k}\mSigma(\mA^{k})^*=\vzero$ for $k\geq\bar{n}$, and $\mLambda_{\bar{n}}\propto\mI$. If the first $\bar{n}$ outputs are discarded, then the MLE in Eq.~\eqref{eq:fulltraj} is equivalent to the estimator in Eq.~\eqref{eq:subtraj}.
\end{prop}
\begin{pf}
  If $\sum_{i=\bar{n}}^k\mG_i\vu_{k-i}=\vzero$, then the sum inside the norm of Eq.~\eqref{eq:fulltraj} is simplified as $\sum_{i=0}^k\mG_i\vu_{k-i}=\sum_{i=0}^{\bar{n}-1}\mG_i\vu_{k-i}$. Lastly if $\mA^{k}\mSigma(\mA^{k})^*=\vzero$ for $k\geq\bar{n}$, then $\mLambda_k =\sum_{i=1}^{\bar{n}}\mH\mA^{i-1}\mSigma(\mH\mA^{i-1})^*+\mGamma=\mLambda_{\bar{n}}$, for $k\geq\bar{n}$. By assumption, $\mLambda_{\bar{n}}\propto\mI$, so the weighted norm is equivalent to the standard $L_2$ norm.\qed
\end{pf}

The assumptions $\sum_{i=\bar{n}}^k\mG_i\vu_{k-i}=\vzero$ and $\mA^{k}\mSigma(\mA^{k})^*=\vzero$ for $k\geq\bar{n}$ can be satisfied if the system has finite impulse response. Alternatively, these two assumptions can be achieved asymptotically under the much weaker assumption that $\rho(\mA)<1$, where $\rho(\cdot)$ denotes the spectral radius of a matrix. Such a result is given in Proposition~\ref{prop:converge}.
\begin{prop}\label{prop:converge}
  Let $\rho(\mA)<1$, the inputs $\vu_k$ be independent realizations of a real-valued random variable, and $\mLambda_{\bar{n}}\propto\mI$. As $\bar{n}\to\infty$, the negative log likelihood of Eq.~\eqref{eq:fulltraj} approaches the subtrajectory LS objective of Eq.~\eqref{eq:subtraj} with probability 1. Moreover, it converges at least linearly.
\end{prop}
\begin{pf}
  The proof is in Appendix~\ref{app:asymptotic}.
\end{pf}

For systems where the assumption $\mLambda_k\propto\mI$ approximately holds for $k>\bar{n}$, this result implies that even if $\mLambda_k$ varies, a good approximation can still be achieved for reasonably small $\bar{n}$ values, especially when $\rho(\mA)$ is smaller. However, there are a number of systems of interest where $\rho(\mA)\geq 1$ such that the conditions of this proposition no longer hold, e.g., periodic systems have $\rho(\mA)=1$.

\subsubsection{Numerical comparison}\label{sec:markov_numeric}
We now perform a comparison between three approaches using the same numerical experiment from~\cite{oymak2019}. The first approach is the least squares (LS) approach used in single rollout in~\cite{oymak2019}. The second approach is one where we assume that the $\mLambda_k$ is given, e.g., by an oracle, so we minimize the same objective as~\cite{oymak2019}, but with a different weighted norm
\begin{equation}\label{eq:gls}
  \hat{\mG}_{0:\bar{n}-1} = \argmin_{\{\mG_i\}_{i=0}^{\bar{n}-1}}\sum_{k=\bar{n}}^{\numObs}\left\lVert \vy_{k} - \sum_{i=0}^{\bar{n}-1}\mG_i\vu_{k-i}\right\rVert^2_{\mLambda_{k}}.
\end{equation}
This objective is henceforth referred to as the generalized least squares (GLS). Finally, we compare with the maximum {\it a-posteriori} (MAP) estimate of the Bayesian approach described in Section~\ref{sec:lti}. The state-space approach is used numerically, though it is theoretically equivalent to the input-output approach.

This experiment begins as follows. A random state-space system was generated by independently sampling the entries of $\mH$ and $\mD$ from $\N(0,1/\dimy)$ and of $\mB$ from $\N(0,1/\dimx)$. The dimensions of the system were $\dimy=2$, $\dimx=5$, and $\dimu=3$, and the noise covariances were $\mSigma=\sigma_{\xi,\eta}^2\mI$ and $\mGamma=\sigma_{\xi,\eta}^2\mI$. To highlight the difference between the standard LS and GLS, $\mA$ was set as the identity matrix to ensure that $\rho(\mA)=1$ and consequently that the covariance $\mLambda_k$ would vary significantly over time. To avoid having priors give the Bayesian algorithm an edge, improper uniform priors were placed on the state-space matrices, and only weakly informative priors of $\halfN(0,1)$ were placed on the parameters $\sigma_{\xi}$ and $\sigma_{\eta}$ to enforce positivity and improve convergence. Then, data were generated by simulating the system with inputs sampled from a standard normal, i.e., $\vu_k\sim\N(\vzero,\mI)$. For this experiment, a subtrajectory length of $\bar{n}=18$ was considered for a total of $\dimy\dimu\bar{n}=108$ parameters. The Markov parameters were estimated using the first $K$ subtrajectories of the simulated trajectory where $K=\dimu\bar{n},\ldots,2000$. Optimizing over the posterior is significantly more expensive than solving a linear least squares problem, so for computational feasibility, this optimization was only performed at $K=\dimu\bar{n},\dimu\bar{n}+100,\ldots,2000$. Since the least squares methods do not use the first $\bar{n}$ data points, these data were also removed from MAP estimation for consistency.

\begin{figure}
  \centering
  \begin{subfigure}{0.32\linewidth}
    \centering
    \includegraphics[trim=95 250 107 249, clip,width=\linewidth]{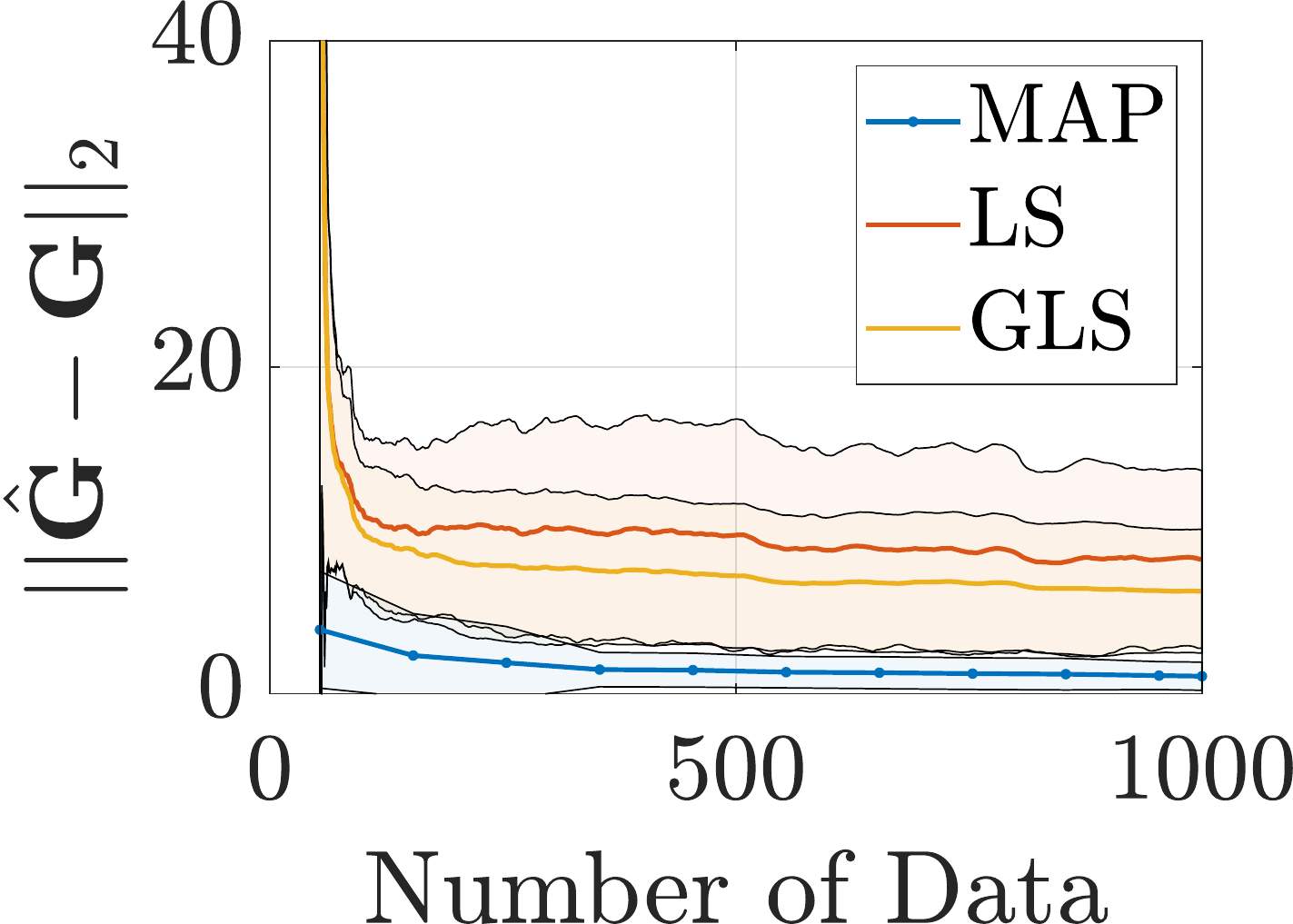}
    \caption{$\sigma_{\xi,\eta}=0.25$,\\\hfill 108 parameters}
    \label{fig:error_small_025}
  \end{subfigure}%
  \begin{subfigure}{0.32\linewidth}
    \centering
    \includegraphics[trim=95 250 107 249, clip,width=\linewidth]{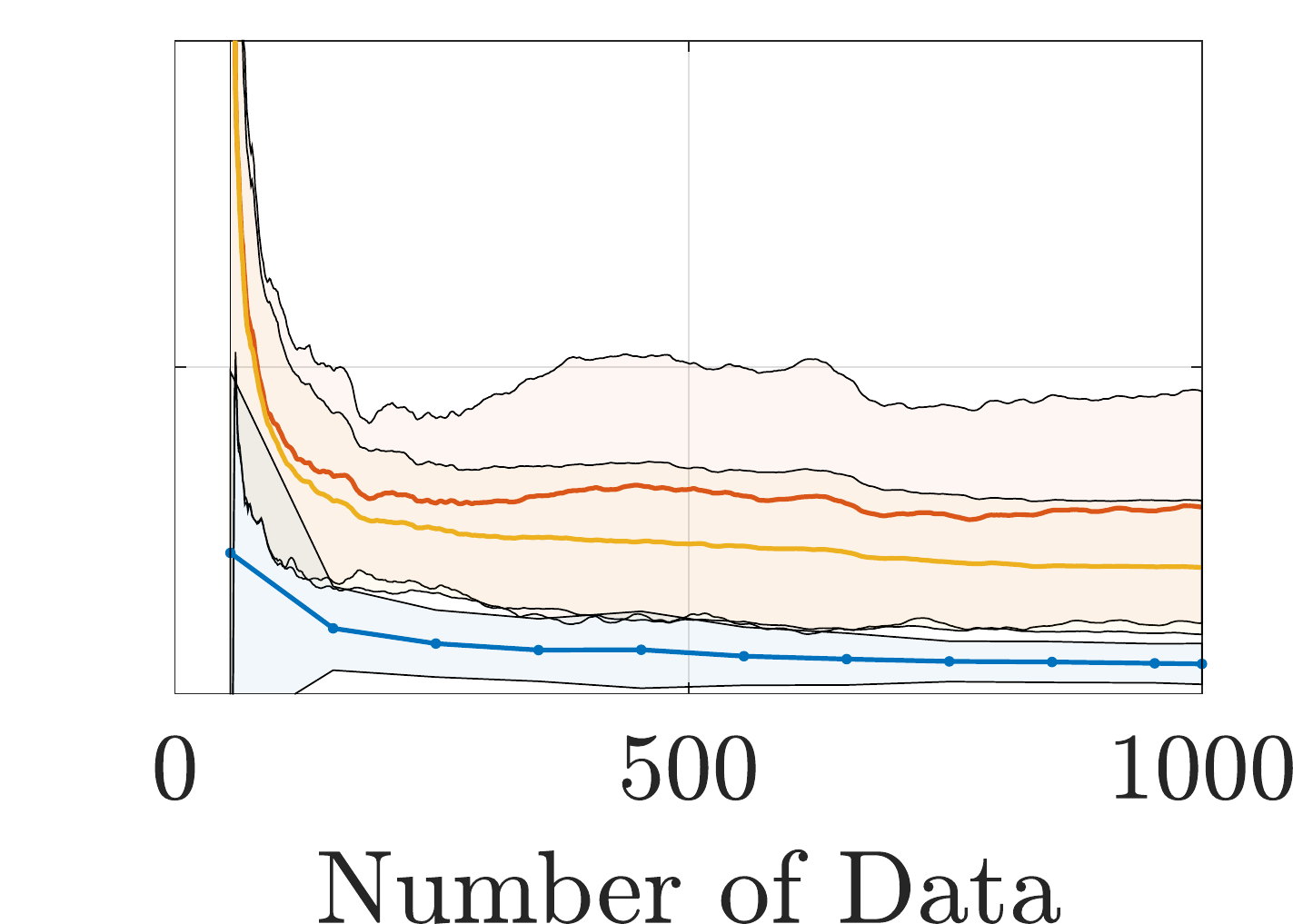}
    \caption{$\sigma_{\xi,\eta}=0.50$,\\108 parameters}
    \label{fig:error_small_05}
  \end{subfigure}%
  \begin{subfigure}{0.32\linewidth}
    \centering
    \includegraphics[trim=95 250 107 249, clip,width=\linewidth]{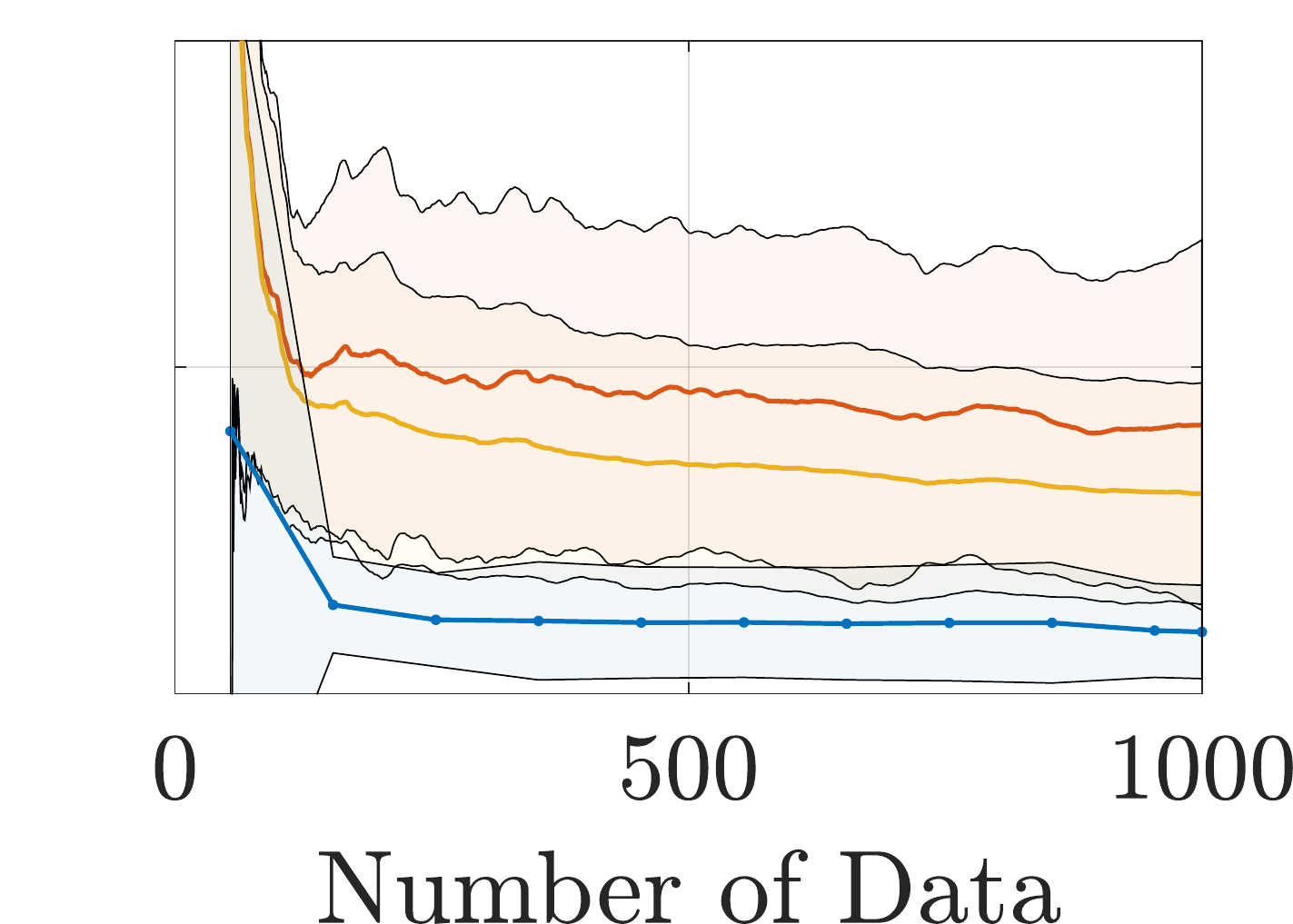}
    \caption{$\sigma_{\xi,\eta}=1.0$,\\108 parameters}
    \label{fig:error_small_1}
  \end{subfigure}%
  \hfill
  \begin{subfigure}{0.32\linewidth}
    \centering
    \includegraphics[trim=95 250 107 249, clip,width=\linewidth]{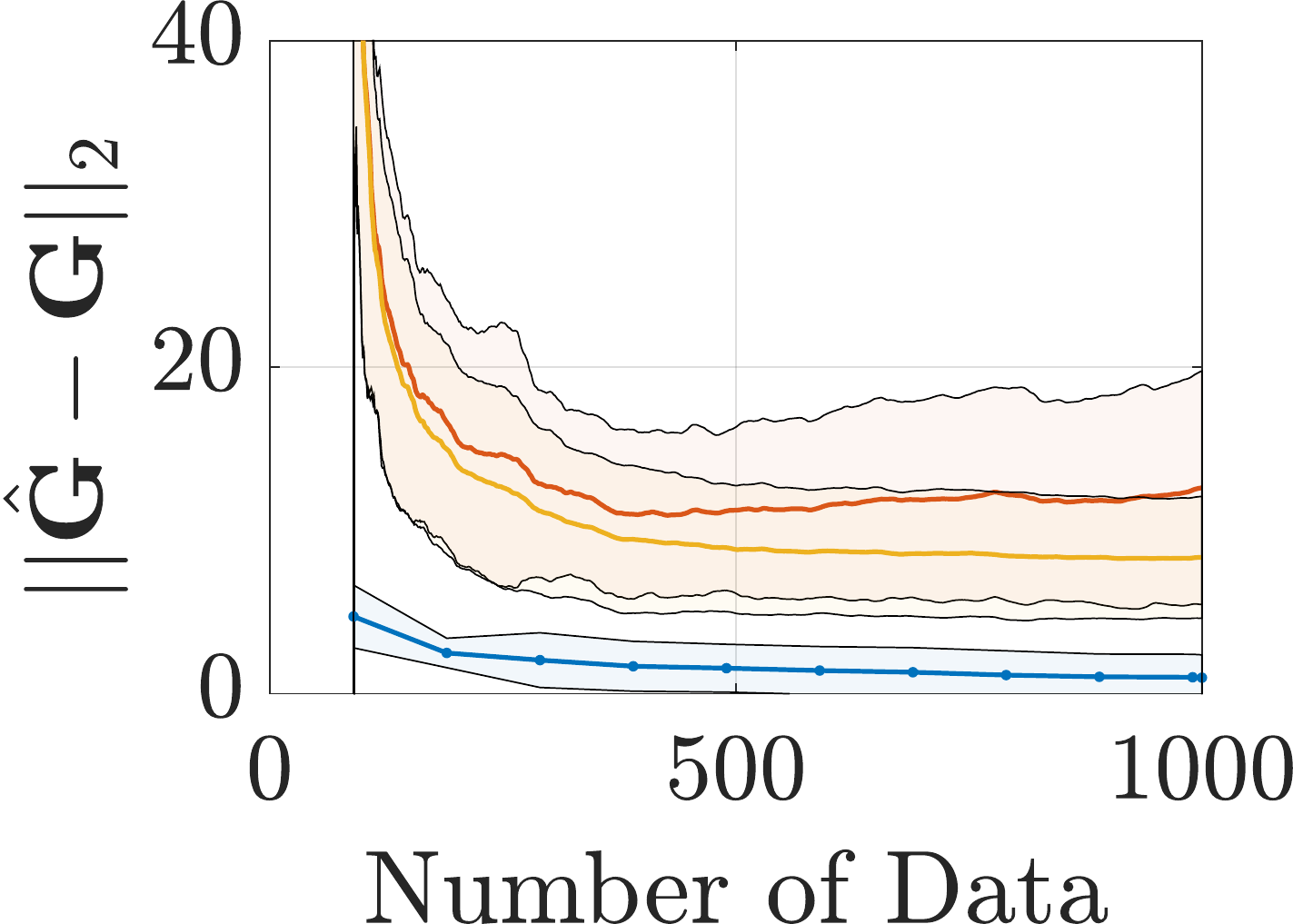}
    \caption{$\sigma_{\xi,\eta}=0.25$,\\720 parameters}
    \label{fig:error_large_025}
  \end{subfigure}%
  \begin{subfigure}{0.32\linewidth}
    \centering
    \includegraphics[trim=95 250 107 249, clip,width=\linewidth]{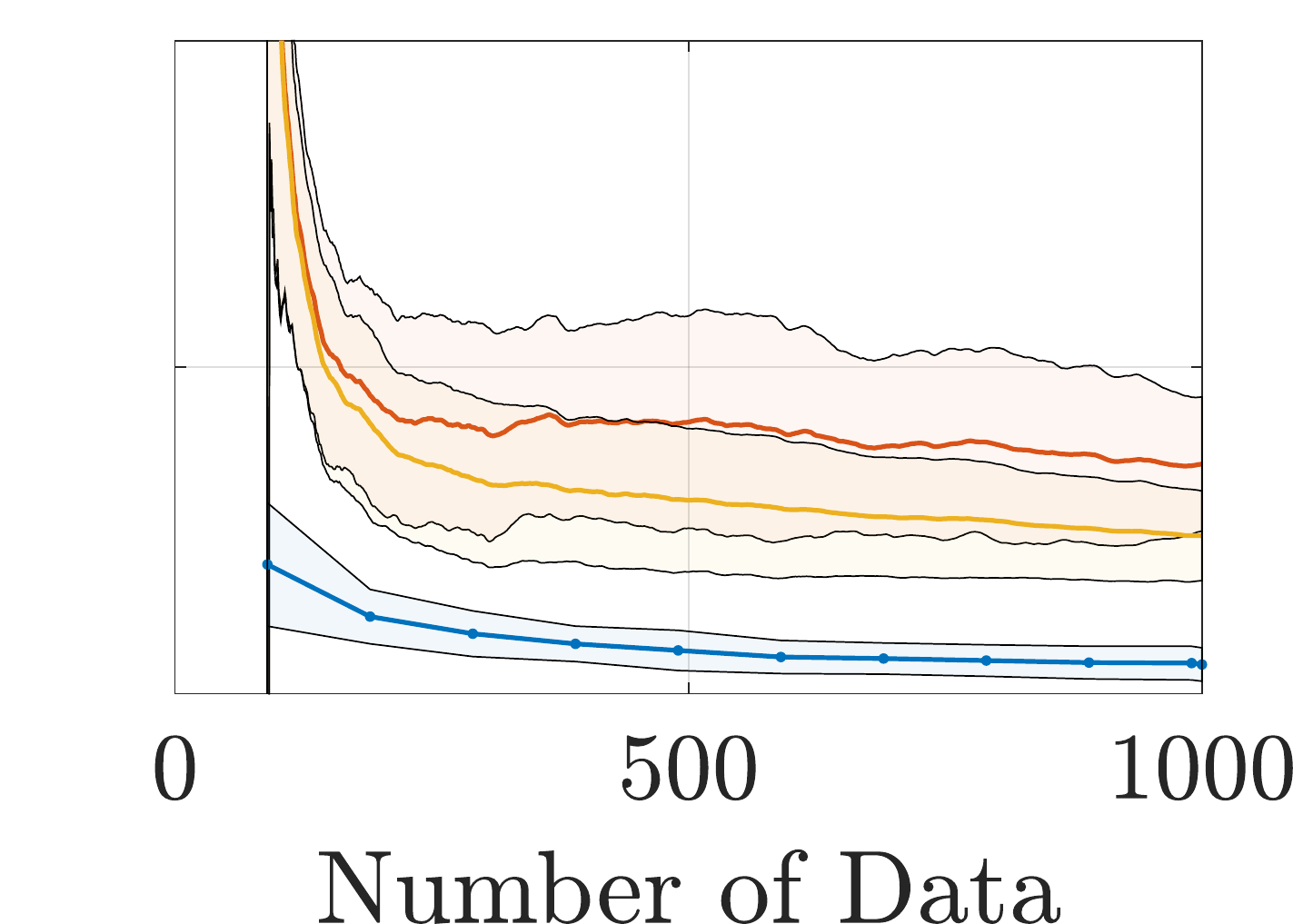}
    \caption{$\sigma_{\xi,\eta}=0.50$,\\720 parameters}
    \label{fig:error_large_05}
  \end{subfigure}%
  \begin{subfigure}{0.32\linewidth}
    \centering
    \includegraphics[trim=95 250 107 249, clip,width=\linewidth]{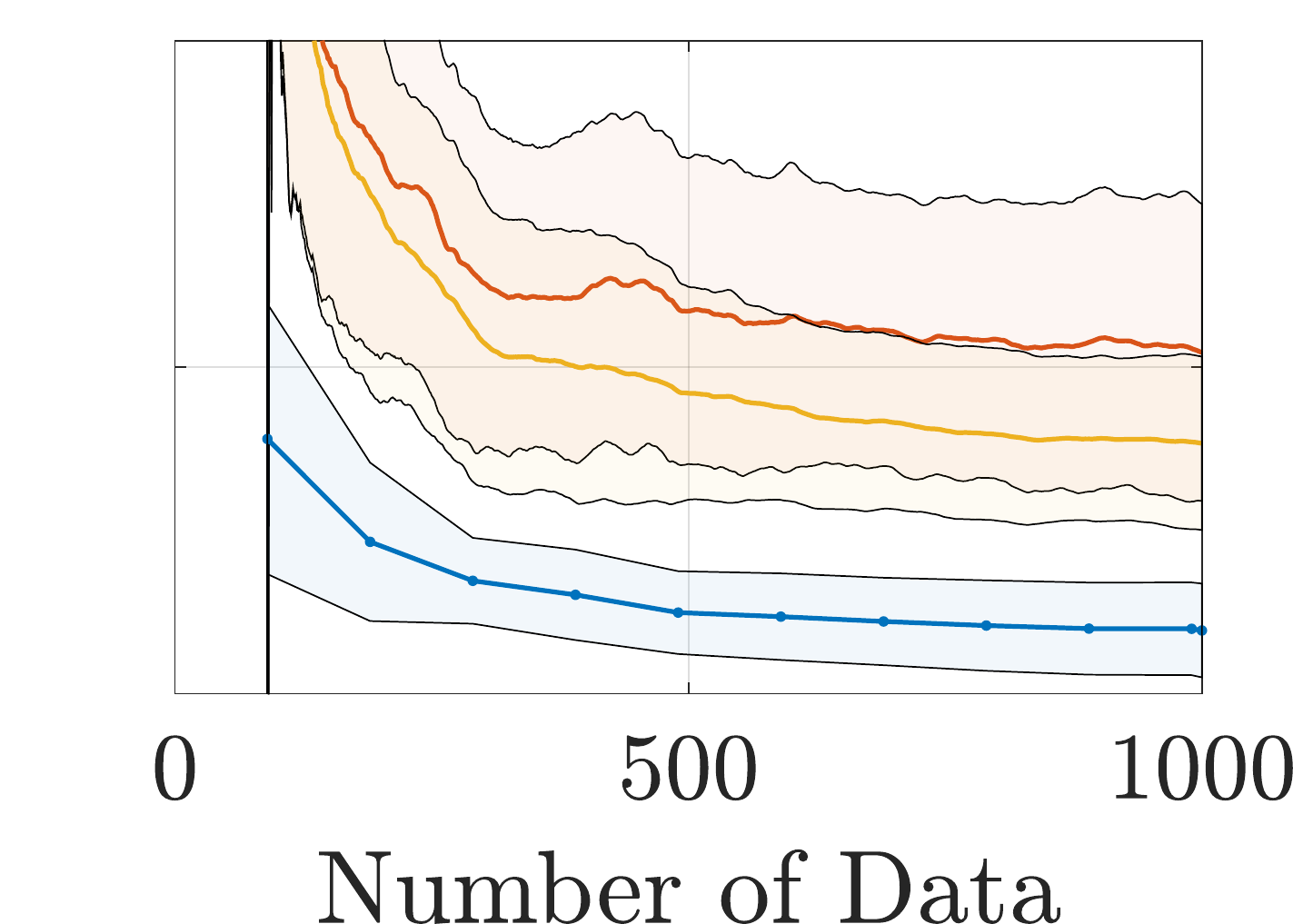}
    \caption{$\sigma_{\xi,\eta}=1.0$,\\720 parameters}
    \label{fig:error_large_1}
  \end{subfigure}%
  \caption{A comparison of the spectral norm of the Markov parameters estimation error for $\bar{n}=18$ using the LS, GLS, and MAP estimates at varying noise levels. The lines represent mean error values and the shaded regions represent plus-minus one standard deviation.}
  \label{fig:markov}
\end{figure}

To assess the accuracy of the estimate, the spectral norm of the error in the Markov parameters $\lVert\hat{\mG}_{0:\bar{n}-1}-\mG_{0:\bar{n}-1}\rVert_2$ was evaluated. This experiment was repeated 50 times, and the top row of Fig.~\ref{fig:markov} shows the average error norm plotted as a solid line with a shaded region representing plus-minus one standard deviation against the number of data used in each estimate. The figure also compares LS, GLS, and the Bayesian MAP estimates at various noise levels $\sigma_{\xi,\eta}=1/4,1/2,1$. The weighting used by the GLS resulted in the GLS estimate having lower error mean and variance at all noise levels than the LS estimate. The MAP estimate yielded the lowest mean error and lowest error variance of all by a significant margin. The same experiment was repeated with larger dimensions of $\dimy=8$, $\dimx=10$, and $\dimu=5$ for a total 720 parameters, and the results are shown in the bottom row of Fig.~\ref{fig:markov}. Again, the ranking of the performance of the estimates is the same, but in the larger system, the improvement yielded by the MAP estimate is even greater as evidenced by the wider gap between the shaded regions of the MAP and GLS estimates. We also observe very little degradation of the MAP estimate when the system dimensions increase other than in convergence rate. The slower convergence can be attributed to the greater number of parameters. The results of these experiments illustrate the performance costs incurred by adding simplifying assumptions into the objective.

\subsection{Estimation of nonlinear state-space models}\label{sec:ms}
We next turn to nonlinear system identification, where the dynamics are parameterized via nonlinear mappings such as neural networks or nonlinear PDEs.

In nonlinear system ID, there are certain behaviors that the models and/or underlying system can exhibit that make many LS objectives unsuitable for estimation. For example, many nonlinear models have the potential for their states to go to infinite values in finite time depending on the model parameters. Therefore, if the objective function requires the model to be simulated over a relatively long period of time, the optimizer will likely run into this diverging behavior for many parameter values. When the states diverge, the objective function value and gradients are undefined, which makes optimization especially difficult.

Another example of nonlinear behavior that makes system ID challenging is chaos. In a chaotic system, arbitrarily small perturbations to the state grow exponentially over time. Thus, even small errors in the model's simulated state can quickly lead to large errors, making it difficult for the objective function to discern good models. Mechanically, these issues lead to objective functions that are multi-modal and difficult to optimize. To address these issues, an MS objective~\cite{bock1981} has been developed. MS introduces simulation length as a design parameter as a means to avoid the excessive error growth common in nonlinear systems. The MS objective has been used in system ID~\cite{voss2004,forgione2021}, state estimation~\cite{alessandri2008}, and control~\cite{giftthaler2018,chen2020} contexts.

\subsubsection{Discussion on the common least squares objective}
Before introducing the MS objective, we first provide a discussion on two fundamental objectives that will motivate the design of the MS objective. An objective function $\mathcal{J}$ in system ID is typically defined as the mean squared difference between the observed output $\vy$ and an estimated output $\hat{\vy}$
\begin{equation}
  \mathcal{J}(\vtheta) = \sum_{k=0}^{\numObs}\lVert\vy_k - \hat{\vy}_k(\vtheta)\rVert_{2}^2,
\end{equation}
possibly paired with a regularization/physics-enforcing term~\cite{pillonetto2016,raissi2019}. Regularization aside, the differences among objective functions typically come from how the estimated output $\hat{\vy}$ is evaluated. For expositional purposes, let us define a function $f(\vx_i,\vu_{i:j},t_j;\vtheta)$ parameterized by $\vtheta$ that maps an initial state $\vx_i$ to the output at time $t_j$ with inputs from time $t_i$ to time $t_j$ denoted as $\vu_{i:j}$, where $i\leq j$. One fundamental objective uses a single initial condition $\vx_0$ and evaluates $\hat{\vy}$ as
\begin{equation}\label{eq:det_ls}
  \hat{\vy}_k = f(\vx_0,\vu_{0:k},t_k;\vtheta),
\end{equation}
for all $k\geq0$. Since the system is simulated without any corrections to the trajectory, we refer to this objective as the deterministic LS. This objective is used to train popular machine learning models such as ODE-nets~\cite{chen2018}, PDE-nets~\cite{long2018}, and universal differential equations (UDE)~\cite{rackauckas2020}. The other fundamental objective estimates the output as
\begin{equation}\label{eq:prop_ls}
  \hat{\vy}_k = f(\vy_{k-1},\vu_{k-1:k},t_k;\vtheta),
\end{equation}
using the most recent data point as the initial condition and simulating the system only until the time at which the next data point is available. We refer to this objective as the propagator LS since in this case, $f$ propagates one data point to the next. The propagator objective is used within algorithms such as dynamic mode decomposition (DMD)~\cite{schmid2010} and sparse identification of nonlinear dynamics (SINDy)~\cite{brunton2016} and for training certain neural networks such as Hamiltonian neural networks~\cite{greydanus2019}. We reinforce the difference between Eq.~\eqref{eq:det_ls} and \eqref{eq:prop_ls} is that Eq.~\eqref{eq:det_ls} uses $\vx_0$ as the initial condition and returns the output after $k$ timesteps whereas Eq.~\eqref{eq:prop_ls} uses $\vy_{k-1}$ as the initial condition and returns the output after only one timestep. Models learned by the deterministic and propagator LS objectives are sometimes known as simulation and prediction models, respectively~\cite{schoukens2019}.

There have been several comparisons between the deterministic and propagator LS objectives~\cite{de2002,aguirre2010,galioto2020,ribeiro2020} that have shown that the deterministic LS yields better estimates when measurement noise is included and process noise is omitted, and the propagator LS yields better estimates when process noise is included and measurement noise is omitted. In addition, the deterministic LS is much more difficult and computationally intensive to optimize since it involves compositions of the dynamics propagator, which allows errors to accumulate and leads to complicated, non-convex surfaces. The differences in the performance of these objectives are primarily caused by the length of un-interrupted simulation, i.e., the value $j-i$. Longer simulation lengths lead to greater error accumulation, but shorter simulation lengths can introduce bias when the data are noisy. Deterministic and propagator LS objectives use the extremal values of possible simulation lengths. The idea of MS is to use an intermediate simulation length, $1<j-i<\numObs$, to mitigate the issues experienced by these other two objectives.

\subsubsection{Multiple shooting objective}
Here, the MS objective is introduced. We then show that the deterministic~\eqref{eq:det_ls} and propagator LS~\eqref{eq:prop_ls} objectives can be seen as special cases of MS. In MS, the output trajectory is divided into $L$ disjoint subtrajectories with initial times $\{t_{\ell_i}\}_{i=1}^{L}$ such that the $i$th subtrajectory has length $\dell_i\coloneqq \ell_{i+1}-\ell_i$. Then the output at time $t_k$ contained within the $i$th subtrajectory is estimated as $\hat{\vy}_k = f(\vx_{\ell_i},\vu_{\ell_i:k},t_k; \vtheta)$. Such an objective function requires the estimation of the set of subtrajectory initial conditions $\mathcal{Z}_L\coloneqq\{\vx_{\ell_i}\}_{i=1}^{L}$, which can be done by adding the initial conditions as parameters~\cite{ribeiro2020}, training an encoder~\cite{masti2021}, or, if the system is fully observed, simply using the data $\vx_{\ell_i}=\vy_{\ell_i}$~\cite{zhong2019}. In effect, this method introduces additional parameters (the initial conditions) as the cost for an improved estimate.

The MS objective function is defined as
\begin{equation}\label{eq:ms}
  \mathcal{J}(\vtheta) = \sum_{i=1}^{L}\sum_{k=\ell_i}^{\ell_{i+1}-1}\lVert \vy_k - \hat{\vy}_k \rVert_{2}^2,
\end{equation}
where $\ell_{L+1}\coloneqq\numObs+1$. Oftentimes, a constant length of $T = \dell_{i}$ for all $i=1,\ldots,L$ is used for simplicity. A model learned with such an objective is sometimes called a $T$-step-ahead predictor~\cite{schoukens2019}. In the case $T=\numObs$, the objective is equivalent to the deterministic LS, and if $T=1$, the objective is equivalent to the propagator LS. Therefore, when $1<T<\numObs$, MS can be seen as a type of combination of these two objectives. In the original paper~\cite{bock1984}, the objective additionally had the constraint that $\vx_{\ell_{i+1}}=\Psi^{\Delta\ell_i}(\vx_{i},\thetdyn)$, where $\Psi^{\Delta\ell_i}$ denotes $\Delta\ell_i$ compositions of the $\Psi$ operator, but this constraint is sometimes removed to simplify optimization. We will distinguish between the objectives with and without the constraints by referring to them as the constrained and unconstrained MS objectives, respectively.

\subsubsection{Relation to probabilistic approach}
From a probabilistic perspective, the MS objective amounts to a joint parameter-state estimation problem. Rather than estimating the state at every timestep, however, only the subset of subtrajectory initial conditions $\mathcal{Z}_L\subseteq\mathcal{X}_{\numObs}$ are estimated. The posterior of such a problem can be factorized with Bayes' rule as
\begin{equation}\label{eq:joint_post}
    \probd(\mathcal{Z}_L,\vtheta | \yn)\propto \like(\vtheta,\mathcal{Z}_L;\yn)\probd(\mathcal{Z}_L,\vtheta).
\end{equation}
The likelihood and prior are defined as
\begin{subequations}
\begin{align}
  \like(\vtheta,\mathcal{Z}_L;\yn) &= \prod_{k=\ell_i}^{\ell_{i+1}-1}\probd(\vy_{k} | \vx_{\ell_i},\vtheta),\\
  \probd(\mathcal{Z}_L,\vtheta) &= \probd(\vtheta)\prod_{i=1}^{L}\probd(\vx_{\ell_i}|\vx_{\ell_{i-1}},\vtheta),
\end{align}
\end{subequations}
where $\probd(\vx_{\ell_1}|\vx_{\ell_0},\vtheta)\coloneqq\probd(\vx_{\ell_1}|\vtheta)$. Each term in the likelihood $\probd(\vy_k|\vx_{\ell_i},\vtheta)$ can still be evaulated with Algorithm~\ref{alg:marg_like} using data from only a single trajectory. The most significant difference is that the prior has the added terms $\probd(\vx_{\ell_i}|\vx_{\ell_{i-1}},\vtheta)$, which can be evaluated as
\begin{equation}\label{eq:prior}
  \begin{split}
    \probd(\vx_{\ell_i}|\vx_{\ell_{i-1}},\vtheta) = \int&\prod_{k=\ell_{i-1}+1}^{\ell_i}\probd(\vx_{k}|\vx_{k-1})\\
    &\times\rmd\vx_{\ell_i-1}\rmd\vx_{\ell_i-2}\ldots \rmd\vx_{\ell_{i-1}+1}.
  \end{split}
\end{equation}
From Eq.~\eqref{eq:prior}, it can be seen that $\probd(\vx_{\ell_i}|\vx_{\ell_{i-1}},\vtheta)$ represents the probability of $\vx_{\ell_i}$ averaged over all trajectories that start at $\vx_{\ell_{i-1}}$ with dynamics determined by $\vtheta$. Another interpretation of this term is as a soft constraint enforcing the estimated initial conditions to be connected by a trajectory determined by the proposed dynamics with initial condition $\vx_{\ell_1}$. Under certain conditions, this constraint is equivalent to the MS constraints. Furthermore, estimators based off of the posterior~\eqref{eq:joint_post} are equivalent to estimators using the constrained/unconstrained MS objectives~\eqref{eq:ms} under certain assumptions. This result is stated in Proposition~\ref{prop:ms}.
\begin{prop}\label{prop:ms}
  Assume an improper uniform prior distribution on the parameters $\vtheta$ and that the process noise $\mSigma$ is zero. Then, the negative log marginal likelihood of the joint parameter-state estimation problem of Eq.~\eqref{eq:joint_post} is equivalent to the unconstrained MS objective~\eqref{eq:ms}. Moreover, the negative log posterior is equivalent to the constrained MS objective.
\end{prop}
\begin{pf}
  According to Theorem 2 in~\cite{galioto2020}, each term in the marginal likelihood $\prod_{k=\ell_i}^{\ell_{i+1}-1}\probd(\vy_{k} | \vx_{\ell_i},\vtheta)$ is equivalent to a deterministic LS objective when there is no process noise. Then, taking the negative log of this product gives the unconstrained MS objective. When the prior over the states is added, each term $\probd(\vx_{\ell_i}|\vx_{\ell_{i-1}},\vtheta)$ approaches the Dirac delta function $\delta_{\Psi^{\Delta_{\ell_i}}(\vx_{\ell_i-1},\thetdyn)}(\vx_{\ell_i})$ as the process noise goes to zero. These Dirac delta functions are equivalent to the constraints in the constrained MS objective. The prior over the parameters is constant, so taking the negative log of the posterior yields the constrained MS objective, completing the proof.\qed
\end{pf}

Depending on the proposed dynamics, the prior term $\probd(\vx_{\ell_i}|\vx_{\ell_i-1},\vtheta)$ can either substantially improve the optimization landscape or hardly affect the posterior at all. As an example, consider an LTI system where the distribution $\probd(\vx_{\ell_i}|\vx_{\ell_i-1},\vtheta)$ is Gaussian with mean $\vmu=\sum_{j=1}^{\dell_{i-1}}\mA^{j-1}\mB\vu_{\ell_{i}-j}+\mA^{\dell_{i-1}}\vx_{\ell_{i-1}}$ and covariance $\mP=\sum_{j=1}^{\dell_{i-1}}\mA^{j-1}\mSigma(\mA^{j-1})^*$. If the system has an attractor, i.e., $\rho(\mA)<1$, then the variance term will shrink over time, and the probability density will be concentrated closely around the mean.  When probability density is more concentrated, the optimization landscape is steeper, theoretically speeding up optimization. If $\rho(\mA)\geq1$, then the variance will grow over time and the probability density will be more spread out, adding little effect to the optimization landscape.

\subsubsection{Comparison of smoothing effects}\label{sec:smooth}
Now that it is understood that the MS objective is equivalent to an objective of a joint parameter-state estimation problem, we demonstrate empirically that the additional expense of inferring the subtrajectory initial conditions is not computationally necessary. More specifically, we demonstrate that the proposed marginal likelihood improves the optimization surface in a similar fashion to the MS objective, and we provide a brief discussion on the advantages of the proposed approach.

First, let us relate the proposed approach to recent theoretical findings. In~\cite{ribeiro2020}, it was proven that the more often the state of the dynamics is readjusted to prevent divergence of the trajectory, the smoother the LS objective function becomes. Within MS, this readjustment comes from the estimation of the state at different times along the trajectory. In the marginal likelihood, state adjustment comes in the form of the update step of a Bayesian filter. Previous work~\cite{galioto2020} found that the update step is a sufficient readjustment to induce improved smoothness of the posterior. Now, we provide a comparison to MS.

To show the similarity of the smoothing effects in MS and the marginal likelihood, the logistic map example from~\cite{ribeiro2020} is considered. The logistic map is a difference equation defined as
\begin{equation}
  y_{k+1} = \theta y_{k}(1-y_k).
\end{equation}
This system exhibits chaotic behavior when the parameter $\theta$ is within the range [3.57, 4]. For this example, $\theta$ was set to 3.78, an initial condition of $y_0=0.5$ was used, and 200 noiseless data points were collected. To show how the objective surfaces vary with $\theta$, all other parameters must be fixed. For the MS objective, the initial conditions were all set to the true values, and for the marginal likelihood, the measurement noise variance was set to a small value of $\mGamma=10^{-16}$ to maintain positive definiteness. Then, the objectives were compared at different time horizons $T$ and process noise to measurement noise variance ratios $\mSigma/\mGamma$. A ratio is used for the variance values because in this problem, the shape of the objective did not appear to change for a fixed ratio, regardless of the $\mSigma$ and $\mGamma$ values. Note that the validity of using this ratio is only possible since the observation function is the identity, meaning that $\mSigma$ and $\mGamma$ are represented in the same coordinate frame.

\begin{figure}
  \centering
  \begin{subfigure}{0.45\linewidth}
    \centering
    \includegraphics[trim=95 250 125 249, clip,width=\linewidth]{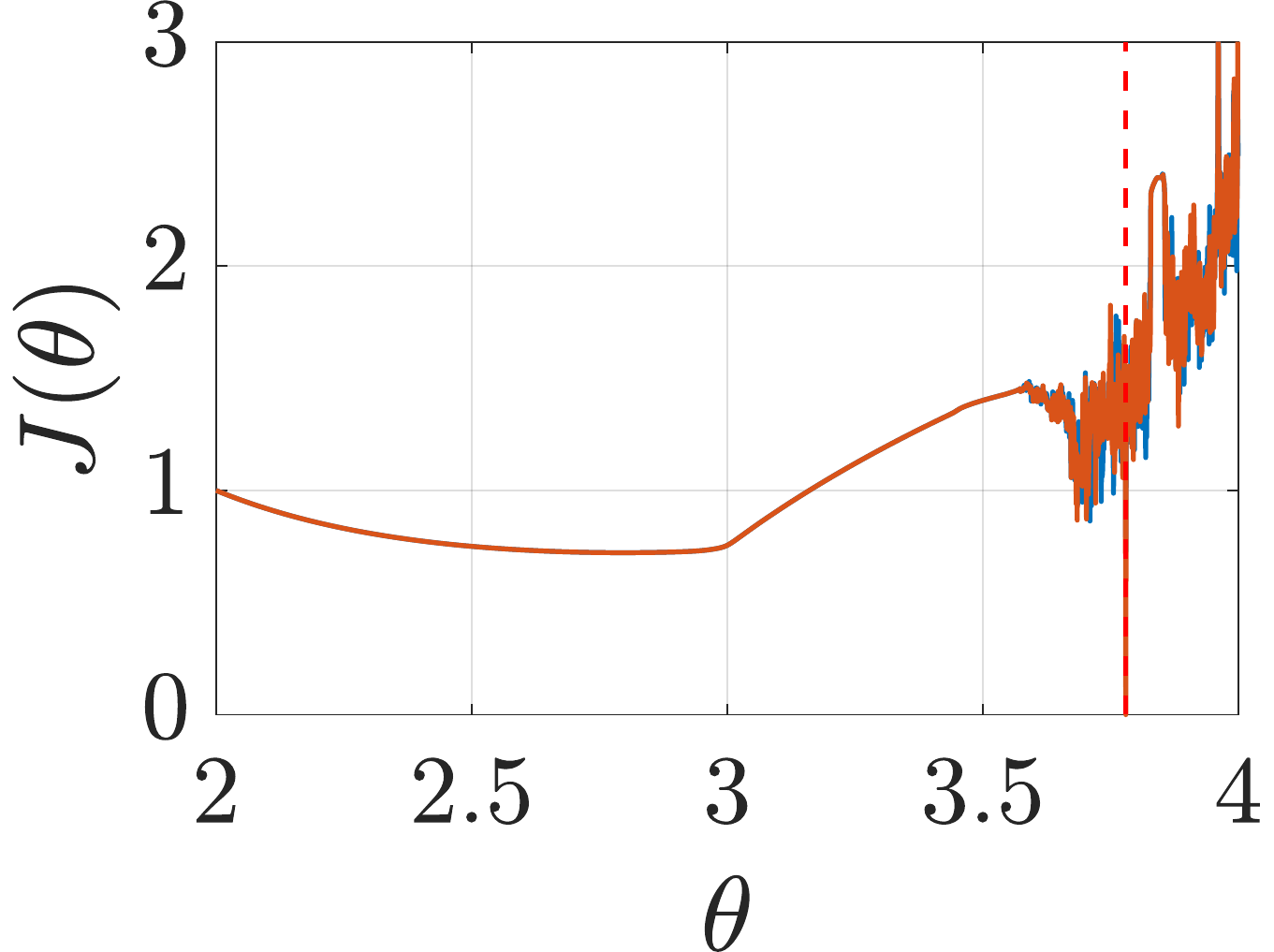}
    \caption{$T=\numObs$, $\mSigma/\mGamma=10^{-10}$}
    \label{fig:objSingle}
  \end{subfigure}%
  \begin{subfigure}{0.45\linewidth}
    \centering
    \includegraphics[trim=95 250 125 249, clip,width=\linewidth]{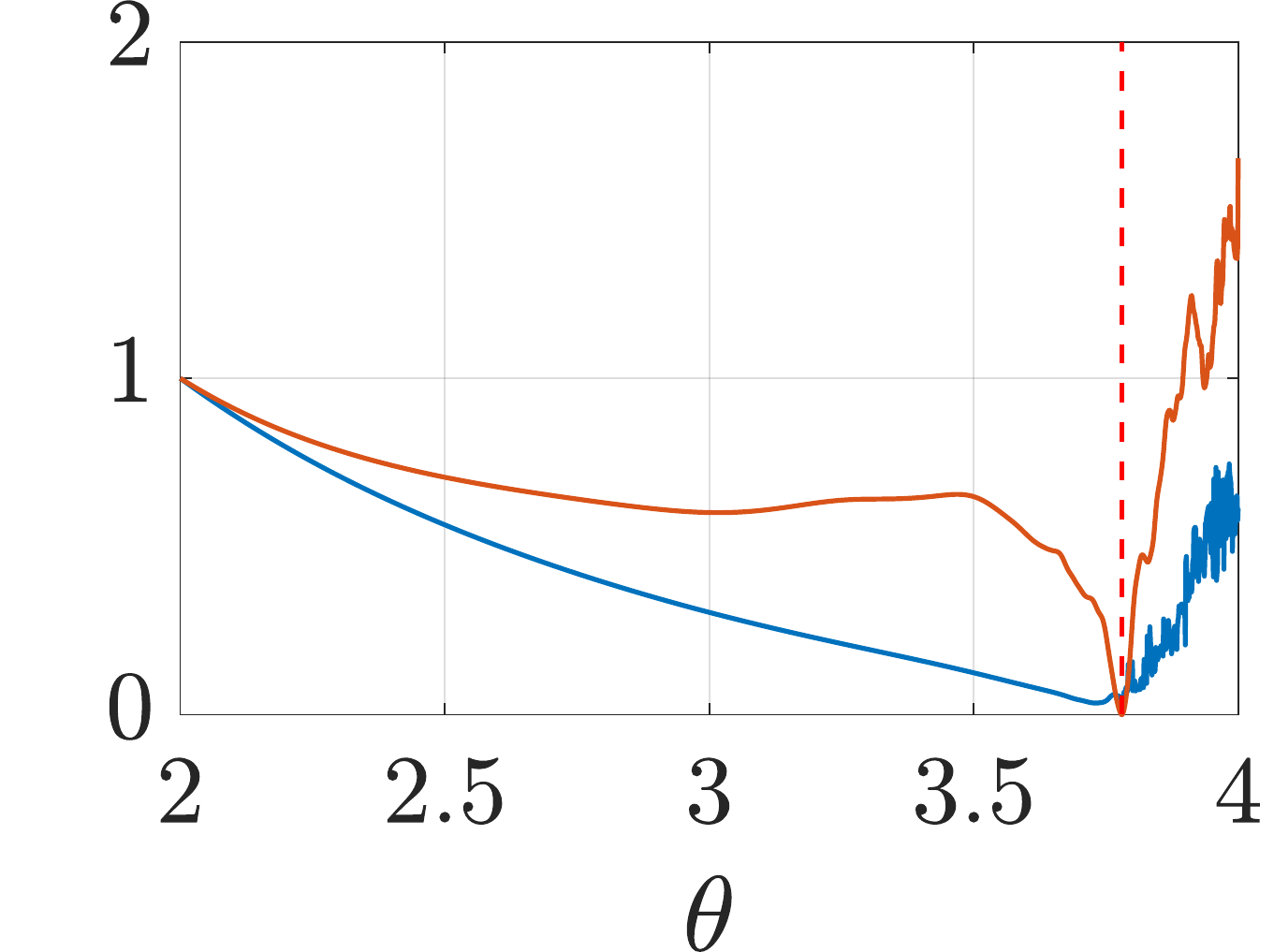}
    \caption{$T=10$, $\mSigma/\mGamma=0.5$}
    \label{fig:obj10}
  \end{subfigure}%
  \hfill
  \begin{subfigure}{0.45\linewidth}
    \centering
    \includegraphics[trim=95 250 125 249, clip,width=\linewidth]{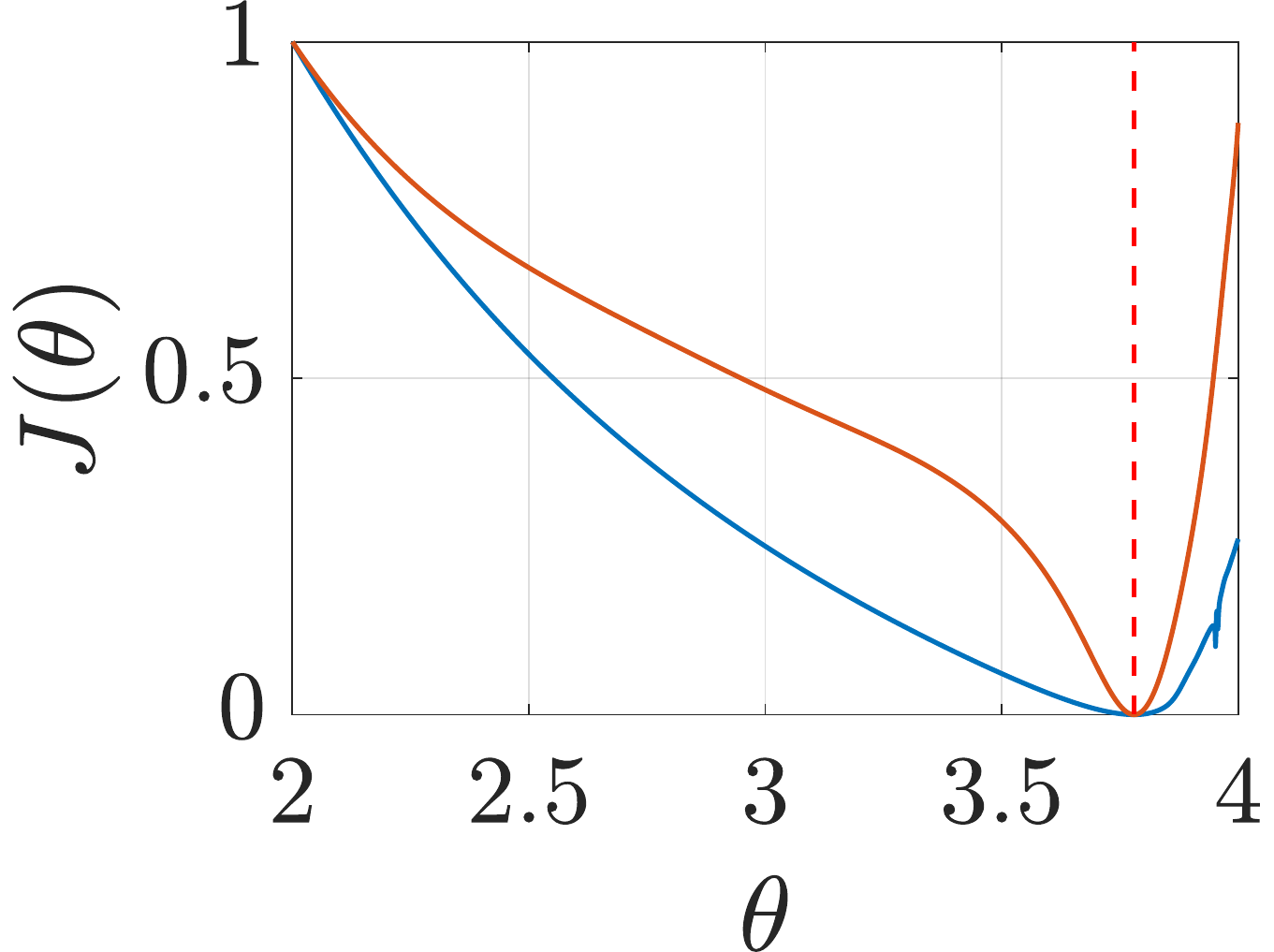}
    \caption{$T=5$, $\mSigma/\mGamma=0.7$}
    \label{fig:obj5}
  \end{subfigure}%
  \begin{subfigure}{0.45\linewidth}
    \centering
    \includegraphics[trim=95 250 125 249, clip,width=\linewidth]{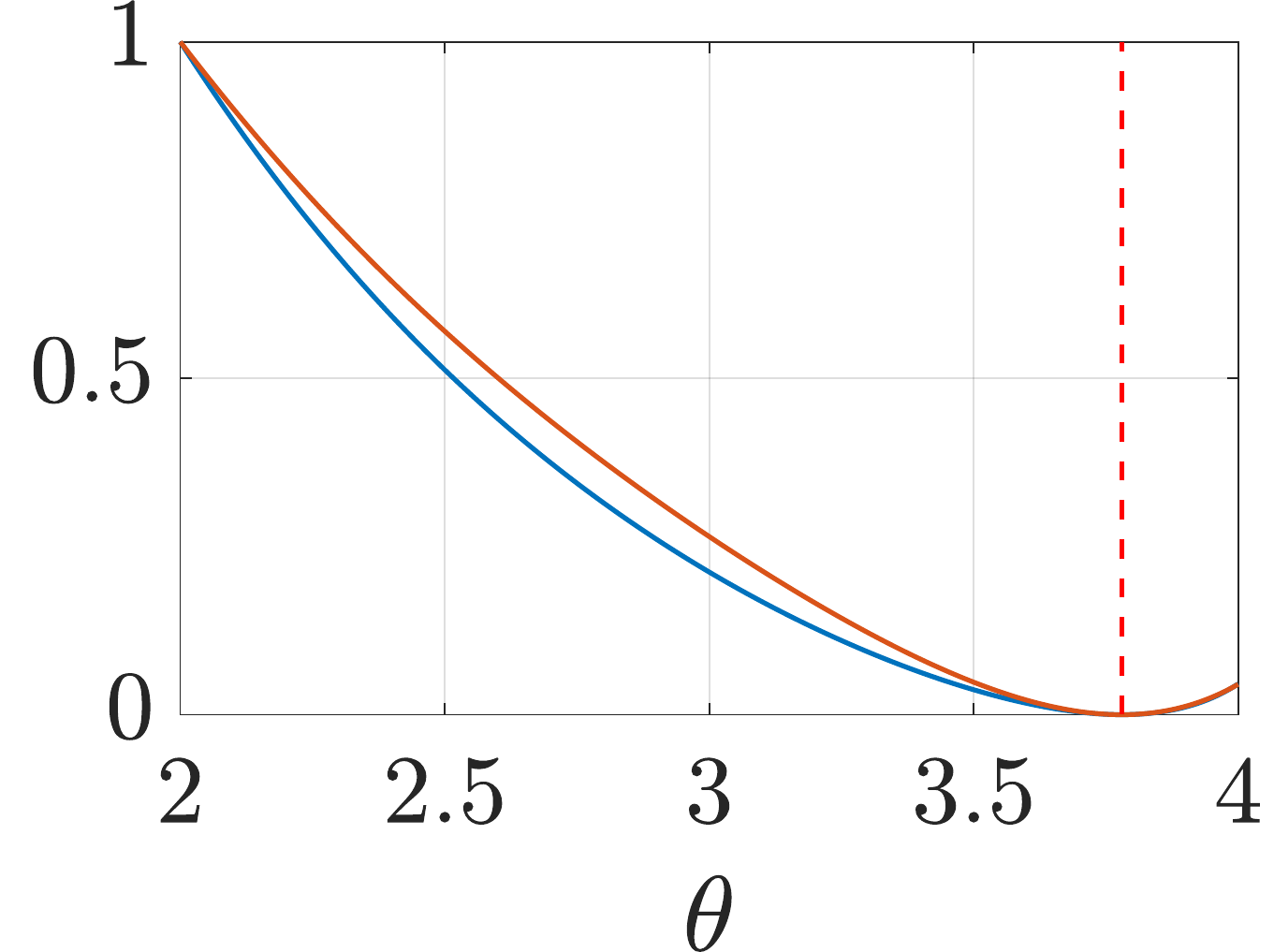}
    \caption{$T=2$, $\mSigma/\mGamma=1.0$}
    \label{fig:obj2}
  \end{subfigure}%
  \hfill
  \includegraphics[trim=185 385 184 380, clip,width=0.55\linewidth]{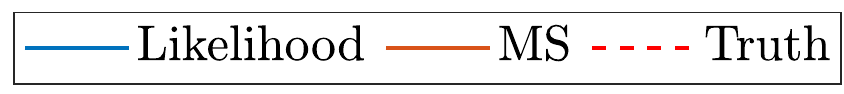}
  \caption{Comparison of the log marginal likelihood and MS objective as $\mSigma$ and $T$, respectively, vary.}
 \label{fig:MS}
\end{figure}

Both surfaces were normalized to equal 1.0 at $\theta=2$, and the results are shown in Fig.~\ref{fig:MS}. There is no exact mapping between the time horizons and the variance ratio, so the values of $\mSigma/\mGamma$ were chosen such that the smoothness of the marginal likelihood roughly matched that of the MS objective by visual comparison. Fig.~\ref{fig:objSingle} shows values of $T$ and $\mSigma/\mGamma$ where MS is equivalent to and the marginal likelihood approximates the deterministic LS. Due to the chaotic nature of this system, the deterministic LS objective is filled with local minima that make optimization extremely difficult. As $T$ is decreased and $\mSigma$ is increased, both surfaces show increasing smoothness, demonstrating the similar effect these variables have on their respective objectives. An important difference to note between $T$ and $\mSigma$ is that $T$ is a discrete scalar variable, whereas $\mSigma$ is a positive definite matrix of continuous values. Therefore, $\mSigma$ gives the user greater flexibility when tuning the marginal likelihood, including the ability to use different variance values for different components of the state.

\section{Numerical experiments}\label{sec:results}

In this section, four numerical experiments are conducted to compare the performance of the Bayesian approach to the LS+ERA~\eqref{eq:subtraj}, deterministic LS~\eqref{eq:det_ls}, and MS~\eqref{eq:ms} objective functions. The first experiment considers a linear pendulum with forcing for various data timesteps and noise levels. This experiment shows that for every timestep and noise level considered, the Bayesian approach can, on average, improve the estimate provided by the LS+ERA algorithm. The second example uses the Wiener-Hammerstein system ID benchmark~\cite{schoukens2009} and shows that the Bayesian approach displays greater robustness than MS when the number of training data is reduced and noise is present in the data. The third example is the forced Duffing oscillator in the chaotic regime and shows that the deterministic LS metric cannot always identify good models in such systems, even when paired with MS. The Bayesian approach learns a model with an attractor very similar to the true system, but is ranked poorly by the LS metric despite how well it captures the system's phase space behavior. The fourth and final example considers the prediction of a PDE quantity of interest and shows that deterministic LS struggles when noise is introduced and there are more model parameters than data, whereas the Bayesian approach can account for both the measurement noise and the expressiveness of the model.

To make predictions with the posterior, we require posterior samples. However, non-identifiability of parameters presents a number of well-known challenges to MCMC sampling. A state-space dynamics model is at best unique up to a change of coordinates transformation, and in overparameterized cases, models are not even unique for a fixed coordinate frame. We attempt to address this issue by fixing the observation parameters $\thetobs$ (and $\mB$ parameters in LTI models) at the MAP and then sampling the remaining parameters. The rationale behind this approach is to constrain the coordinate frame as a means to mitigate one of these sources of non-identifiability. Fixing parameters runs the risk of neglecting uncertainty, but we are primarily concerned with the uncertainty in the output behavior, not in the parameters. This constraint should theoretically not restrict the behavior of the model dynamics nor the uncertainty in the output, so we consider this an acceptable tradeoff.

\begin{figure}
  \centering
  \begin{subfigure}{0.45\linewidth}
    \centering
    \includegraphics[trim=95 244 107 248, clip,width=\linewidth]{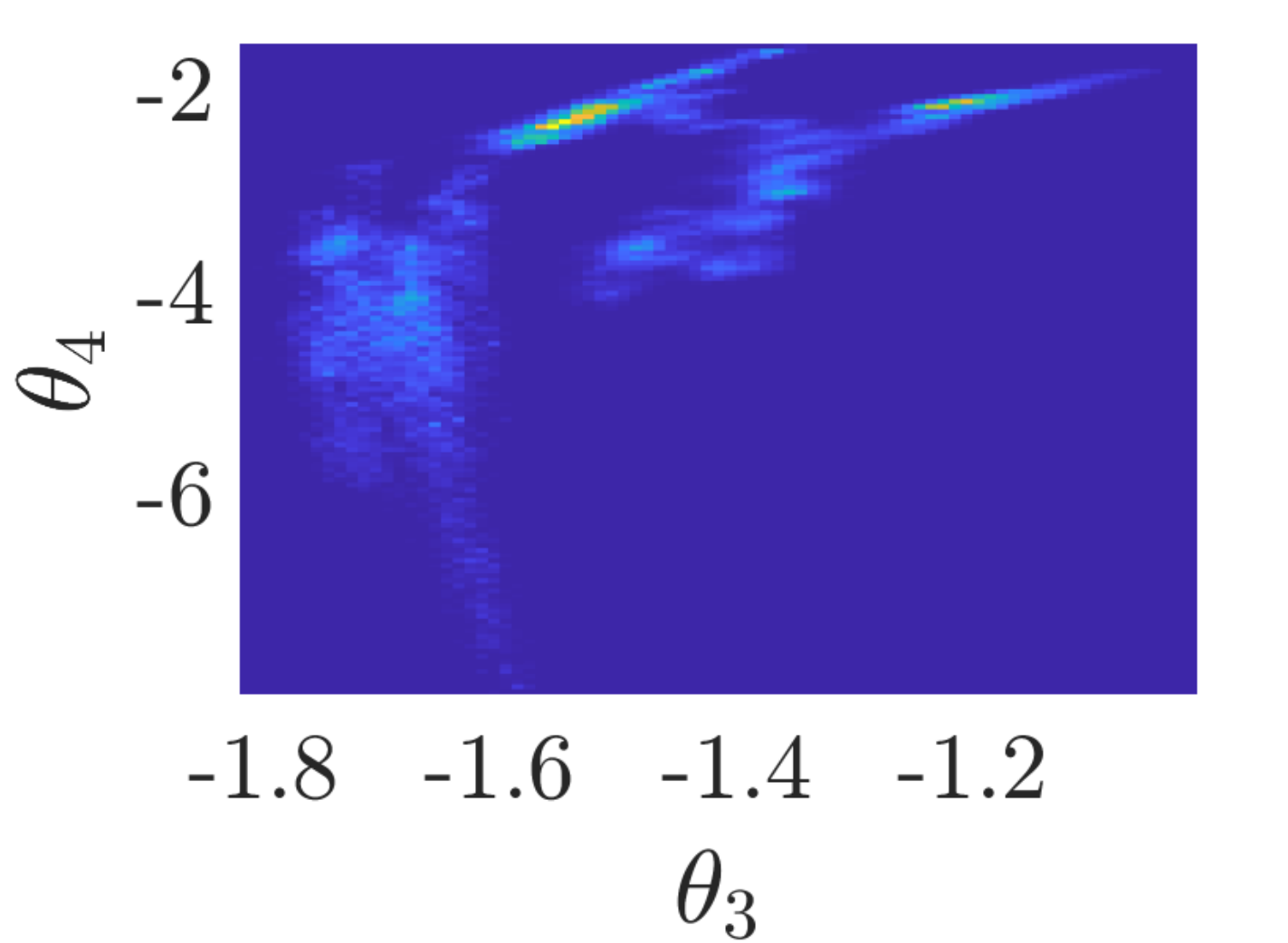}
    \caption{}
    \label{fig:sampling_2d_bad}
  \end{subfigure}%
  \begin{subfigure}{0.45\linewidth}
    \centering
    \includegraphics[trim=95 244 107 248, clip,width=\linewidth]{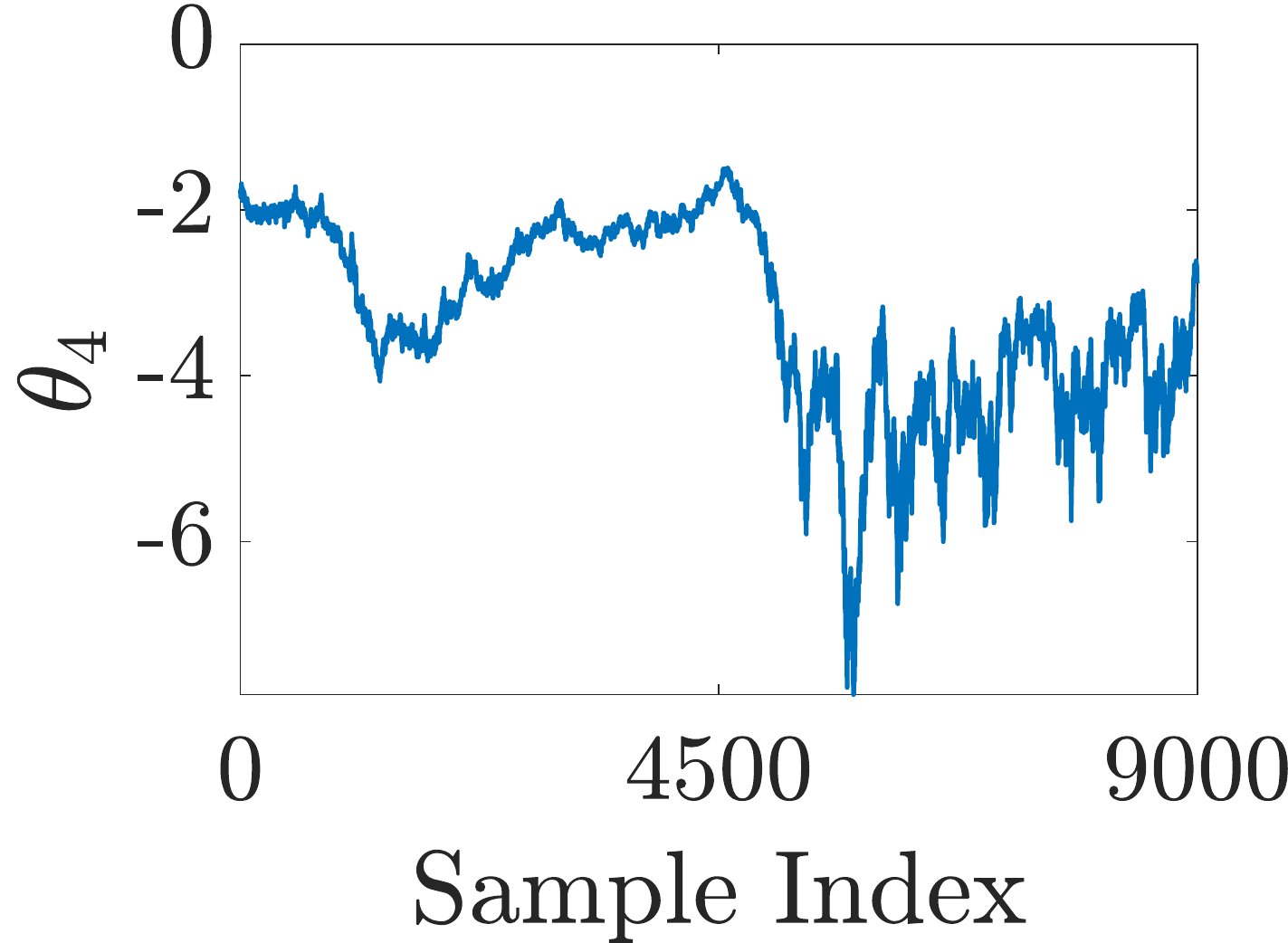}
    \caption{}
    \label{fig:sampling_chain_bad}
  \end{subfigure}%
  \hfill
  \begin{subfigure}{0.45\linewidth}
    \centering
    \includegraphics[trim=95 244 107 248, clip,width=\linewidth]{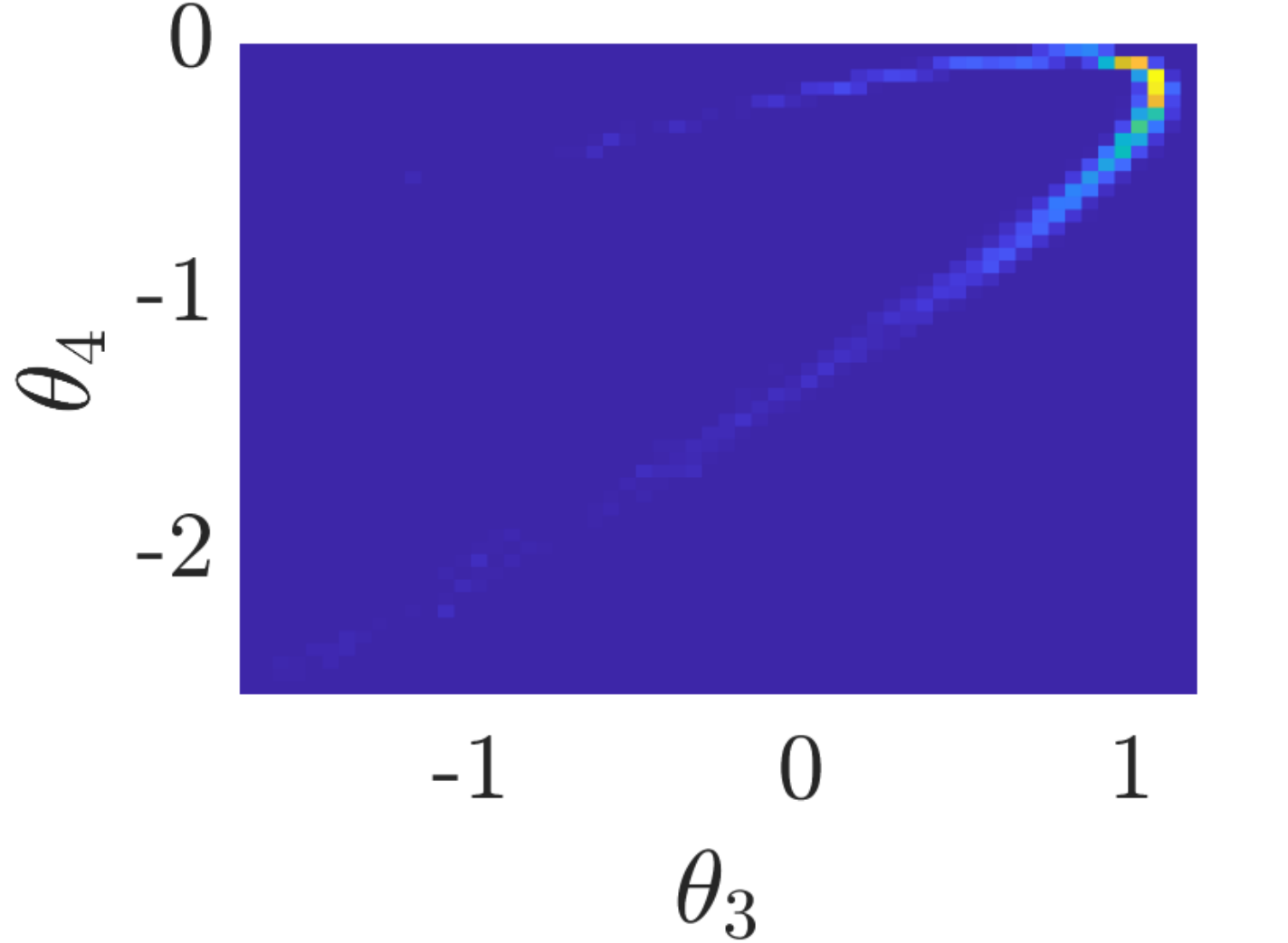}
    \caption{}
    \label{fig:sampling_2d_good}
  \end{subfigure}%
  \begin{subfigure}{0.45\linewidth}
    \centering
    \includegraphics[trim=95 244 107 248, clip,width=\linewidth]{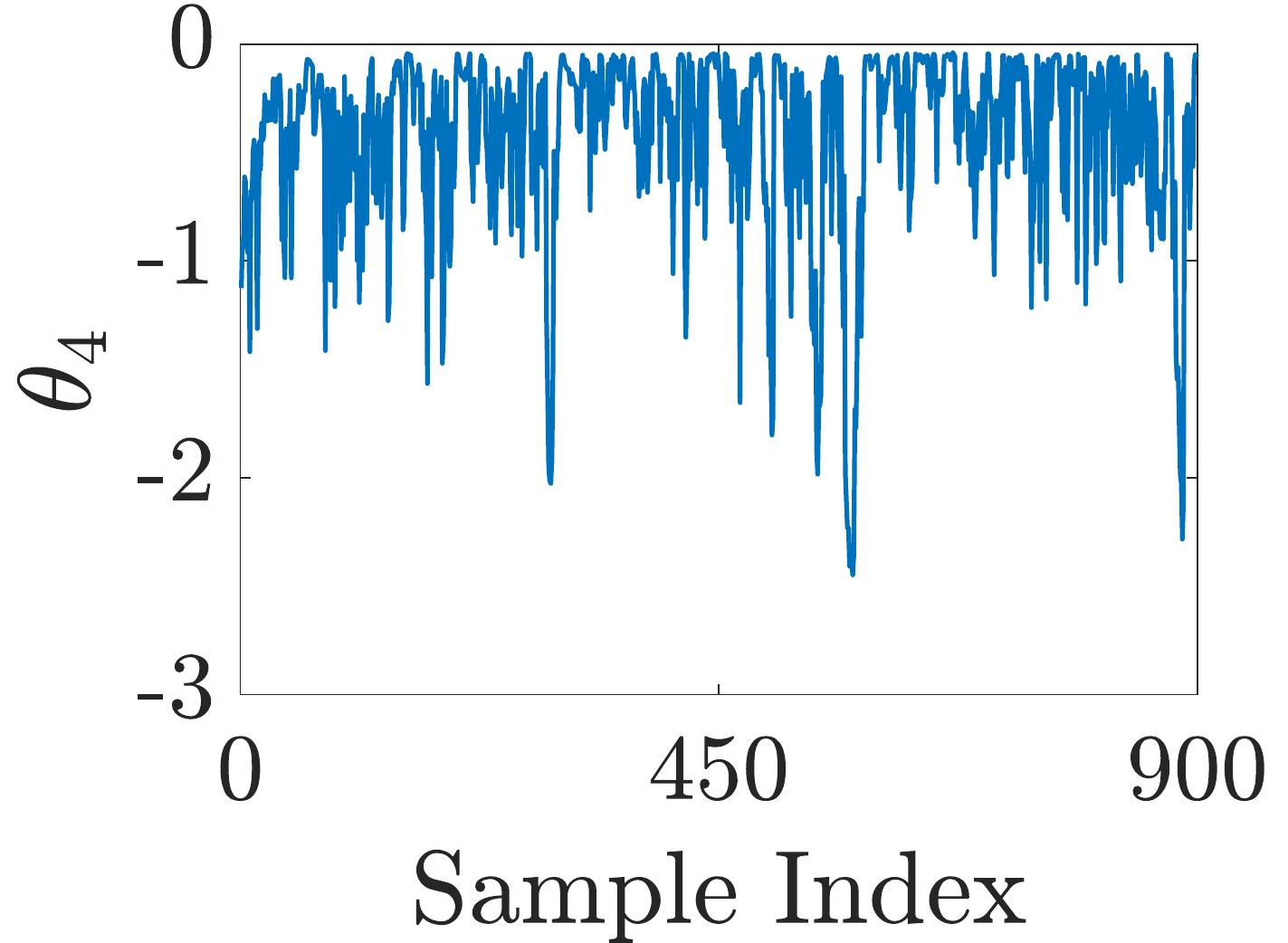}
    \caption{}
    \label{fig:sampling_chain_good}
  \end{subfigure}%
  \caption{2D marginal distributions and chains from MCMC sampling System~\eqref{eq:pend_model}. Figs.~\ref{fig:sampling_2d_bad} and \ref{fig:sampling_chain_bad} show samples drawn using DRAM, and Figs.~\ref{fig:sampling_2d_good} and \ref{fig:sampling_chain_good} show samples drawn using a DRAM within Gibbs procedure.}
  \label{fig:sampling}
\end{figure}

To perform sampling, we use an MCMC within Gibbs sampling scheme. In the Gibbs sampler, the parameters are separated into groups $\{\thetic\}$, $\{\thetdyn\}$, and $\{\thetsig,\thetgam\}$ and sampled sequentially from the corresponding conditional distribution of each parameter group using the delayed rejection adaptive Metropolis (DRAM) algorithm~\cite{haario2006}. To show how fixing $\thetobs$ and using the DRAM within Gibbs approach improves sampling compared to basic DRAM, Fig.~\ref{fig:sampling} shows samples drawn from the $\mA$ matrix of the linear system in Section~\ref{sec:pend} using both approaches. Figs.~\ref{fig:sampling_2d_bad} and \ref{fig:sampling_chain_bad} show the samples drawn when using DRAM without Gibbs and without fixing any parameters at the MAP. With this approach, $10^7$ samples were drawn, the first $10^6$ discarded as burn-in, and every 1,000th remaining sample was plotted. Figs.~\ref{fig:sampling_2d_good} and \ref{fig:sampling_chain_good} show samples drawn with the DRAM within Gibbs approach with matrices $\mB$ and $\mH$ fixed at the MAP. For this method, $10^6$ samples were drawn, $10^5$ were discarded as burn-in, and every 1,000th remaining sample was plotted. Despite the former approach drawing 10 times as many samples, we observe that the latter approach covers more of the posterior. Moreover, the mixing of the chain in the DRAM within Gibbs approach appears much better.

For each experiment, we compare the proposed approach to the others by looking at forecasted predictions. In the Bayesian case, the forecast sometimes uses the full stochastic model~\eqref{eq:dynamics}, where the Bayesian filter is used to update the estimate during periods where data are available. When data are not available, the forecast evolves according to Eq.~\eqref{eq:dynamics} with random realizations of the process noise. Other times, a deterministic simulation is used to show the forecast ability without state updates to better compare to the other methods. 

The following experiments were run on MATLAB R2020a, and the code for these experiments can be found at \url{github.com/ngalioto/BayesID}. In all experiments, optimization was performed using MATLAB's \texttt{fmincon} function and integration was performed with MATLAB's \texttt{ode45} function.

\subsection{Linear pendulum with control}\label{sec:pend}
The first example demonstrates that as the noisiness and sparsity of data increase, the performance of the Bayesian method decays at a slower rate than the LS+ERA method from Section~\ref{sec:era}. This example builds upon that presented in~\cite{galioto2021} by comparing to the subtrajectory LS objective~\eqref{eq:subtraj} rather than the exactly determined objective~\eqref{eq:fulltraj}. In addition, posterior predictive sampling is added to make probabilistic forecasts that quantify uncertainty.

For this example, a pendulum with unit length and mass with damping and random inputs is considered. The dynamics of such a system are given: 
\begin{equation}
  \begin{aligned}
  \vx_{k+1} &= \expm\left(\begin{bmatrix}0 & 1 \\ -9.81 & -1 \end{bmatrix}\Delta t\right)\vx_k+\begin{bmatrix}0\\1\end{bmatrix}u_k, \\
  \vy_k &= \begin{bmatrix}1 & 0\end{bmatrix}\vx_k + \veta_k; \qquad \vx_0 = \vzero.
  \end{aligned}
  \label{eq:lpend_damped}
\end{equation}
where $\expm$ is the matrix exponential and the inputs are Gaussian-distributed as $u_k\sim\N(0,\Delta t)$. The damping term is included to ensure that the $\mA$ matrix is asymptotically stable at all $\Delta t$ considered. This damping term ensures $\rho(\mA)<1$, and since $\dimy=1$, the conditions of Proposition~\ref{prop:converge} are satisfied. Therefore, LS+ERA should be able to give a decent estimate for finite $\bar{n}$.

Data were collected from this system over a 20s training period at various timesteps and noise levels. For this experiment, timesteps of $\Delta t=0.10,0.15,\ldots,0.50$ and noise ratios of $\sigma=0.00,0.025,\ldots,0.200$ were considered. Here, the noise ratio is defined as $\sigma\coloneqq \sigma_{\veta} / \max(\vx[1])$, where $\sigma_{\veta}$ is the standard deviation of the measurement noise. For each noise-timestep pair, 100 realizations of data were generated, and every realization was trained on separately such that each method estimated a set of 100 models per pair. Note that since the inputs are random, the system behavior is also random and the noise ratio of each dataset will therefore vary, even within a given noise-timestep pair.

The model parameterization is
\begin{equation}\label{eq:pend_model}
\begin{aligned}
  \vx_0 = \begin{bmatrix}\vtheta_1\\\vtheta_2\end{bmatrix}; \hspace{2mm} \vx_{k+1} &= \begin{bmatrix}\vtheta_3 & \vtheta_5 \\ \vtheta_4 & \vtheta_6 \end{bmatrix}\vx_{k}+\begin{bmatrix}\vtheta_7\\\vtheta_8\end{bmatrix}u_k + \vxi_k, \\
    \vy_k &= \begin{bmatrix}\vtheta_9 & \vtheta_{10}\end{bmatrix}\vx_k + \veta_k, \\
    &\hspace{-2cm}\vxi_k\sim\mathcal{N}\left(\vzero, \begin{bmatrix}\vtheta_{11}&0\\0&\vtheta_{12}\end{bmatrix}\right),\quad\veta_k\sim\mathcal{N}(0,\vtheta_{13}).
\end{aligned}
\end{equation}
Half-normal priors of $\halfN(0,10^{-6})$ and $\halfN(0,1)$ were placed on the process and measurement noise variance parameters, respectively, and an improper uniform prior was placed on the remaining 10 parameters. The MAP was estimated by optimizing the negative log posterior with a random initial point. The LS+ERA and MAP estimates were compared with respect to the $\log_{10}$ of the average mean squared error (MSE) at each noise-timestep pair. For a given value of the noise ratio and timestep, let $i=1,\ldots,100$ index the data realizations. Then, the MSE on the $i$th dataset is defined as $\frac{1}{\numObs}\sum_{k=1}^{\numObs}(x_k[1]-\hat{y}_k(\vtheta_i))^2$, where $\vtheta_i$ represents the parameter estimate on the given dataset. In addition to the MSE during the training period, the MSE on a testing period of 20s beyond the training data was also calculated over time indices $k=\numObs+1,\ldots,2\numObs$. There were a handful of outliers in the LS+ERA MSE that significantly skewed the average value, so only the lowest 99 MSE values were used to compute the average MSE of the LS+ERA estimate. The average MSE of the MAP estimate retains all 100 MSE values.

\begin{figure}
  \centering
  \begin{subfigure}{0.45\linewidth}
    \centering
    \includegraphics[trim=100 250 100 240, clip,width=\linewidth]{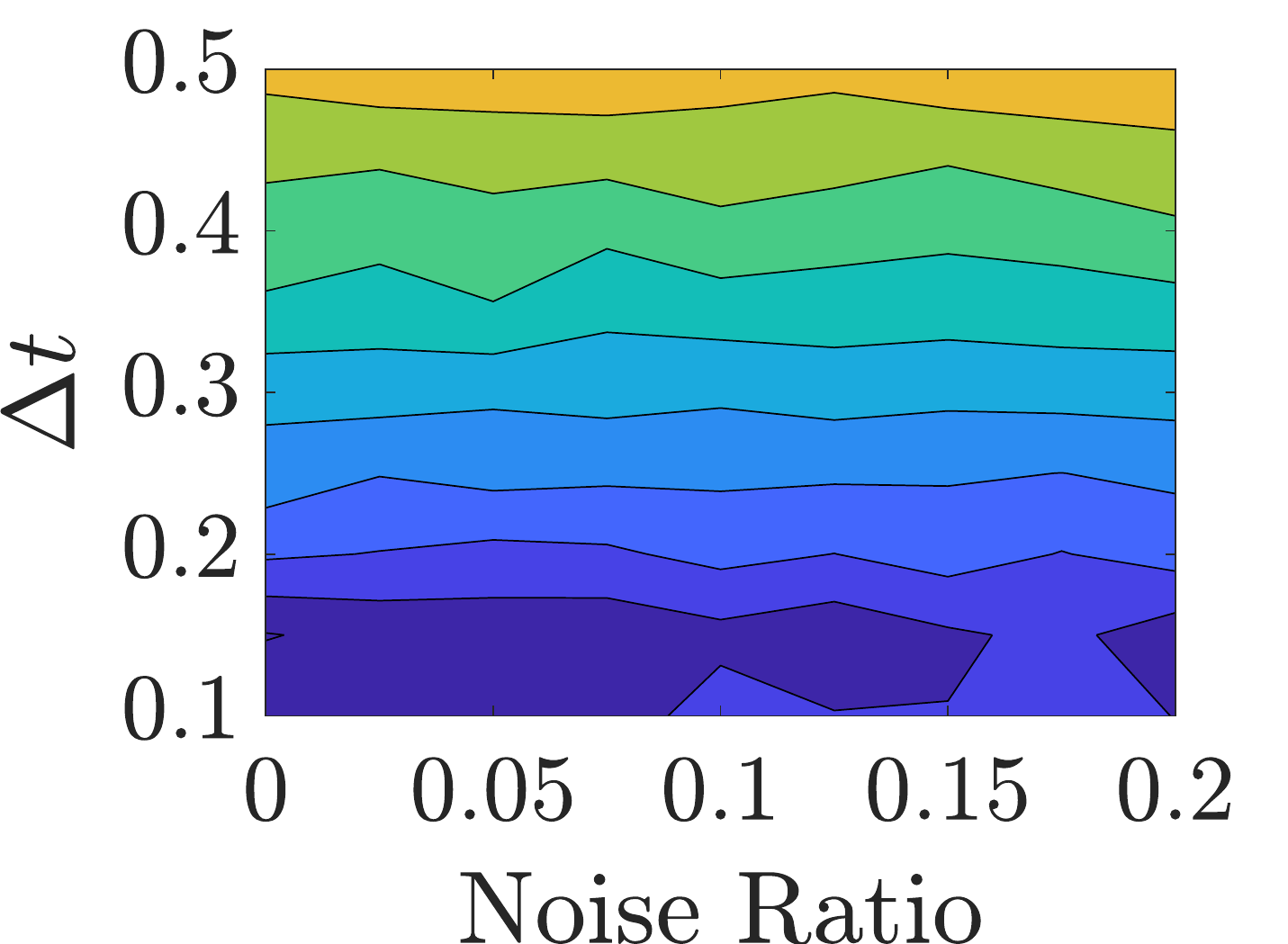}
    \caption{LS+ERA Training}
     \label{fig:contourFitRandomLS}
  \end{subfigure}%
  \begin{subfigure}{0.45\linewidth}
    \centering
    \includegraphics[trim=100 250 100 240, clip,width=\linewidth]{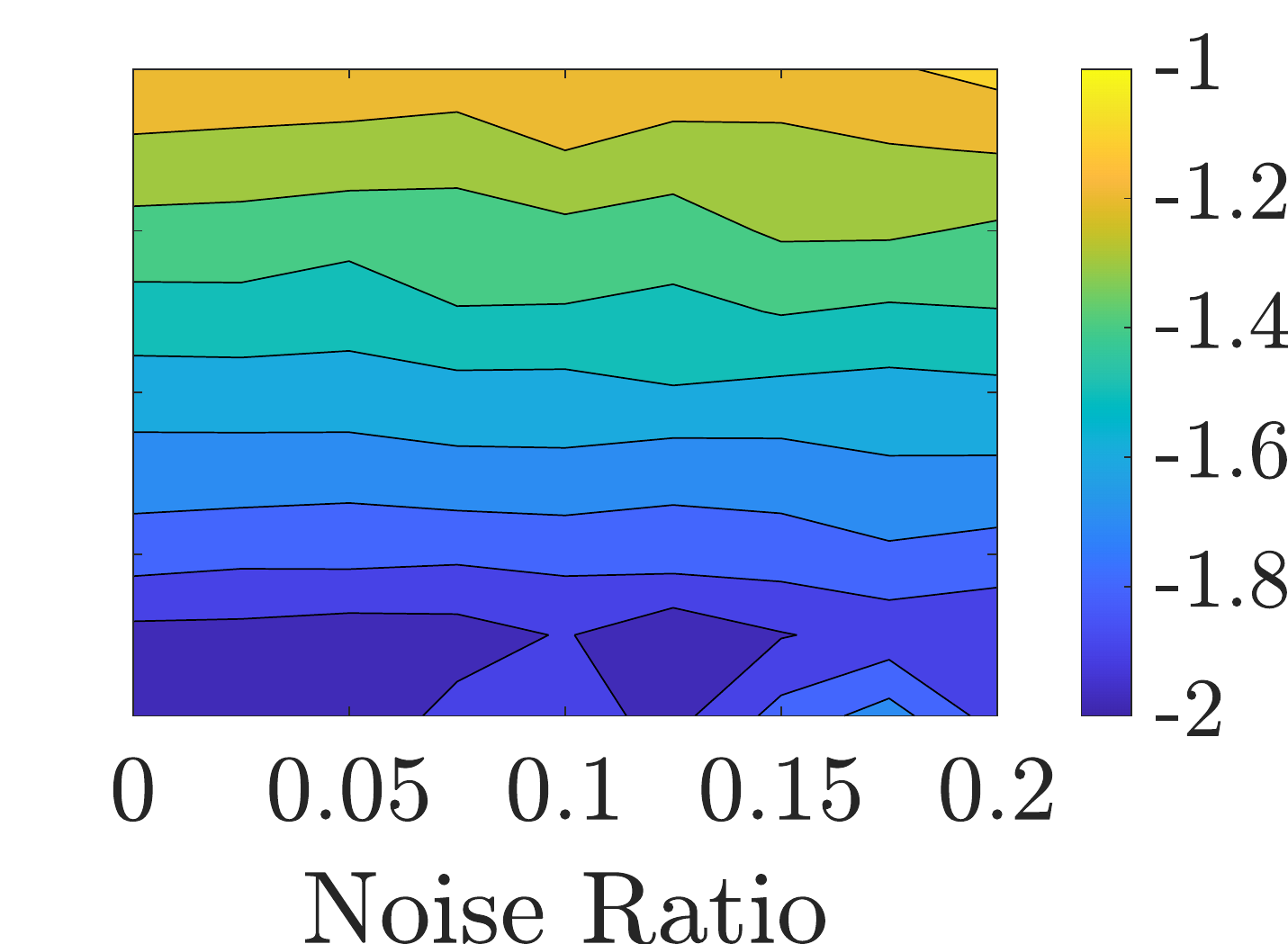}
    \caption{LS+ERA Testing}
    \label{fig:contourPredRandomLS}
  \end{subfigure}%
  \hfill
  \begin{subfigure}{0.45\linewidth}
    \centering
    \includegraphics[trim=100 250 100 240, clip,width=\linewidth]{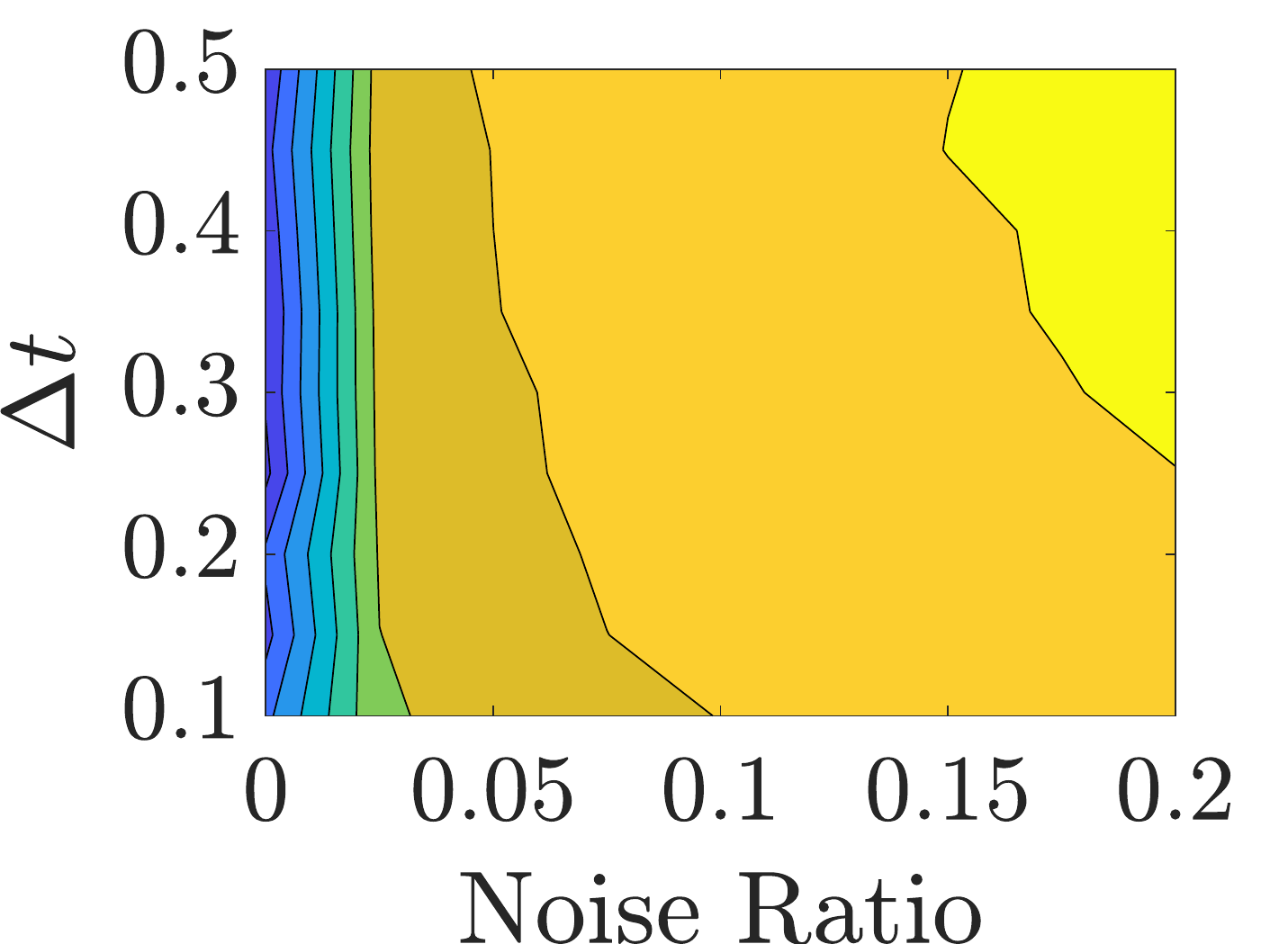}
    \caption{MAP Training}
     \label{fig:contourFitRandomBayes}
  \end{subfigure}%
  \begin{subfigure}{0.45\linewidth}
    \centering
    \includegraphics[trim=100 250 100 240, clip,width=\linewidth]{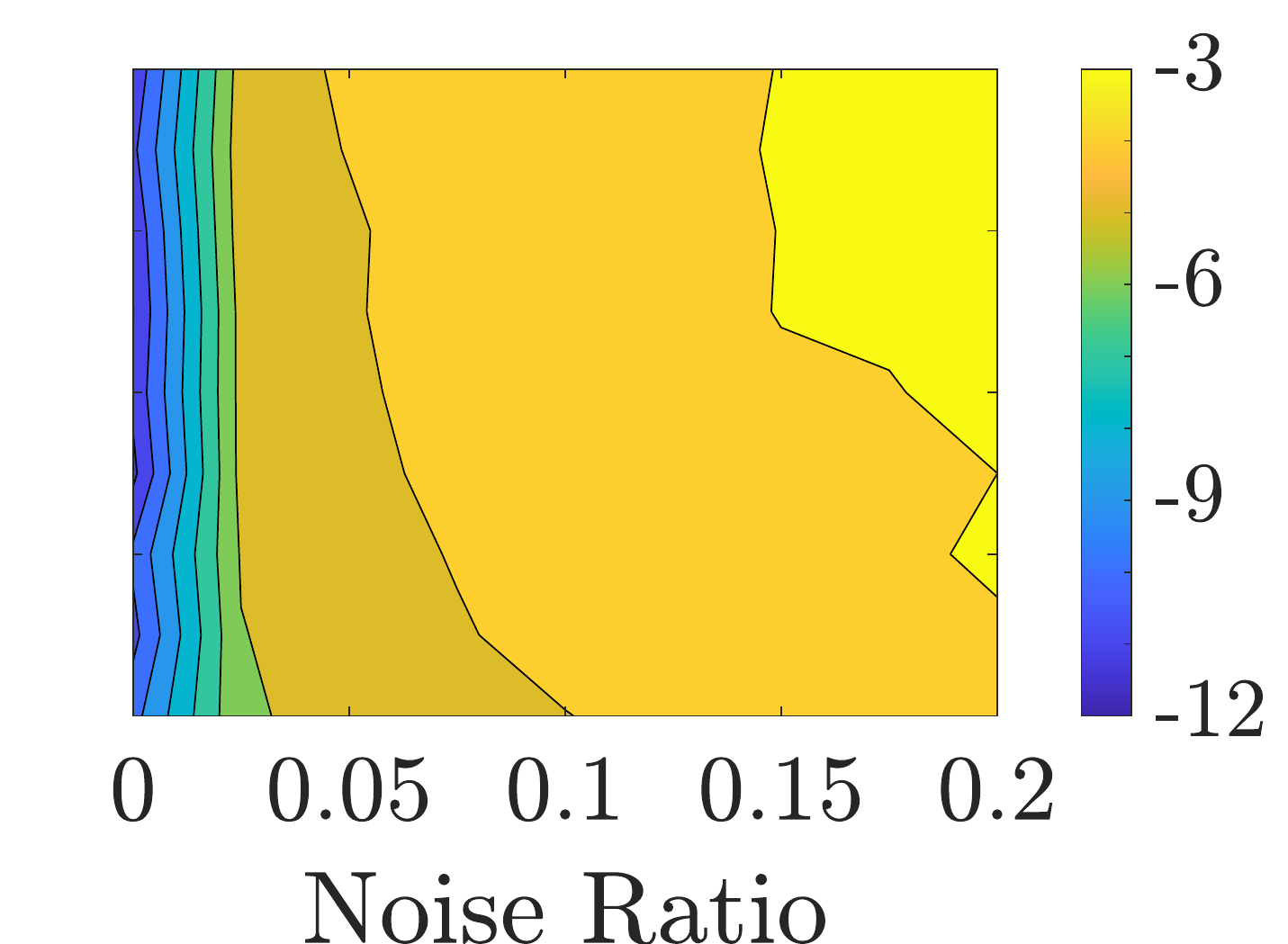}
    \caption{MAP Testing}
    \label{fig:contourPredRandomBayes}
  \end{subfigure}%
  \caption{Contour plots of the $\log_{10}(\text{MSE})$ of the LS+ERA and MAP estimates on System~\ref{eq:lpend_damped}. Fig.~\ref{fig:contourFitRandomLS} and \ref{fig:contourFitRandomBayes} are the training MSE and Figs.~\ref{fig:contourPredRandomLS} and \ref{fig:contourPredRandomBayes} are the testing MSE.}
  \label{fig:contourStable}
\end{figure}

Contour plots of the $\log_{10}$ average MSE of the LS+ERA and MAP estimates are given in Fig.~\ref{fig:contourStable}. We observe that the LS+ERA performance degrades most significantly as the timestep increases as a consequence of having fewer data available for estimation. The MAP estimate, on the other hand, appears to degrade more as the noise ratio is decreased, but its degradation due to increasing timestep is of similar magnitude. The slower degradation along the timestep axis suggests that the Bayesian approach has low data requirements, especially when the data are low-noise. Based on the colorbars of each set of plots, the MAP estimate gives at least an order of magnitude of improvement over the LS+ERA estimate.

\begin{figure}
  \centering
  \begin{subfigure}{0.45\linewidth}
    \centering
    \includegraphics[trim=95 240 122 243, clip,width=\linewidth]{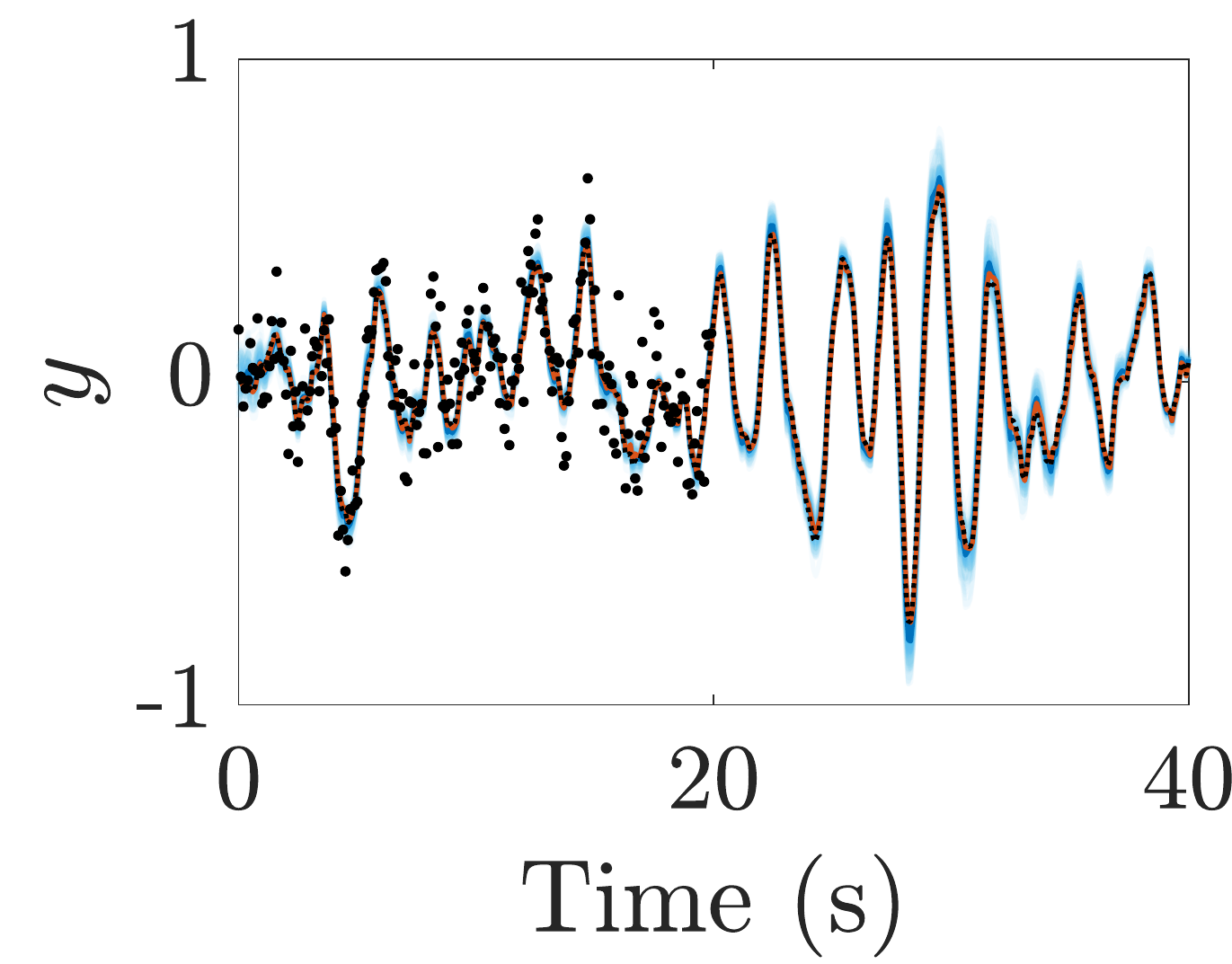}
    \caption{$\sigma=0.2$, $\Delta t=0.1$}
    \label{fig:outputPosteriorNoiseStable}
  \end{subfigure}%
  \begin{subfigure}{0.45\linewidth}
    \centering
    \includegraphics[trim=95 240 122 243, clip,width=\linewidth]{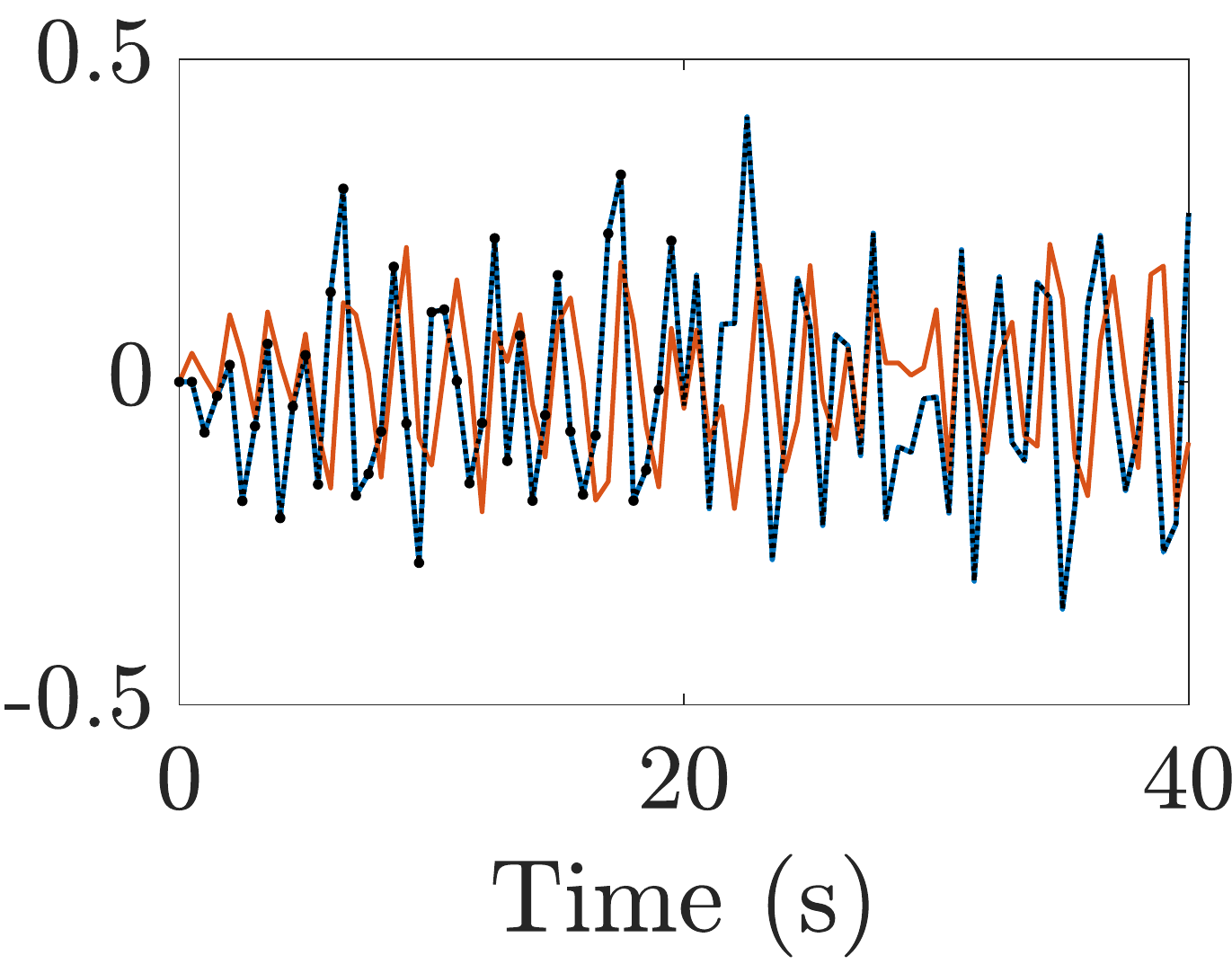}
    \caption{$\sigma=0.0$, $\Delta t=0.5$}
    \label{fig:outputPosteriorTimestepStable}
  \end{subfigure}%
  \hfill
  \begin{subfigure}{0.45\linewidth}
    \centering
    \includegraphics[trim=95 240 122 243, clip,width=\linewidth]{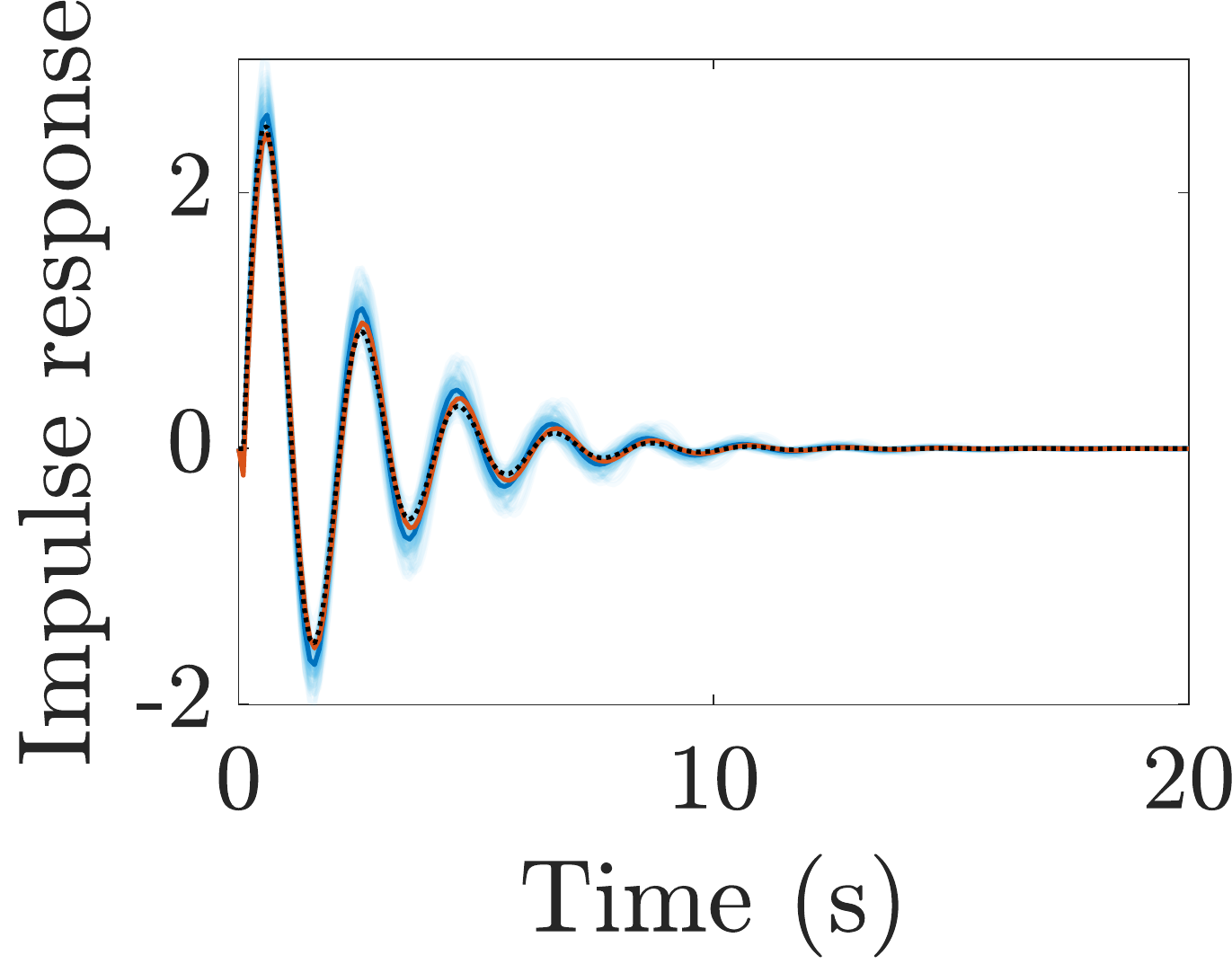}
    \caption{$\sigma=0.2$, $\Delta t=0.1$}
    \label{fig:impulsePosteriorNoiseStable}
  \end{subfigure}%
  \begin{subfigure}{0.45\linewidth}
    \centering
    \includegraphics[trim=95 240 122 243, clip,width=\linewidth]{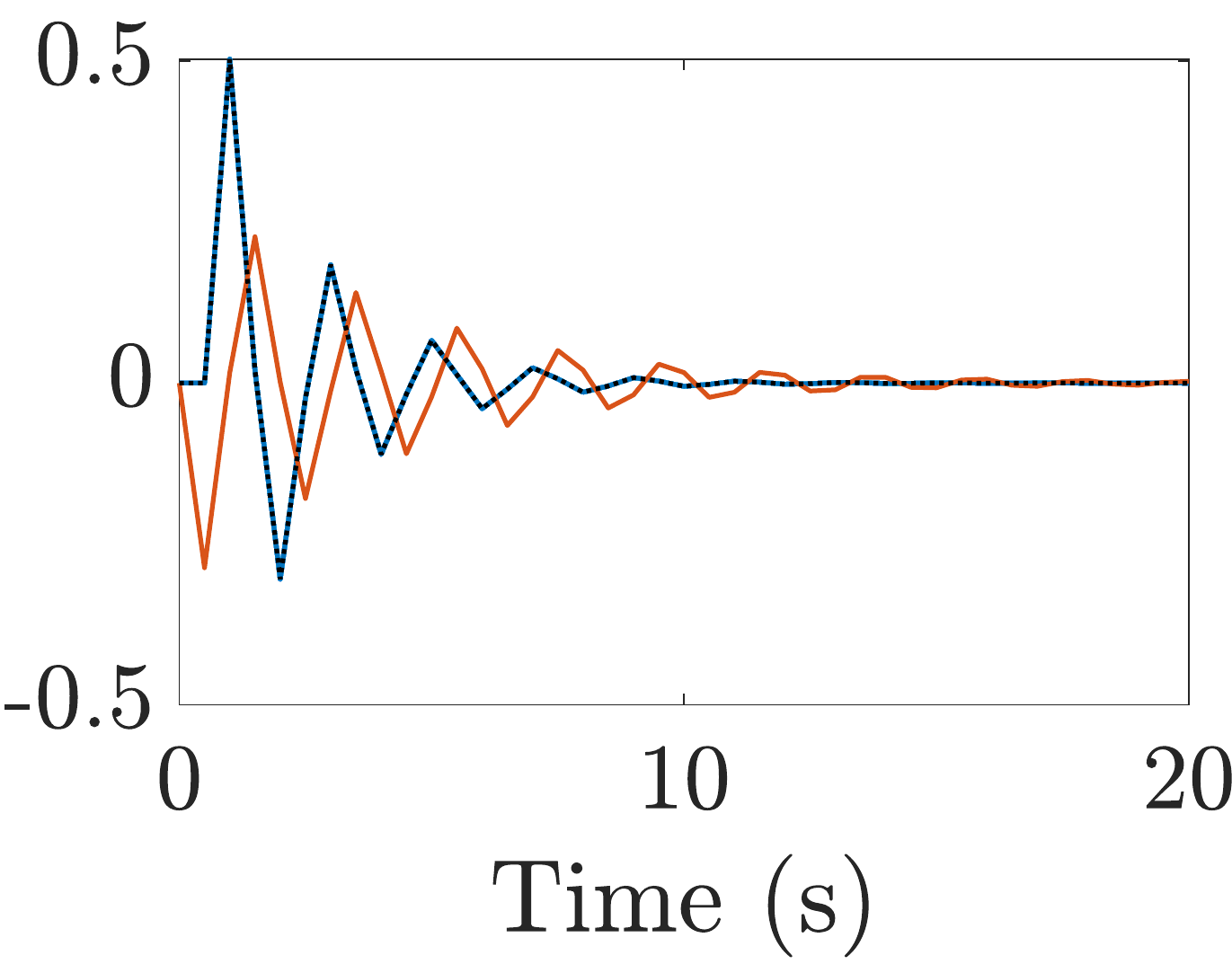}
    \caption{$\sigma=0.0$, $\Delta t=0.5$}
    \label{fig:impulsePosteriorTimestepStable}
  \end{subfigure}%
  \hfill
  \includegraphics[trim=105 395 95 370, clip,width=0.9\linewidth]{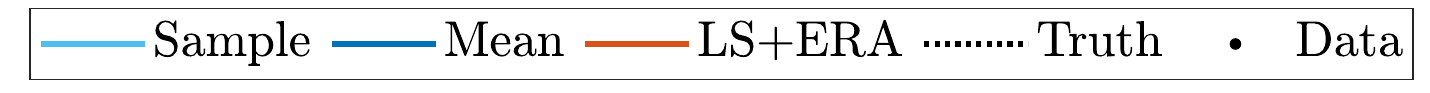}
  \caption{The Bayesian estimate is compared to the LS+ERA with $\bar{n}=18$. The top row shows the LS+ERA estimate, deterministic simulations of 100 posterior samples, and the mean of the sample outputs. The bottom row shows the impulse response of each of these estimates. The left column shows the high noise, low sparsity case. The right column is the low noise, high sparsity case.}
 \label{fig:pendOutputStable}
\end{figure}

To get a better understanding of how each method performs when the data are noisy/sparse, two points were selected from the contour plots in Fig.~\ref{fig:contourStable}, a data realization was chosen from each point, and the estimated output from each method was examined. The selected (noise, timestep) pairs are the high noise, low sparsity case (0.20, 0.1), and the low noise, high sparsity case (0.00, 0.5). These two points were chosen so that the effect of high noise and high sparsity could be studied separately. Within each point, the data realization chosen was the dataset for which the LS+ERA method had the lowest training MSE so that its peformance was fairly represented. The estimated outputs of the two estimates are shown in Figs.~\ref{fig:outputPosteriorNoiseStable} and \ref{fig:outputPosteriorTimestepStable} for the high noise and high sparsity cases, respectively.

Furthermore, the impulse response is examined since it can completely describe the response of an LTI system. Due to model over-parameterization, it is possible the estimated model only learned how to produce sinusoids at a given frequency. The impulse response will reveal if the estimated system actually approximates a realization of the state-space matrices as desired. The impulse response for the models in the high noise and high sparsity cases are shown in Figs.~\ref{fig:impulsePosteriorNoiseStable} and \ref{fig:impulsePosteriorTimestepStable}, respectively.

To represent the posterior predictive distribution, $10^6$ samples were drawn from the posterior using the MCMC procedure described earlier in Section~\ref{sec:results}, the first $10^5$ were discarded as burn-in, and 100 samples selected at regular intervals were simulated and plotted. The blue `mean' line indicates the mean of these posterior predictive samples.

In the high noise, low sparsity case (Figs.~\ref{fig:outputPosteriorNoiseStable} and \ref{fig:impulsePosteriorNoiseStable}), the mean estimate and LS+ERA estimate both appear to fit the truth fairly well, despite the noisiness of the data. The MSE of the LS+ERA estimate over the full 40s is $2.86\times10^{-4}$ and the MSE of the mean estimate is $4.74\times10^{-4}$. This was the only dataset on which the LS+ERA MSE was less than that of the MAP estimate. In fact, the next lowest LS+ERA training MSE was $4.68\times 10^{-4}$, which is larger than the average MAP training MSE of $4.47\times10^{-4}$. Furthermore, the standard deviations of the LS+ERA training and testing MSE over all 100 datasets are $1.72\times10^{-2}$ and $2.58\times10^{-2}$, respectively, while the training and testing standard deviations of the MAP are $3.65\times10^{-4}$ and $5.56\times10^{-4}$, respectively. The fact that the standard deviation of the LS+ERA MSE is about 100 times greater than that of the MAP MSE indicates that there are far worse LS+ERA estimates than the one presented here, but the MAP estimates likely all resemble the one shown in the figure. In the impulse response, the LS+ERA estimate is also closer to the truth than the mean estimate, but the posterior is wide and encompasses the truth. Therefore, the Bayesian method is `aware' of the error and gives a reasonable quantification of the estimate uncertainty. In the low noise, high sparsity case (Figs.~\ref{fig:outputPosteriorTimestepStable} and \ref{fig:impulsePosteriorTimestepStable}), the posterior is so narrow that it visually appears as a single line, indicating low uncertainty. In both the output and impulse response plots, the mean is directly on top of the truth, and the LS+ERA estimate has large discrepancies between its output and the truth. In contrast to the Bayesian estimate, the LS+ERA method has no way to identify this larger error/uncertainty.

\subsection{Wiener-Hammerstein benchmark}
Next, experimental data collected from a nonlinear system is considered. For this example, the proposed Bayesian method is tested on the Wiener-Hammerstein benchmark~\cite{schoukens2009}, which is a standard dataset that has been used to compare the performances of different nonlinear system ID methods. The underlying system consists of electronic components combined into a Wiener-Hammerstein structure. For more details on how the system was built, see~\cite{schoukens2009}. The benchmark dataset is composed of 188,000 low-noise input-output data points, with a suggested training/testing split of 100,000/88,000. The best performance to date on this benchmark to the authors' knowledge comes from~\cite{beintema2021}, which achieved an RMSE value of 0.241 mV on the testing data using the MS objective. In this experiment, the method of~\cite{beintema2021} will be compared to the Bayesian method.

Because the dataset has such a high number of data points with low measurement noise, the identification problem has relatively low uncertainty, and the advantages of the Bayesian approach are not nearly as evident in cases with large and not noisy data. It will be shown, however, that methods that work well with a large amount of low-noise data are not necessarily best-suited for estimation when the data are few and noisy. To this end, only the first 1,000 data points of the original 100,000 point training set were used for training. Furthermore, zero-mean Gaussian noise with standard deviation $\sigma=0.0178$ was added to these training data. This standard deviation is equal to $1\%$ of $(\vy_{max}-\vy_{min})$.

The nonlinear model follows the form of Eq.~\eqref{eq:hmm} with latent space dimension $\dimx=6$, where the dynamics operator $\Psi$ and observation operator $h$ are now parameterized as neural networks. Following the approach of~\cite{beintema2021}, each neural network has a single hidden layer with 15 nodes and tanh activation functions. Additionally, a linear transformation from the input of the network directly to the output is included such that the network is of the form:
\begin{equation}\label{eq:nn}
  \begin{split}
    \vz_{out} &= \mA_1(\vtheta)\tanh(\mA_2(\vtheta)\vz_{in} + \vb_2(\vtheta))\\
    &\quad+ \mA_3(\vtheta)\vz_{in} + \vb_3(\vtheta),
  \end{split}
\end{equation}
where $\vz_{in}=\begin{bmatrix}\vx_k^* & u_k\end{bmatrix}^*$. For the dynamics network $\Psi$, $\vz_{out}=\vx_{k+1}$, and for the observation network $h$, $\vz_{out}=\vy_k$. The combined number of parameters in $\Psi$ and $h$ is 401. The one difference between our model and that of~\cite{beintema2021} is that rather than learning an encoder function to estimate the current state, the initial condition is estimated directly. The priors used for this model were $\text{half--}\N(0,10)$ on the process noise variance parameters, $\text{half--}\N(0,0.01)$ on the measurement noise variance parameters, and $\N(0, 0.2)$ on the remaining parameters.

Before training, the input and output data were both normalized to have zero means and standard deviations of one. The comparison method was trained with Adam batch optimization using the available code in the repository linked by~\cite{beintema2021}: \url{https://github.com/GerbenBeintema/SS-encoder-WH-Silver}. The batch size was reduced from 1,024 to 256 to handle the smaller dataset, but the number of epochs was kept at 100,000. The time horizon was also kept at $T=80$ since it was chosen according to the time scale of the system, which does not change. The Bayesian method was trained for 10,000 iterations. Then, $10^5$ samples were drawn from the posterior, and $2\times10^{4}$ were discarded as burn-in.

\begin{figure}
  \centering
  \begin{subfigure}{0.425\linewidth}
    \centering
    \includegraphics[trim=95 240 100 240, clip,width=\linewidth]{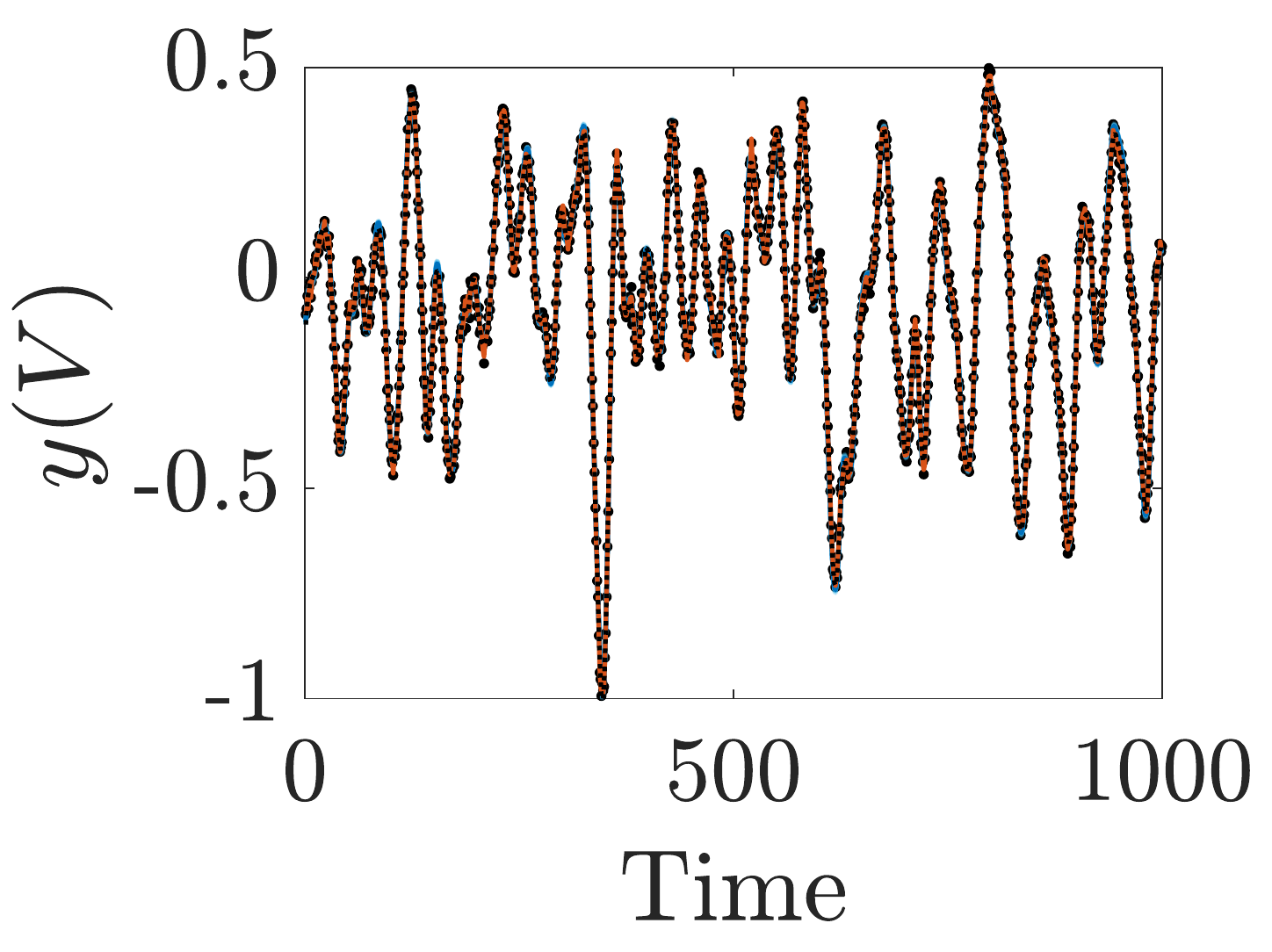}
    \caption{Training period}
    \label{fig:postTrain}
  \end{subfigure}%
  \begin{subfigure}{0.425\linewidth}
    \centering
    \includegraphics[trim=95 240 100 240, clip,width=\linewidth]{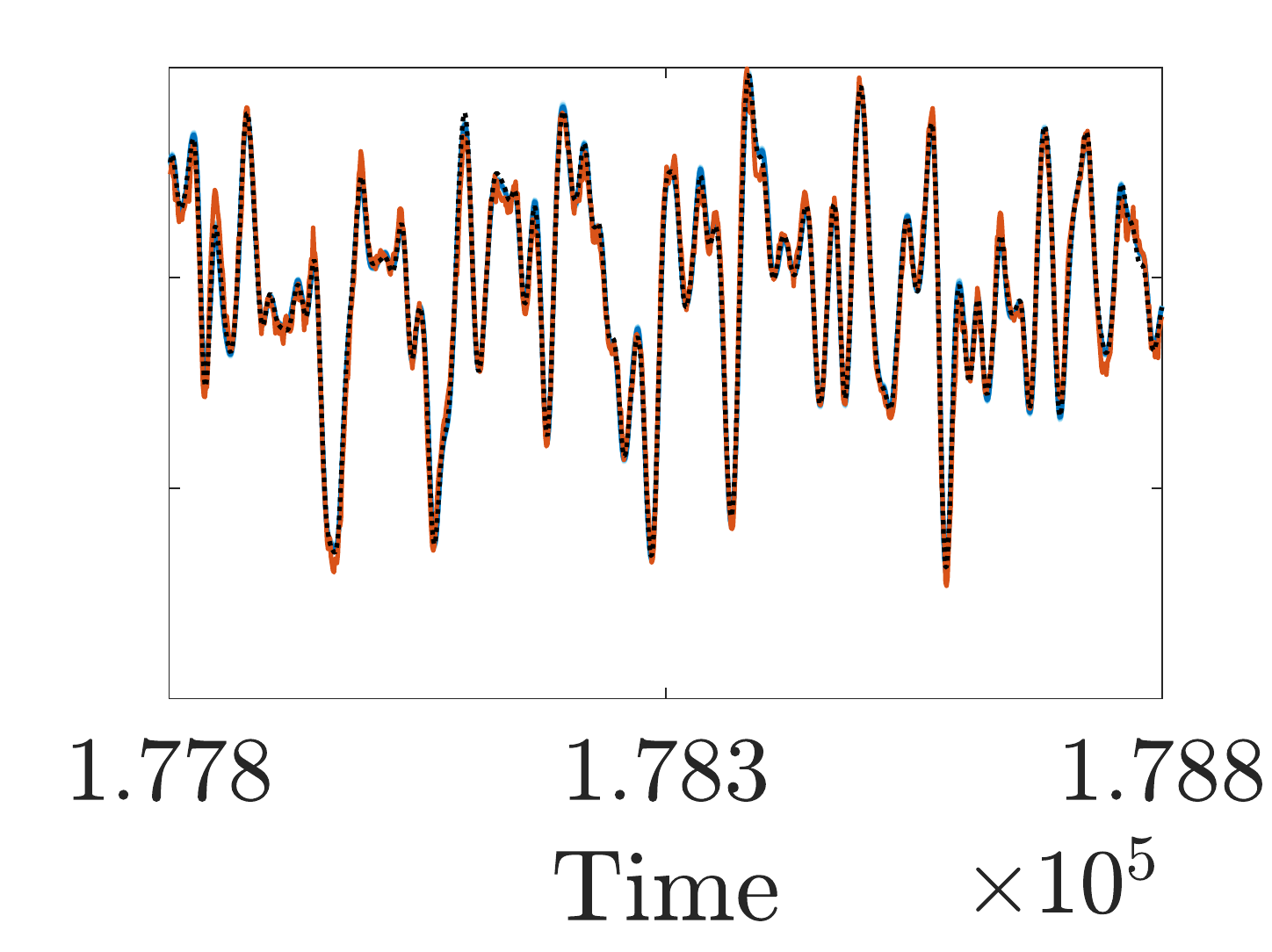}
    \caption{Testing period subset}
    \label{fig:postTest}
  \end{subfigure}%
  \includegraphics[trim=240 325 230 295, clip,width=0.15\linewidth]{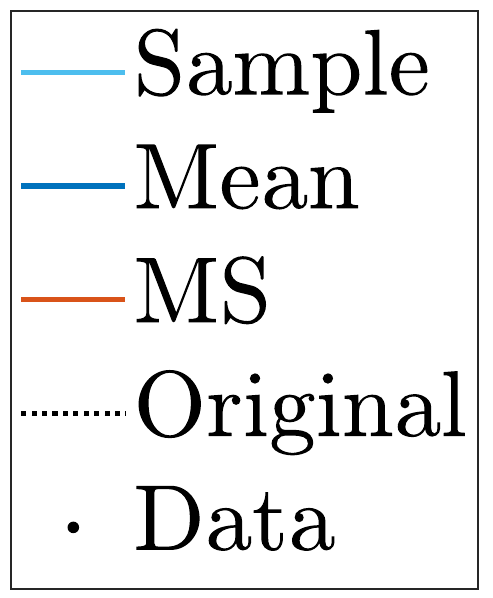}
  \hfill
  \begin{subfigure}{0.425\linewidth}
    \centering
    \includegraphics[trim=95 240 100 240, clip,width=\linewidth]{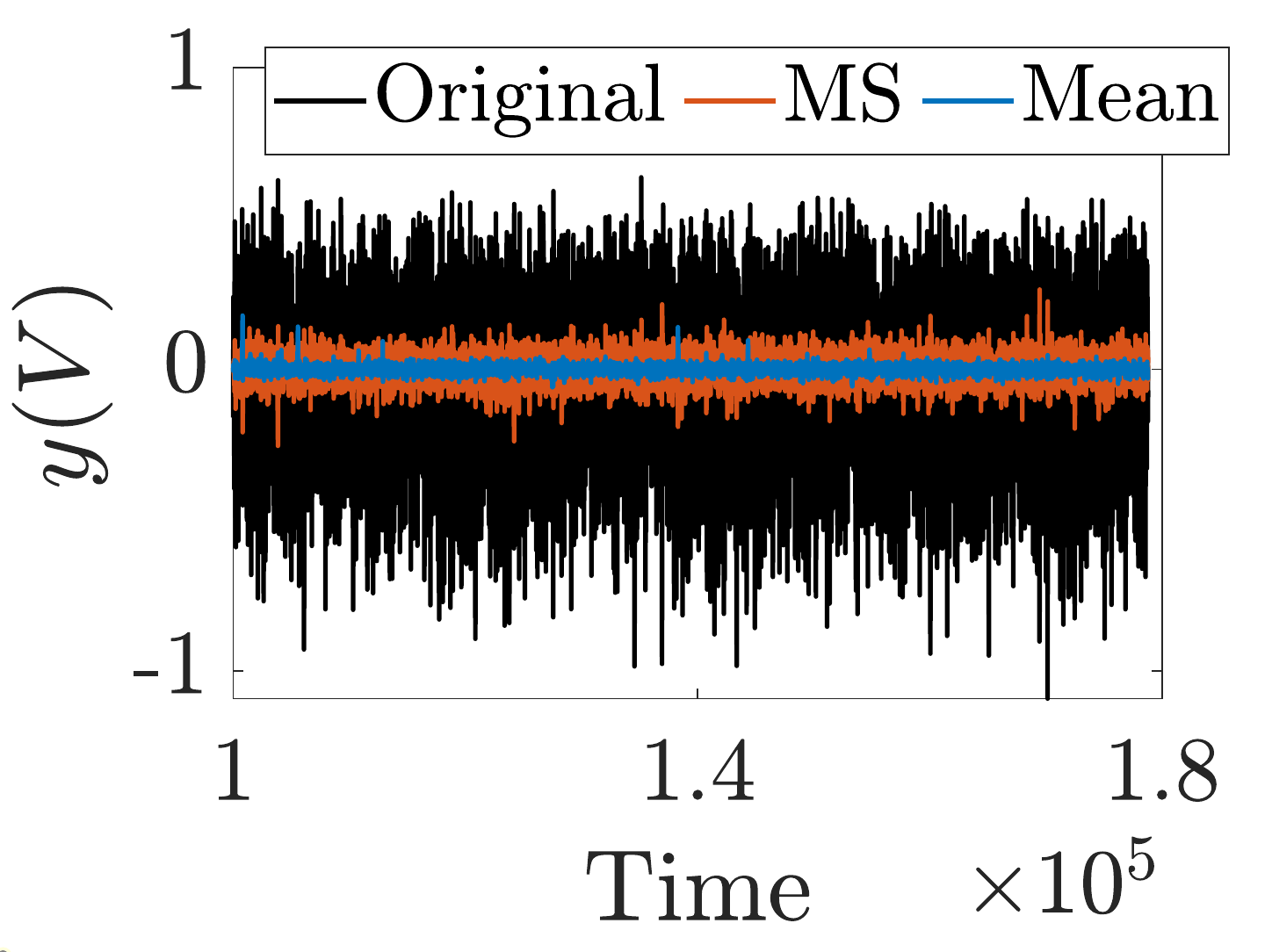}
    \caption{Time domain error}
    \label{fig:errorTime}
  \end{subfigure}%
  \begin{subfigure}{0.425\linewidth}
    \centering
    \includegraphics[trim=95 240 100 240, clip,width=\linewidth]{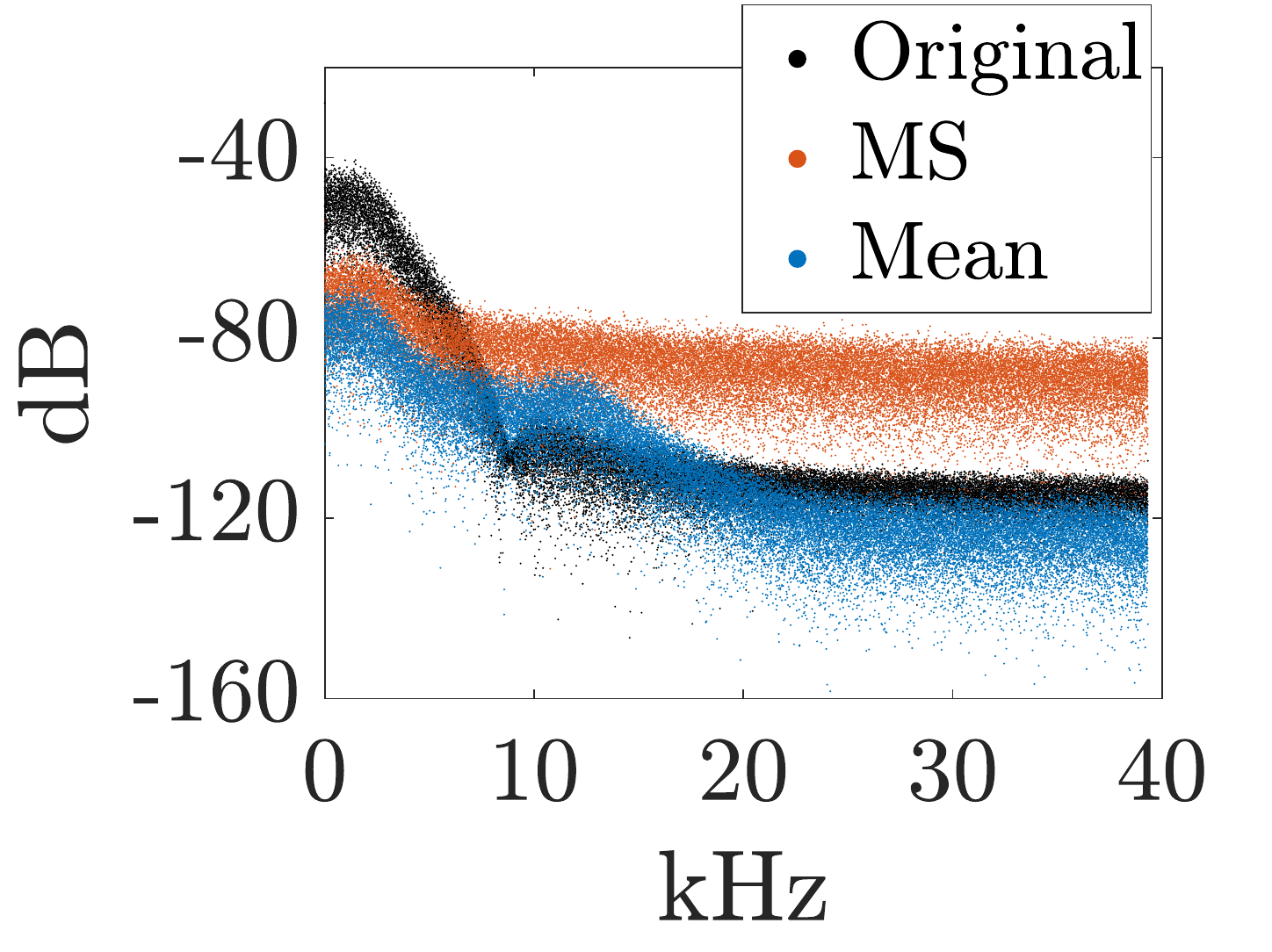}
    \caption{Freq. domain error}
    \label{fig:errorFreq}
  \end{subfigure}%
  \hspace{0.10\linewidth}
  \caption{The top row shows the trajectory estimates of the MS and Bayesian methods over the duration of the training data in Fig.~\ref{fig:postTrain} and over the last 1,000 testing data in Fig.~\ref{fig:postTest}. The bottom row shows the error between the unaltered data and the estimates in the time, Fig.~\ref{fig:errorTime}, and frequency, Fig.~\ref{fig:errorFreq}, domains.}
 \label{fig:WH}
\end{figure}

The results are shown in Fig.~\ref{fig:WH}. Figs.~\ref{fig:postTrain} and \ref{fig:postTest} show the estimated output of the Bayesian and MS estimates in the time domain during the training period and during the last 1,000 iterations of the testing period, respectively. The posterior predictive distribution is represented by 100 samples drawn at regular intervals from the collected samples and simulated deterministically. `Mean' refers to the mean of these 100 posterior predictive samples. In Fig.~\ref{fig:postTrain}, the estimates look nearly identical, but in Fig.~\ref{fig:postTest}, some noisiness has appeared in the MS estimate indicative of overfitting while the Bayesian estimate remains smooth due to its inherent regularization. Figs.~\ref{fig:errorTime} and \ref{fig:errorFreq} show the errors during the testing period in the time and frequency domain, respectively. The MSE values of the MS and mean estimates on the testing data are $1.0948\times10^{-3}$ and $1.2546\times10^{-4}$, respectively. The posterior predictive mean MSE is over 8.7 times lower than that of MS.

\subsection{Forced Duffing oscillator}\label{sec:duffing}
Next, the utility of the proposed Bayesian method for learning chaotic behavior will be demonstrated. For this example, the Duffing oscillator is considered. This system was also considered in one of the author's previous works~\cite{galioto2021}, but the solution was periodic and a linear model was used for estimation. This work, on the other hand, considers a chaotic solution and therefore uses a nonlinear model parameterization.

The Duffing oscillator takes the form of a damped harmonic oscillator where the linear restoring force is now replaced by negative linear and positive cubic restoring terms.  The governing equation is given as
\begin{equation}
  \begin{bmatrix}\dot{x}\\\ddot{x}\end{bmatrix} = \begin{bmatrix}0&1\\\alpha&\delta\end{bmatrix}\begin{bmatrix}x\\\dot{x}\end{bmatrix} + \beta\begin{bmatrix}0\\x^3\end{bmatrix} + \begin{bmatrix}0\\1\end{bmatrix}\gamma\cos(\omega t).
\end{equation}
Depending on the value of the parameters, the solution of this system can be periodic or chaotic, and there has been substantial study of the system's period-doubling cascade as it transitions to a chaotic regime. In this example, the parameter values are set as $\alpha=1$, $\delta = -0.3$, $\beta=-1$, $\omega=1.2$, and $\gamma=0.65$ following an example in~\cite{jordan2007} that yields chaotic behavior. To generate the data for this problem, an initial condition of $(x,\dot{x})=(0,0)$ is used, and the system is simulated for 600 seconds before beginning data collection to eliminate any initial transient behavior. After this initial period, the position $x$ of the system is measured every $\Delta t=0.25$s for 300s for a total of $1200$ data points, each with additive Gaussian noise with standard deviation $\sigma=10^{-3}$. The observation operator $h$ and measurement noise covariance $\Gamma$ are assumed to be known, and the dynamics model is the neural network architecture from the previous example defined in Eq.~\eqref{eq:nn} with latent space dimension $\dimx=2$. The priors are $\text{half--}\N(0,10^{-4})$ on the process noise variance parameters, and $\N(0,0.2)$ on the remaining parameters. For sampling, $10^{6}$ samples are drawn starting at the estimated MAP point, and half are discarded as a conservative burn-in.

\begin{figure}
  \centering
  \begin{subfigure}{0.3\linewidth}
    \centering
    \includegraphics[trim=95 240 114 249, clip,width=\linewidth]{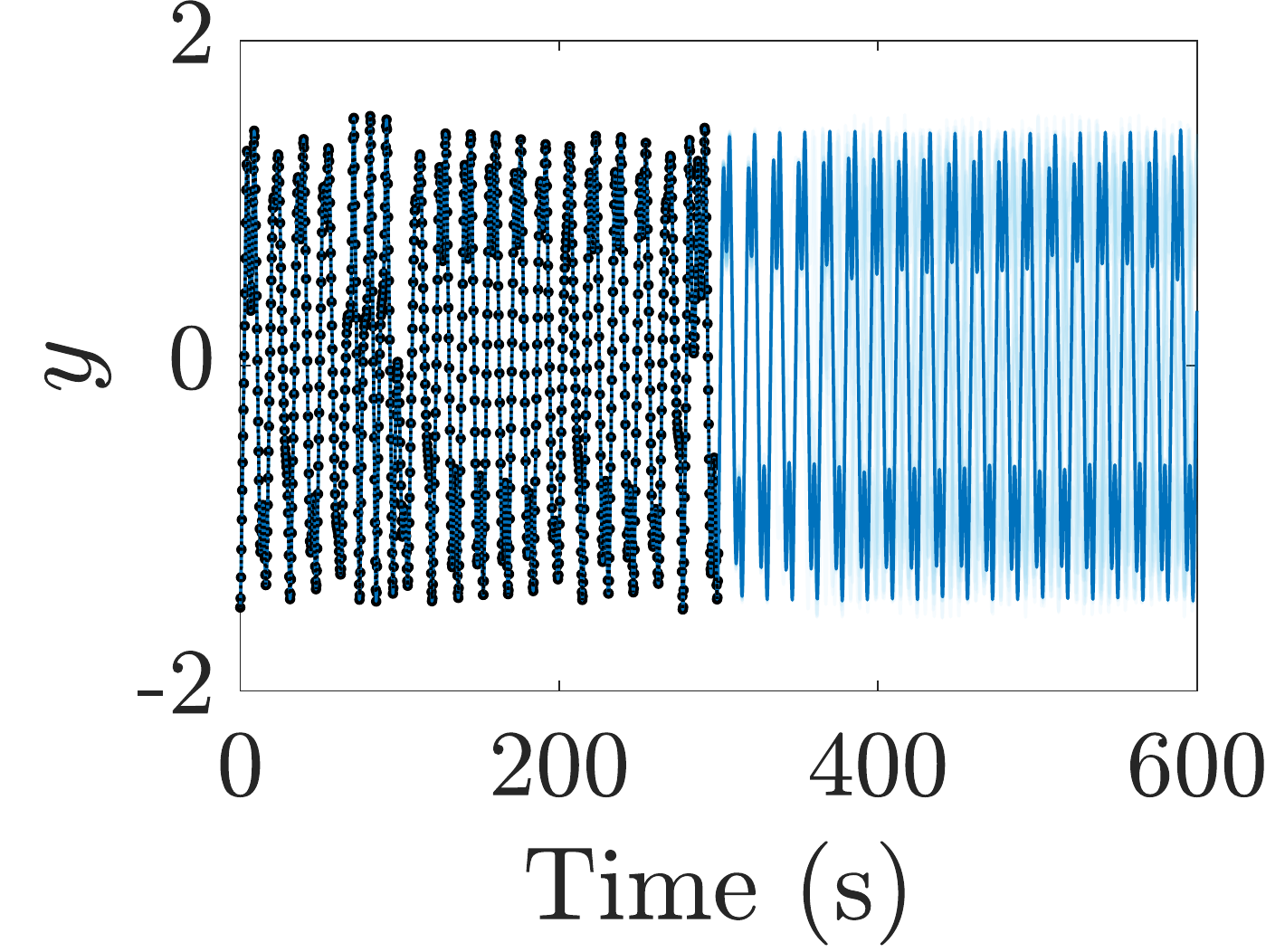}
    \caption{Stoch. sim.}
    \label{fig:timeUpdate}
  \end{subfigure}%
  \begin{subfigure}{0.3\linewidth}
    \centering
    \includegraphics[trim=95 240 114 249, clip,width=\linewidth]{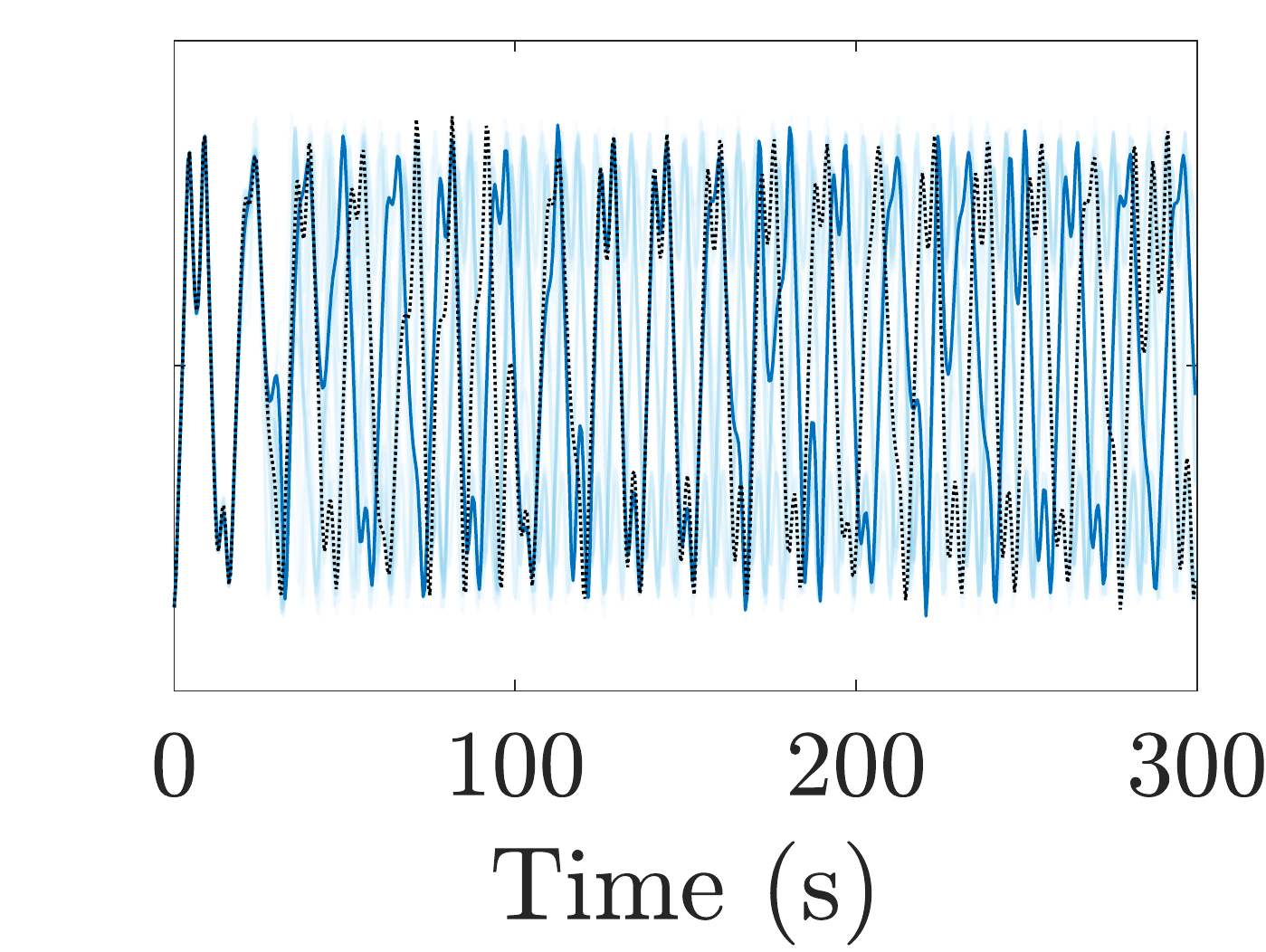}
    \caption{Det. sim.}
    \label{fig:timePredict}
  \end{subfigure}%
  \begin{subfigure}{0.3\linewidth}
    \centering
    \includegraphics[trim=95 240 114 249, clip,width=\linewidth]{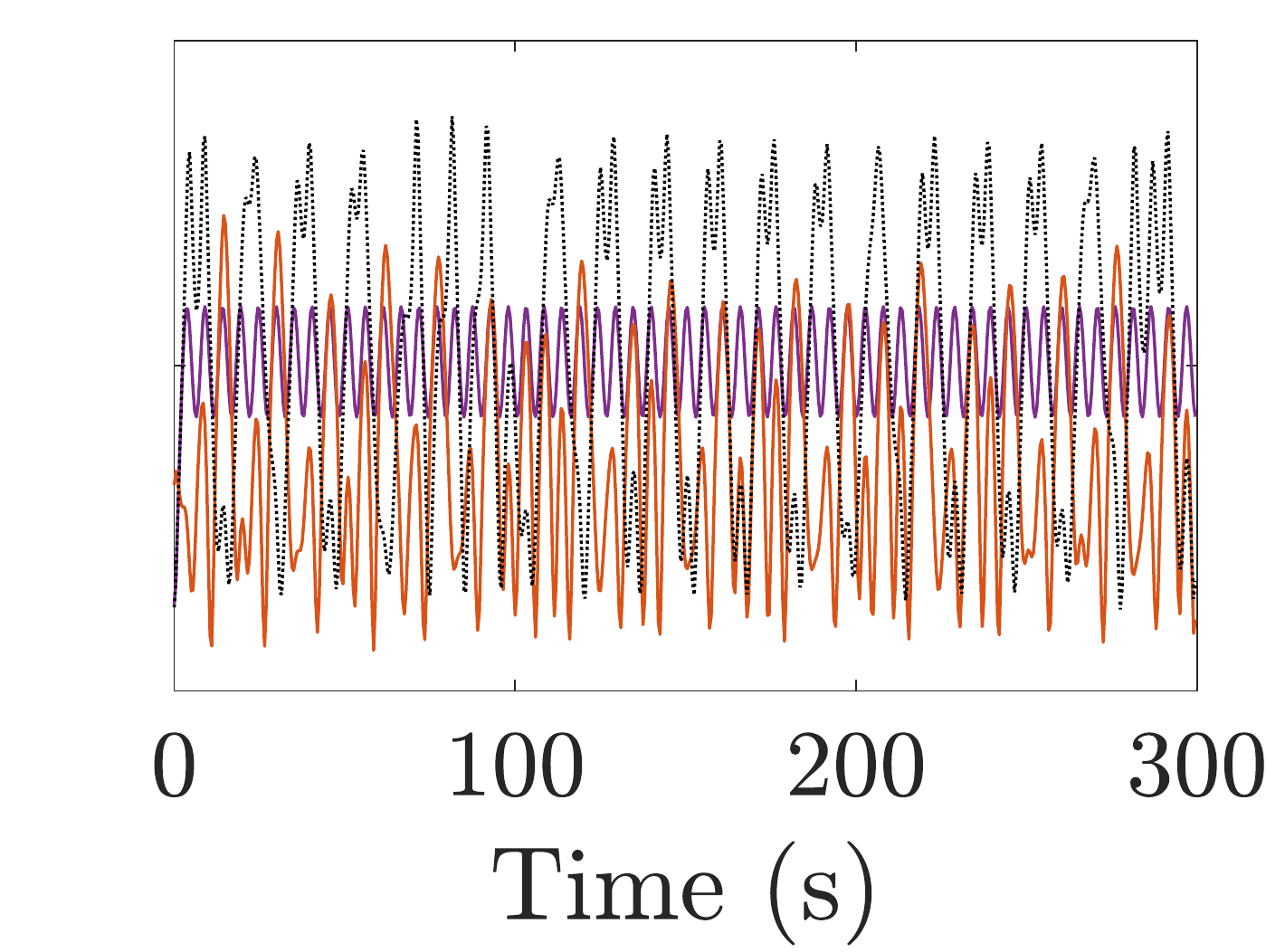}
    \caption{Det. sim.}
    \label{fig:timePredictLS}
  \end{subfigure}%
  \includegraphics[trim=260 355 255 325, clip,width=0.10\linewidth]{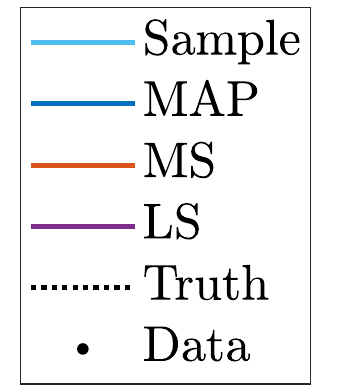}
  \hfill
  \begin{subfigure}{0.3\linewidth}
    \centering
    \includegraphics[trim=95 240 114 249, clip,width=\linewidth]{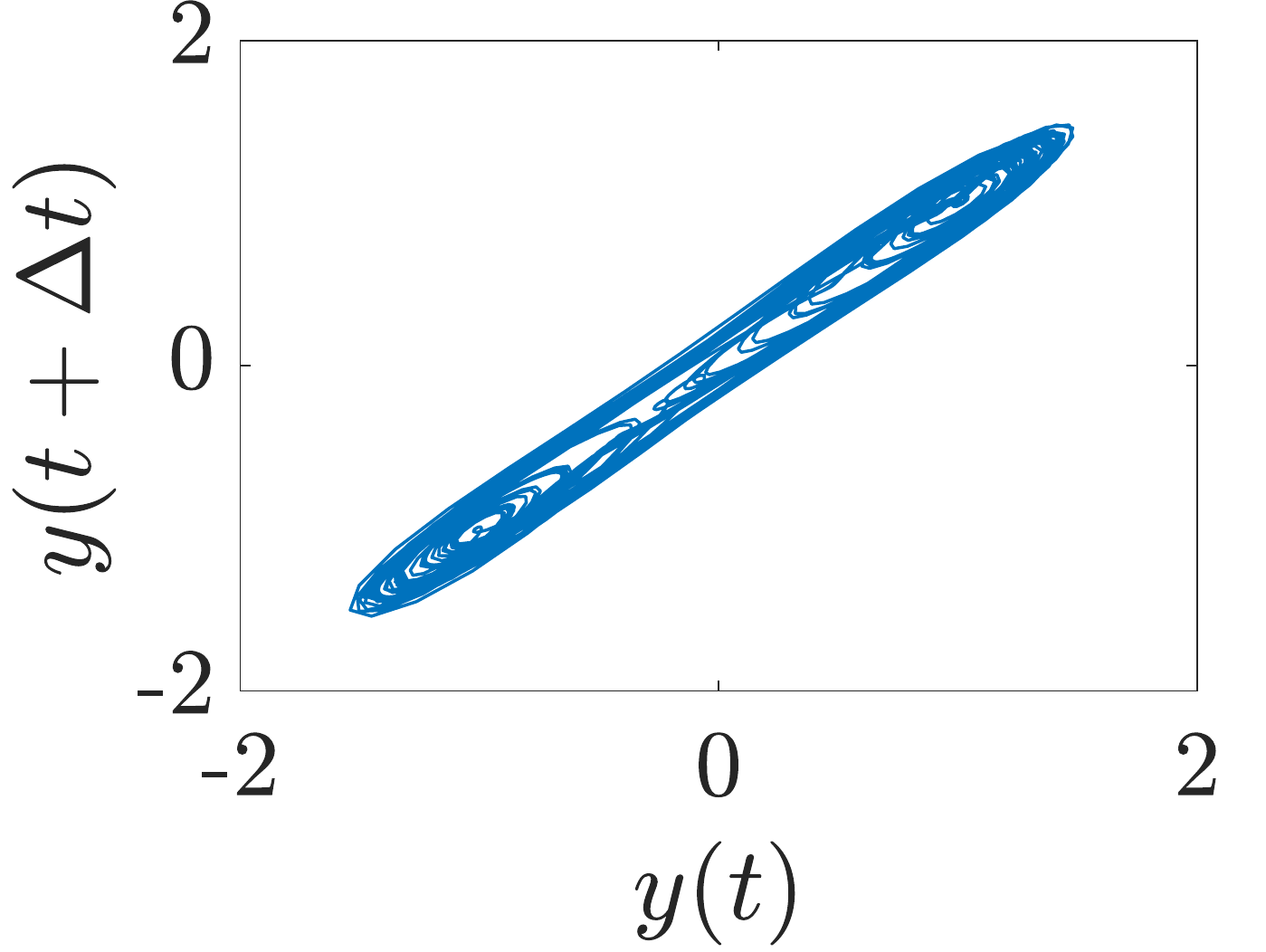}
    \caption{MAP phase}
    \label{fig:phaseMAP}
  \end{subfigure}%
  \begin{subfigure}{0.3\linewidth}
    \centering
    \includegraphics[trim=95 240 114 249, clip,width=\linewidth]{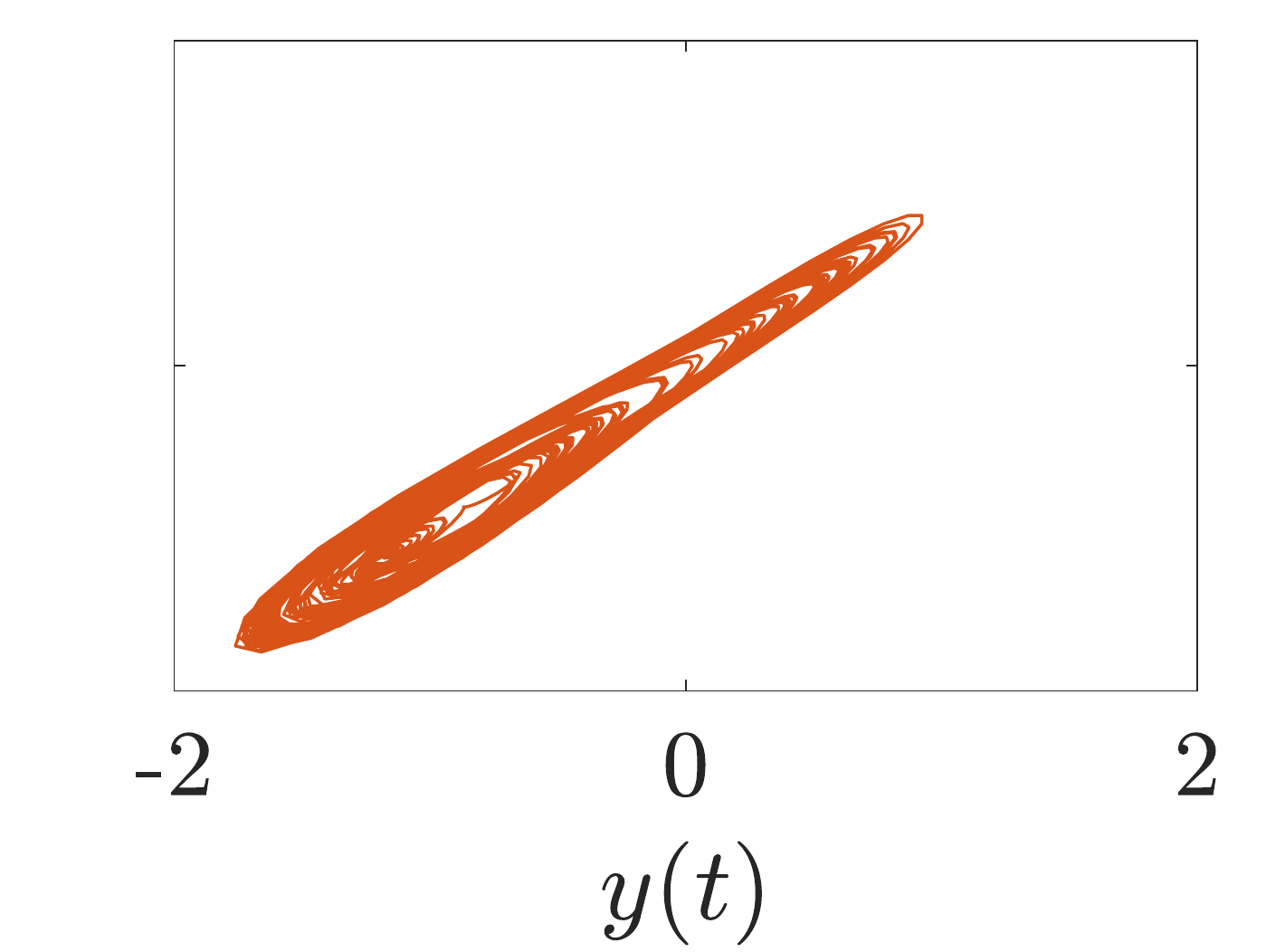}
    \caption{MS phase}
    \label{fig:phaseMS}
  \end{subfigure}%
  \begin{subfigure}{0.3\linewidth}
    \centering
    \includegraphics[trim=95 240 114 249, clip,width=\linewidth]{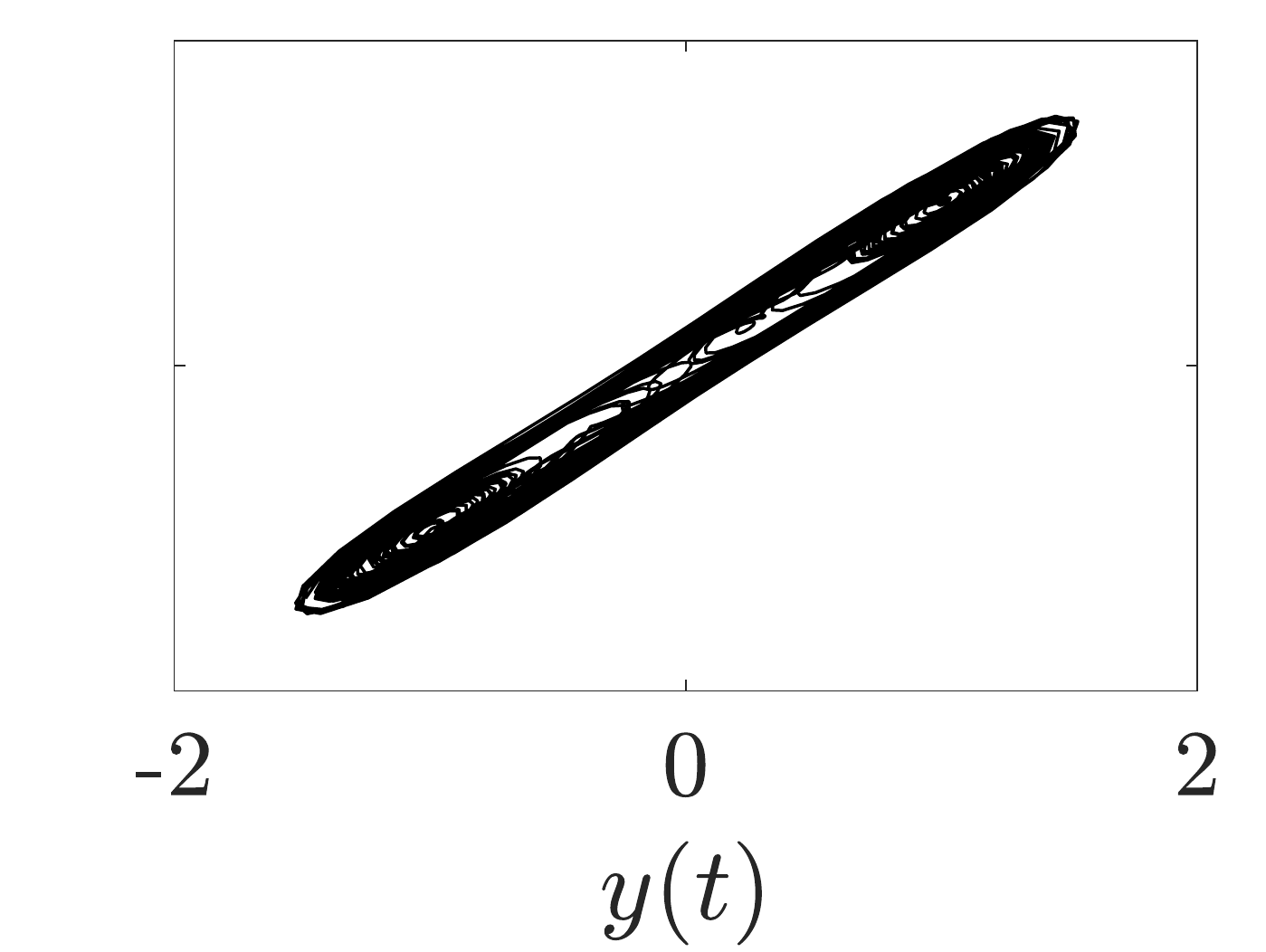}
    \caption{Truth phase}
    \label{fig:phaseTruth}
  \end{subfigure}%
  \includegraphics[trim=270 375 265 345, clip,width=0.10\linewidth]{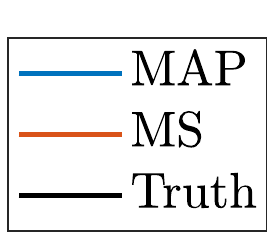}
  \hfill
  \caption{Figs.~\ref{fig:timeUpdate} and \ref{fig:timePredict} compare the posterior predictive distribution and MAP to the truth, where the posterior samples and MAP are generated using stochastic and deterministic simulation, respectively, and Fig.~\ref{fig:timePredictLS} shows deterministic simulations of the MS and LS estimates. Figs.~\ref{fig:phaseMAP}, \ref{fig:phaseMS}, and \ref{fig:phaseTruth} show the phase space of the MAP, MS, and truth, respectively, over 600 seconds.}
 \label{fig:duffing}
\end{figure}

The Bayesian algorithm is compared to the deterministic LS and MS objectives, and the results are shown in Fig.~\ref{fig:duffing}. For the MS objective, a time horizon of $T=200$ was used. Smaller values of $T$ in the range [30, 80] were tried but were found to give worse estimates. Fig.~\ref{fig:timeUpdate} shows 25 posterior samples and the estimated MAP point simulated stochastically and plotted alongside the deterministic LS estimate and the data. In Fig.~\ref{fig:timePredict}, the same samples and MAP point are simulated deterministically and plotted next to the LS estimate, the MS estimate, and the truth for a direct comparison. The LS estimate clearly looks much worse in these two figures, but the MSE of the LS estimate is actually lower than that of the MAP estimate. The LS estimate has an MSE of $0.7419$, and the MAP estimate has an MSE of $1.2791$. This shows that for certain system ID problems, especially ones including chaotic behavior, the squared error metric induces a nonsensical ranking within the model space.

Although it is difficult to identify any sort of structure in the behavior of chaotic systems in the time domain, the Duffing oscillator possesses an invariant set known as an attractor in phase space. Therefore, one way to assess how similar a model is to the underlying system is by comparing the phase space of the two systems. Figs.~\ref{fig:phaseMAP}, \ref{fig:phaseMS}, and \ref{fig:phaseTruth} show the phase space of the MAP model, the MS model, and the truth system, respectively, over 600 seconds. The MS and MAP models have similar shapes to the truth attractor, and both have foci near $\pm(1,1)$ around which their outputs rotate. The MS model's attractor, however, becomes larger around its $-(1,1)$ focus compared to both its $+(1,1)$ focus and the truth attractor. The structure of the MAP model, on the other hand, visibly appears consistent with the truth attractor, suggesting that it is a more accurate representation of the truth despite its high MSE.

\subsection{Allen-Cahn equation with forcing}\label{sec:allencahn}
In certain applications involving PDEs, one is not interested in the full-field solution, but only in certain statistics of the full field~\cite{migliorati2013,castrillon2016}. In this experiment, the goal is to learn a dynamical model of a PDE quantity of interest (QoI) that can be used for forecasting. Continuing with a focus on non-autonomous systems, we consider an example of the Allen-Cahn equation with forcing that was used in~\cite{dolgov2021}. The system uses Neumann boundary conditions, and its dynamics are given as
  \begin{equation}\label{eq:allencahn}
    \frac{\partial}{\partial t} w(\xi,t) = \sigma\frac{\partial^2w}{\partial\xi^2} + w(1-w^2) + \chi_{\delta}(\xi)u(t),
  \end{equation}
  where $w$ is the flow, $\xi\in[-1,1]$ is the spatial coordinate, $t\in[0,\infty)$ is the time coordinate, $u$ is the control input, and $\chi$ is an indicator function that takes the value one when $\xi\in\delta=[-0.5,0.2]$ and zero otherwise. The control inputs are sampled as $u(t_k)\sim\N(0,10^{-2})$ for $k=0,\ldots,\numObs$, and a zero-order hold is assumed for intermediate time values. To generate the data for this system, a spatial mesh with 256 cells and a time discretization with $\Delta t=0.1$ are used. Then at each timestep $t_k$, Eq.~\eqref{eq:allencahn} is solved for $w(\xi_i,t_k)$ at each vertex $\xi_i$ using the \texttt{solve} function in FEniCS~\cite{logg2010,logg2012}. The output of this system is chosen to be the second moment of the flow, which is approximated as
  \begin{equation}
    y_k = \frac{1}{257}\sum_{i=0}^{256}w^2(\xi_i,t_k) + \eta_k,
  \end{equation}
  where $\eta_k\sim\N(0,0.2^2)$ represents sensor noise. The system is simulated using an initial condition of $w(\xi_i,0)=0$ $\forall i=0,\ldots,256$, and training data collection begins at $t=20$ since the first 20 seconds contain a transient period where the system moves toward a stable equilibrium at $\pm1$. After this initial period, data are collected for 10s at $\Delta t=0.1$ intervals for a total $101$ data points.
  
\begin{figure}
  \centering
  \begin{subfigure}{0.499\linewidth}
    \centering
    \includegraphics[trim=8 291 39 293, clip,width=\linewidth]{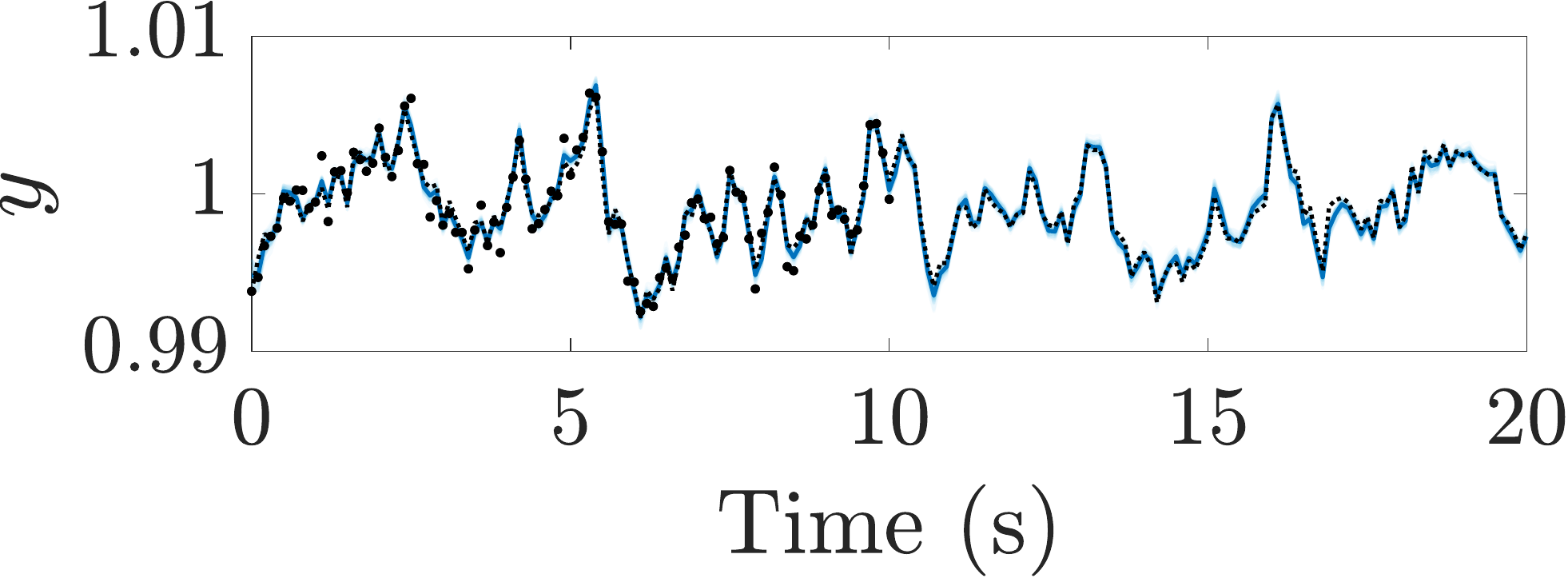}
    \caption{Stochastic simulation}
    \label{fig:postFilter}
  \end{subfigure}%
  \begin{subfigure}{0.499\linewidth}
    \centering
    \includegraphics[trim=8 291 39 293, clip,width=\linewidth]{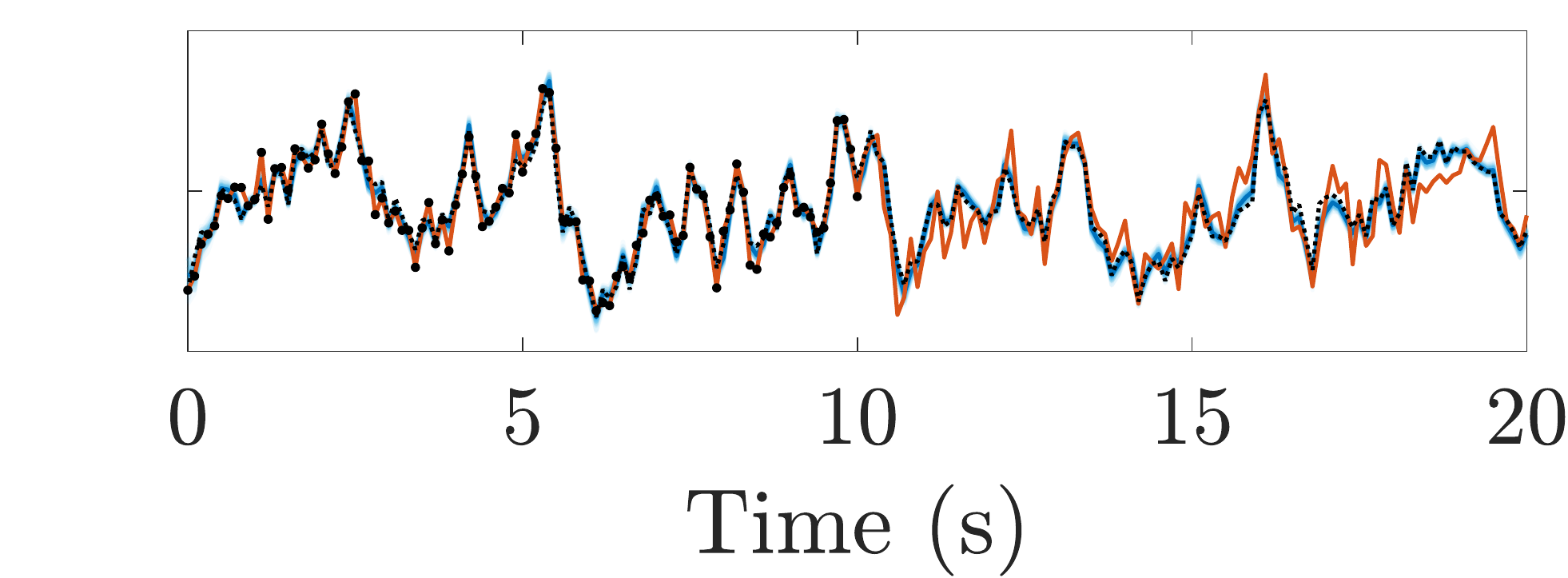}
    \caption{Deterministic simulation}
    \label{fig:postSim}
  \end{subfigure}%
  \hfill
  \includegraphics[trim=65 400 41 350, clip,width=0.7\linewidth]{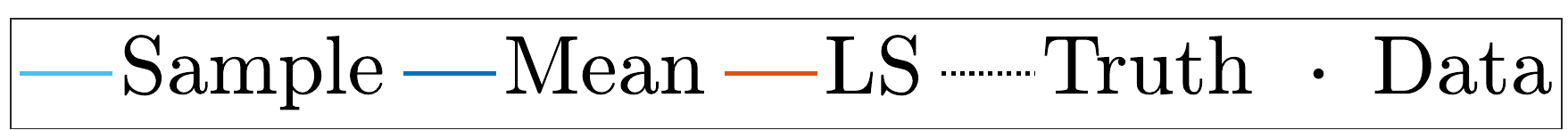}
  \caption{The estimates of the Allen-Cahn QoI over the 10s training period and the subsequent 10s testing period.}
 \label{fig:AC}
\end{figure}

The posterior predictive distribution found with the Bayesian algorithm is compared to a model trained using the deterministic LS objective in Fig.~\ref{fig:AC}. The dynamics model used by both algorithms is the neural network used in previous examples and defined in Eq.~\eqref{eq:nn} with $\dimx=8$. The neural network has 350 parameters and there are only 101 data points, which puts many system ID algorithms at risk of overfitting. The observation operator is fixed as $\mH=\begin{bmatrix}1 & \vzero_{1\times 7}\end{bmatrix}$. The priors are $\text{half--}\N(0,10^{-6})$ on the process noise variance parameters, $\text{half--}\N(0,1)$ on the measurement noise variance parameters, and $\N(0,4)$ on the remaining parameters. The data were normalized before training to have zero mean and standard deviation of one. In this experiment, $10^{5}$ samples were drawn from the posterior, and half were discarded as burn-in.

Fig.~\ref{fig:postFilter} shows 100 samples simulated using stochastic simulation, and Fig.~\ref{fig:postSim} shows the samples simulated deterministically. In both figures, the mean represents the mean of these sample trajectories. The LS estimate matches the data closely during the training period, but performs poorly beyond this period due to overfitting. The posterior predictive mean, on the other hand, provides a good estimate of the truth throughout the 20s time period.

The RMSE values of the posterior predictive mean and LS estimate on the training data and on the noiseless QoI values during the training and testing periods are given in Table~\ref{tab:acrmse}. The LS estimate has a training data RMSE two orders of magnitude smaller than that of the mean estimate, but the noiseless training QoI RMSE of the LS estimate is actually worse than that of the mean estimate. Moreover, the noiseless testing QoI RMSE of the LS estimate is an order of magnitude worse than the mean estimate. Also note that the RMSE of the mean estimate on the noiseless training and testing QoIs are very similar, indicating good generalizability of the estimate. Recall that the prior used on the dynamics parameters was only weakly informative, so the improved generalizability of the mean estimate over the LS estimate comes nearly entirely from the inclusion of process noise in the likelihood. The LS objective, on the other hand, does not account for model uncertainty and implicitly assumes that the model that most closely fits the data is the best,  making it prone to overfit when the model form is very expressive, as is the case here.

\begin{table}[]
  \caption{RMSE values of the posterior predictive mean (Bayes) and deterministic LS estimate (LS) on the training data, the noiseless QoI values during the training period, and the noiseless QoI values during the testing period.}
  \label{tab:acrmse}
\begin{tabular}{|l|l|l|l|}
\hline
& Training data  & Training QoI & Testing QoI \\
& RMSE           & RMSE           & RMSE \\\hline
Bayes & $7.10\times10^{-4}$ & $3.74\times10^{-4}$ & $3.77\times10^{-4}$   \\ \hline
LS    & $3.43\times10^{-6}$ & $8.18\times10^{-4}$ & $1.46\times10^{-3}$   \\ \hline
\end{tabular}
\end{table}

\section{Conclusion}\label{sec:conclusion}
Theoretical and experimental comparisons were made between the Bayesian system ID method of~\cite{galioto2020} and two other system ID algorithms: the LS+ERA method of~\cite{oymak2019} and the multiple shooting method. Proposition~\ref{prop:blue} showed that many Markov parameter estimation methods that rely on a least squares objective implicitly assume conditional independence of the data. We considered one such method that breaks single rollout data into multiple subtrajectories and proved in Proposition~\ref{prop:converge} that this method is asymptotically equivalent to the MLE estimator as the subtrajectory length gets arbitrarily large for conditionally independent data. With regards to multiple shooting, Proposition~\ref{prop:ms} showed that the unconstrained and constrained objectives are equivalent to the MLE and MAP estimators, respectively, of the Bayesian posterior if zero model error is assumed.

Experimental results showed that the Bayesian method can produce more accurate and generalizable results than the comparison methods when the data are noisy and/or sparse, when the number of data is low, and/or when the underlying system exhibits chaotic behavior. In separate experiments, these results showed orders of magnitude improvement in terms of MSE of the MAP over the LS+ERA estimate, 8.7 times lower MSE of the posterior predictive mean compared to multiple shooting, and 3.9 times lower MSE of the MAP compared to deterministic least squares.  Additionally, it was shown that the deterministic least squares objective is not fit to identify good models of chaotic systems, and the Bayesian method is viable even when there are more parameters than data due to the inherent regularization in the likelihood. Future work will look toward applying the Bayesian method to higher dimensional systems.

\begin{ack}
This work was funded by the AFOSR Computational Mathematics Program (P.M Fariba
Fahroo).
\end{ack}

\bibliographystyle{plain}
\bibliography{bib}

\appendix
\section{Eigensystem realization algorithm}\label{app:era}
The ERA is a subspace identification algorithm that is commonly paired with Markov parameter estimation methods, such as those described in Section~\ref{sec:era}, to procure an estimated realization of the state-space matrices. The implementation of this algorithm is detailed here.

Assume that a subset of the Markov parameters $\{\mG_i\}_{i=0}^{\numObs}$ are available and the state dimension $\dimx$ is known. The first step of the ERA is to form these $\numObs+1$ Markov parameters into a Hankel matrix as shown
\begin{equation}
  \mE =\begin{bmatrix}
  \mG_0 & \mG_1 & \cdots & \mG_{d_2} \\
  \mG_1 & \mG_2 & \cdots & \mG_{d_2+1} \\
  \vdots & \vdots & \vdots & \vdots \\
  \mG_{d_1-1} & \mG_{d_1} & \cdots & \mG_{d_1+d_2}
  \end{bmatrix},
\end{equation}
where $d_1,d_2\in\mathbb{N}$ determine the shape of the Hankel matrix and must satisfy the inequalities $d_1+d_2\leq\numObs$ and $\min\{\dimy d_1,\dimu d_2\}\geq\dimx$. Choosing balanced dimensions $\dimy d_1\approx\dimu d_2$ can possibly improve noise robustness.

The ERA uses this Hankel matrix to construct an estimate of the system's observability and controllability matrices from which a realization of the state-space matrices can be extracted. Define the observability and controllability matrices $\mO=\begin{bmatrix}\mH^* & (\mH\mA)^* & \cdots & (\mH\mA^{d_2})^*\end{bmatrix}^*$ and $\mC=\begin{bmatrix}\mB & \mA\mB & \cdots & \mA^{d_1-1}\mB\end{bmatrix}$, respectively. Define the Hankel submatrices $\mE^-=\mE[:,1:d_2\dimu]$ and $\mE^+=\mE[:,\dimu+1:(d_2+1)\dimu]$. Taking the SVD of the Hankel submatrix $\mE^-$ yields the decomposition $\mE^-=\mU\mS\mV^*$. The rank-$\dimx$ approximation of $\mE^-$ can be decomposed as $\tilde{\mU}\tilde{\mS}\tilde{\mV}^*$, where $\tilde{\mU}=\mU[:,1:\dimx]$, $\tilde{\mS}=\mS[1:\dimx,1:\dimx]$, and $\tilde{\mV}=\mV[:,1:\dimx]$. Next, $\mO$ and $\mC$ can be estimated as $\hat{\mO}=\tilde{\mU}\tilde{\mS}^{\frac{1}{2}}$ and $\hat{\mC}=\tilde{\mS}^{\frac{1}{2}}\tilde{\mV}^*$. The ERA estimate of the state-space matrices are then $\hat{\mB} = \hat{\mC}[:,1:\dimu]$, $\hat{\mH} = \hat{\mO}[1:\dimy,:]$, and $\hat{\mA} = \hat{\mO}^{\dagger}\hat{\mE}^+\hat{\mC}$. 
\begin{algorithm}
  \caption{Least squares + eigensystem realization algorithm (LS+ERA)~\cite{oymak2019}}
  \label{alg:lsera}
  \begin{algorithmic}[1]
    \Require Observations $\mY$, inputs $\mU$, state-space dimension $\dimx$, and Hankel shape parameters $d_1$,$d_2$
    \Ensure $\hat{\mA}$, $\hat{\mB}$, $\hat{\mH}$
    \State Estimate Markov parameters: $\hat{\mG} = (\mU^\dagger\mY)^*$
    \State $\hat{\mE}^- = \hat{\mG}[:,1:d_2\dimu]$ and 
    \Statex $\hat{\mE}^+ = \hat{\mG}[:,(\dimu+1):(d_2+1)\dimu]$
    \State $\mU,\mS,\mV = \text{SVD}(\hat{\mE}^-)$
    \State $\tilde{\mU}=\mU[:,1:\dimx]$, $\tilde{\mS}=\mS[1:\dimx,1:\dimx]$, and
    \Statex $\tilde{\mV}=\mV[:,1:\dimx]$
    \State $\hat{\mO} = \tilde{\mU}\tilde{\mS}^{\frac{1}{2}}$ and $\hat{\mC} = \tilde{\mS}^{\frac{1}{2}}\tilde{\mV}^*$
    \State $\hat{\mB} = \hat{\mC}[:,1:\dimu]$, $\hat{\mH} = \hat{\mO}[1:\dimy, :]$, and $\hat{\mA} = \hat{\mO}^{\dagger}\hat{\mE}^+\hat{\mC}$
  \end{algorithmic}
\end{algorithm}

\section{Proof of Proposition~\ref{prop:converge}}\label{app:asymptotic}
Our goal is to show that the assumptions introduced in Proposition~\ref{prop:equiv} that lead to equivalency hold asymptotically. That is, we want to show $\lim_{\bar{n}\to\infty}\sum_{i=\bar{n}}^{k}\mG_i\vu_{k-i}=\vzero$ and $\lim_{\bar{n}\to\infty}\mA^{k}\mSigma(\mA^k)^*=\vzero$ for $k\geq\bar{n}$. Let $N=k-\bar{n}+1$ be the number of terms in the sum. For this proof, we assume $N$ is bounded such that it cannot grow arbitrarily large.

  To show $\lim_{\bar{n}\to\infty}\sum_{i=\bar{n}}^{k}\mG_i\vu_{k-i}=\vzero$, it suffices to show $\lim_{\bar{n}\to\infty}\sum_{i=\bar{n}}^k\lvert\mG_i[j,:]\vu_{k-i}\rvert=0$ for any $j\in\{1,\ldots,\dimy\}$. By assumption, the inputs $\vu_i[j]\in\reals$ are independent realizations of the random variable $\vu$ for $i=0,1,\ldots$ and $j=1,\ldots,\dimu$. Recall that for any real-valued random variable $z$, $\lim_{a\to\infty}\int_{-a}^{a}\probd(z)\rmd z=1$, where the integral represents the probability that $\lvert z\rvert<a$. This implies that for any $0<\varepsilon<1$, $\exists \va\in\reals^{\dimu}$ such that $\lvert\vu[j]\rvert < \va[j]$ for $j=1,\dots,\dimu$ with probability (w.p.) $1-\varepsilon$. Then, upper and lower bounds can be established as
  \begin{equation}0 \leq \sum_{i=\bar{n}}^k\left\lvert\mG_i[j,:]\vu_{k-i}\right\rvert < \sum_{i=\bar{n}}^k\left\lvert\mG_i[j,:]\right\rvert\va,
  \end{equation}
  w.p. at least $(1-\varepsilon)^{N}$. Next we prove that the upper bound goes to zero as $\bar{n}\to\infty$ regardless of $\va$ by showing that $\lim_{\bar{n}\to\infty}\sum_{i=\bar{n}}^k\mG_i[j,:]=\vzero$. Recall that when $\rho(\mA)<1$, the LTI system is exponentially stable. Specifically, there exist constants $c>0$ and $\lambda\in(0,1)$ such that $\lVert\mA^{k}\vx_0\rVert_2\leq c\lambda^{k}\lVert\vx_0\rVert_2$ for any initial condition $\vx_0$. We can therefore place the following bounds on the norm of the columns of $\mG_i$:
  \begin{equation}
    \begin{split}
      \lVert\mG_i[:,j]\rVert_2 &= \lVert\mH\mA^{i-1}\mB[:,j]\rVert_2 \\
      &\leq \lVert\mH\rVert_2\lVert\mA^{i-1}\mB[:,j]\rVert_2 \\
      &\leq \lVert\mH\rVert_2c\lambda^{i-1}\lVert\mB[:,j]\rVert_2.
    \end{split}
  \end{equation}
  Noting that the quantity $c\lVert\mH\rVert_2\lVert\mB[:,j]\rVert_2$ is constant, we see that $\lVert\mG_i\rVert_2$ is bounded above by an exponentially decaying function of timestep $i$. This leads us to the following bound on the norm of the sum:
  \begin{equation}
    \begin{split}
      \left\lVert\sum_{i=\bar{n}}^{k}\mG_i[:,j]\right\rVert_2 &\leq \sum_{i=\bar{n}}^{k}\lVert\mG_i[:,j]\rVert_2 \\
      &\leq \sum_{i=\bar{n}}^{k}c\lambda^{i-1}\lVert\mH\rVert_2\lVert\mB[:,j]\rVert_2 \\
      &\leq Nc\lambda^{\bar{n}-1}\lVert\mH\rVert_2\lVert\mB[:,j]\rVert_2.
    \end{split}
  \end{equation}
  Since $0\leq\lambda<1$, $\lim_{\bar{n}\to\infty}\lVert\sum_{i=\bar{n}}^{k}\mG_i[:,j]\rVert_2=0$, and consequently $\lim_{\bar{n}\to\infty}\sum_{i=\bar{n}}^{k}\mG_i[:,j]=\vzero$ as well, each w.p. at least $(1-\varepsilon)^N$. The variable $\varepsilon$ can be made arbitrarily small, so the sum $\sum_{i=\bar{n}}^{k}\mG_i\vu_{k-i}$ converges to $\vzero$ w.p. 1 as $\bar{n}\to\infty$. Moreover, the rate of convergence of the upper bound can be found as follows:
  \begin{equation}
    \lim_{\bar{n}\to\infty}\frac{Nc\lambda^{\bar{n}}\lVert\mH\rVert_2\lVert\mB[:,j]\rVert_2}{Nc\lambda^{\bar{n}-1}\lVert\mH\rVert_2\lVert\mB[:,j]\rVert_2} = \lim_{\bar{n}\to\infty}\frac{\lambda^{\bar{n}}}{\lambda^{\bar{n}-1}} = \lambda.
  \end{equation}
  Therefore, the upper bound converges to zero linearly with rate $\lambda$, and the sum $\sum_{i=\bar{n}}^{k}\mG_i\vu_{k-i}$ must converge at least as fast. When $\mA$ is diagonalizable, $\lambda$ can be chosen to be $\rho(\mA)$ such that the convergence rate is bounded above by the maximum eigenvalue of $\mA$.

  The proof for the covariance term is similar.  To show $\lim_{\bar{n}\to\infty}\mA^{\bar{n}}\mSigma(\mA^{\bar{n}})^*=\vzero$, we begin by decomposing the covariance matrix $\mSigma=\mSigma^{\frac{1}{2}}\left(\mSigma^{\frac{1}{2}}\right)^*$. Then, the norm $\mA^{\bar{n}}\mSigma(\mA^{\bar{n}})^*$ is bounded above as follows:
  \begin{equation}
    \lVert\mA^{\bar{n}}\mSigma(\mA^{\bar{n}})^*\rVert_2 \leq \left\lVert\mA^{\bar{n}}\mSigma^{\frac{1}{2}}\right\rVert_2\left\lVert\left(\mA^{\bar{n}}\mSigma^{\frac{1}{2}}\right)^*\right\rVert_2 = \left\lVert\mA^{\bar{n}}\mSigma^{\frac{1}{2}}\right\rVert_2^2.
  \end{equation}
  It then suffices to show $\lim_{\bar{n}\to\infty}\mA^{\bar{n}}\mSigma^{\frac{1}{2}}[:,j]=\vzero$ for any column $j\in\{1,\ldots,\dimx\}$. We start by deriving the following upper bound
  \begin{equation}
    \left\lVert\mA^{\bar{n}}\mSigma^{\frac{1}{2}}[:,j]\right\rVert^2_2 \leq \left(c\lambda^{\bar{n}}\left\lVert\mSigma^{\frac{1}{2}}[:,j]\right\rVert_2\right)^2.
  \end{equation}
  Then, the fact that $\lim_{\bar{n}\to\infty}c^2\lambda^{2\bar{n}}\left\lVert\mSigma^{\frac{1}{2}}[:,j]\right\rVert_2^2=0$ implies that $\lim_{\bar{n}\to\infty}\mA^{\bar{n}}\mSigma^{\frac{1}{2}}[:,j]=\vzero$. Lastly, the rate of convergence of the upper bound is as follows
  \begin{equation}
  \lim_{\bar{n}\to\infty}\frac{c^2\lambda^{2\bar{n}}\left\lVert\mSigma^{\frac{1}{2}}[:,j]\right\rVert_2^2}{c^2\lambda^{2(\bar{n}-1)}\left\lVert\mSigma^{\frac{1}{2}}[:,j]\right\rVert_2^2} = \lim_{\bar{n}\to\infty}\frac{\lambda^{2\bar{n}}}{\lambda^{2\bar{n}-2}} = \lambda^2.
  \end{equation}
  Then $\left\lVert\mA^{\bar{n}}\mSigma^{\frac{1}{2}}\right\rVert_2^2$, and consequently $\lVert\mA^{\bar{n}}\mSigma(\mA^{\bar{n}})^*\rVert_2$, converges at least as fast. Therefore, we have shown that $\sum_{i=\bar{n}}^{k}\mG_i\vu_{k-i}$ and $\mA^{\bar{n}}\mSigma(\mA^{\bar{n}})^*$ both converge to zero as $\bar{n}\to\infty$ with rate no greater than $\lambda$.
\end{document}

%% file: macros.tex
\usepackage{amsmath,amsfonts,bm}

\newcommand{\reals}{\mathbb{R}}

\newcommand{\probd}{p}
\newcommand{\like}{\mathcal{L}}

\newcommand{\N}{\mathcal{N}}
\newcommand{\halfN}{\text{half-}\mathcal{N}}

\newcommand{\dell}{\Delta\ell}

\newcommand{\yn}{\mathcal{Y}_n}
\newcommand{\xn}{\mathcal{X}_n}

\newcommand{\yk}{\mathcal{Y}_k}

\newcommand{\ykk}{\mathcal{Y}_{k-1}}

\newcommand{\thetsig}{\boldsymbol{\theta}_{\mSigma}}
\newcommand{\thetgam}{\boldsymbol{\theta}_{\mGamma}}
\newcommand{\thetdyn}{\boldsymbol{\theta}_{\Psi}} 
\newcommand{\thetobs}{\boldsymbol{\theta}_{h}}
\newcommand{\thetic}{\boldsymbol{\theta}_{\vx_0}}

\newcommand{\dimx}{d_x}
\newcommand{\dimpar}{d_{\theta}}
\newcommand{\dimy}{d_y}
\newcommand{\dimu}{d_u}
\newcommand{\numObs}{n}

\def\vzero{{\mathbf{0}}}

\def\vmu{{\boldsymbol{\mu}}}
\def\vtheta{{\boldsymbol{\theta}}}

\def\vxi{{\boldsymbol{\xi}}}
\def\veta{{\boldsymbol{\eta}}}
\def\vnu{{\boldsymbol{\nu}}}
\def\va{{\mathbf{a}}}
\def\vb{{\mathbf{b}}}

\def\vm{{\mathbf{m}}}

\def\vu{{\mathbf{u}}}

\def\vx{{\mathbf{x}}}
\def\vy{{\mathbf{y}}}
\def\vz{{\mathbf{z}}}

\def\mA{{\mathbf{A}}}
\def\mB{{\mathbf{B}}}
\def\mC{{\mathbf{C}}}
\def\mD{{\mathbf{D}}}
\def\mE{{\mathbf{E}}}

\def\mG{{\mathbf{G}}}
\def\mH{{\mathbf{H}}}
\def\mI{{\mathbf{I}}}

\def\mO{{\mathbf{O}}}
\def\mP{{\mathbf{P}}}

\def\mS{{\mathbf{S}}}

\def\mU{{\mathbf{U}}}
\def\mV{{\mathbf{V}}}
\def\mW{{\mathbf{W}}}

\def\mY{{\mathbf{Y}}}

\def\mLambda{{\boldsymbol{\Lambda}}}
\def\mSigma{{\boldsymbol{\Sigma}}}
\def\mGamma{{\boldsymbol{\Gamma}}}

\DeclareMathOperator*{\argmin}{arg\,min}

%% file: main.bbl
\begin{thebibliography}{10}

\bibitem{abhishek2012}
Kumar Abhishek, MP~Singh, Saswata Ghosh, and Abhishek Anand.
\newblock Weather forecasting model using artificial neural network.
\newblock {\em Procedia Technology}, 4:311--318, 2012.

\bibitem{aguirre2010}
Luis~A Aguirre, Bruno~HG Barbosa, and Ant{\^o}nio~P Braga.
\newblock Prediction and simulation errors in parameter estimation for
  nonlinear systems.
\newblock {\em Mechanical Systems and Signal Processing}, 24(8):2855--2867,
  2010.

\bibitem{alessandri2008}
Angelo Alessandri, Marco Baglietto, and Giorgio Battistelli.
\newblock Moving-horizon state estimation for nonlinear discrete-time systems:
  New stability results and approximation schemes.
\newblock {\em Automatica}, 44(7):1753--1765, 2008.

\bibitem{almunif2020}
Anas Almunif, Lingling Fan, and Zhixin Miao.
\newblock A tutorial on data-driven eigenvalue identification: Prony analysis,
  matrix pencil, and eigensystem realization algorithm.
\newblock {\em International Transactions on Electrical Energy Systems},
  30(4):e12283, 2020.

\bibitem{baros2022}
Stefanos Baros, Chin-Yao Chang, Gabriel~E Colon-Reyes, and Andrey Bernstein.
\newblock Online data-enabled predictive control.
\newblock {\em Automatica}, 138:109926, 2022.

\bibitem{beck2010}
James~L Beck.
\newblock Bayesian system identification based on probability logic.
\newblock {\em Structural Control and Health Monitoring}, 17(7):825--847, 2010.

\bibitem{beintema2021}
Gerben Beintema, Roland Toth, and Maarten Schoukens.
\newblock Nonlinear state-space identification using deep encoder networks.
\newblock In {\em Learning for Dynamics and Control}, pages 241--250. PMLR,
  2021.

\bibitem{bock1981}
Hans~Georg Bock.
\newblock Numerical treatment of inverse problems in chemical reaction
  kinetics.
\newblock In {\em Modelling of chemical reaction systems}, pages 102--125.
  Springer, 1981.

\bibitem{bock1984}
Hans~Georg Bock and Karl-Josef Plitt.
\newblock A multiple shooting algorithm for direct solution of optimal control
  problems.
\newblock {\em IFAC Proceedings Volumes}, 17(2):1603--1608, 1984.

\bibitem{brunton2016}
Steven~L. Brunton, Joshua~L. Proctor, and J.~Nathan Kutz.
\newblock Discovering governing equations from data by sparse identification of
  nonlinear dynamical systems.
\newblock {\em Proceedings of the National Academy of Sciences},
  113(15):3932--3937, 2016.

\bibitem{castrillon2016}
Julio~E Castrillon-Candas, Fabio Nobile, and Raul~F Tempone.
\newblock Analytic regularity and collocation approximation for elliptic pdes
  with random domain deformations.
\newblock {\em Computers \& Mathematics with Applications}, 71(6):1173--1197,
  2016.

\bibitem{chen2018}
Ricky~TQ Chen, Yulia Rubanova, Jesse Bettencourt, and David~K Duvenaud.
\newblock Neural ordinary differential equations.
\newblock {\em Advances in neural information processing systems}, 31, 2018.

\bibitem{chen2012}
Tianshi Chen, Henrik Ohlsson, and Lennart Ljung.
\newblock On the estimation of transfer functions, regularizations and gaussian
  processes—revisited.
\newblock {\em Automatica}, 48(8):1525--1535, 2012.

\bibitem{chen2020}
Yutao Chen, Nicol{\`o} Scarabottolo, Mattia Bruschetta, and Alessandro Beghi.
\newblock Efficient move blocking strategy for multiple shooting-based
  non-linear model predictive control.
\newblock {\em IET Control Theory \& Applications}, 14(2):343--351, 2020.

\bibitem{de2002}
Perry De~Valpine and Alan Hastings.
\newblock Fitting population models incorporating process noise and observation
  error.
\newblock {\em Ecological Monographs}, 72(1):57--76, 2002.

\bibitem{dolgov2021}
Sergey Dolgov, Dante Kalise, and Karl~K Kunisch.
\newblock Tensor decomposition methods for high-dimensional
  hamilton--jacobi--bellman equations.
\newblock {\em SIAM Journal on Scientific Computing}, 43(3):A1625--A1650, 2021.

\bibitem{drineas2016}
Petros Drineas and Michael~W Mahoney.
\newblock Randnla: randomized numerical linear algebra.
\newblock {\em Communications of the ACM}, 59(6):80--90, 2016.

\bibitem{elliott2008}
Robert~J Elliott, Lakhdar Aggoun, and John~B Moore.
\newblock {\em Hidden Markov models: estimation and control}, volume~29.
\newblock Springer Science \& Business Media, 2008.

\bibitem{fallah2018}
Seyedeh~Narjes Fallah, Ravinesh~Chand Deo, Mohammad Shojafar, Mauro Conti, and
  Shahaboddin Shamshirband.
\newblock Computational intelligence approaches for energy load forecasting in
  smart energy management grids: state of the art, future challenges, and
  research directions.
\newblock {\em Energies}, 11(3):596, 2018.

\bibitem{fang2020}
Yaqing Fang, Yiting Nie, and Marshare Penny.
\newblock Transmission dynamics of the covid-19 outbreak and effectiveness of
  government interventions: A data-driven analysis.
\newblock {\em Journal of medical virology}, 92(6):645--659, 2020.

\bibitem{forgione2021}
Marco Forgione and Dario Piga.
\newblock Continuous-time system identification with neural networks: Model
  structures and fitting criteria.
\newblock {\em European Journal of Control}, 59:69--81, 2021.

\bibitem{fox2018}
Ian Fox, Lynn Ang, Mamta Jaiswal, Rodica Pop-Busui, and Jenna Wiens.
\newblock Deep multi-output forecasting: Learning to accurately predict blood
  glucose trajectories.
\newblock In {\em Proceedings of the 24th ACM SIGKDD international conference
  on knowledge discovery \& data mining}, pages 1387--1395, 2018.

\bibitem{galioto2020}
Nicholas Galioto and Alex~Arkady Gorodetsky.
\newblock Bayesian system id: optimal management of parameter, model, and
  measurement uncertainty.
\newblock {\em Nonlinear Dynamics}, 102:241--267, Sep 2020.

\bibitem{galioto2021}
Nicholas Galioto and Alex~Arkady Gorodetsky.
\newblock A new objective for identification of partially observed linear
  time-invariant dynamical systems from input-output data.
\newblock In {\em Learning for Dynamics and Control}, pages 1180--1191. PMLR,
  2021.

\bibitem{giftthaler2018}
Markus Giftthaler, Michael Neunert, Markus St{\"a}uble, Jonas Buchli, and
  Moritz Diehl.
\newblock A family of iterative gauss-newton shooting methods for nonlinear
  optimal control.
\newblock In {\em 2018 IEEE/RSJ International Conference on Intelligent Robots
  and Systems (IROS)}, pages 1--9. IEEE, 2018.

\bibitem{gneiting2005}
Tilmann Gneiting and Adrian~E Raftery.
\newblock Weather forecasting with ensemble methods.
\newblock {\em Science}, 310(5746):248--249, 2005.

\bibitem{green-sa2015}
Peter~L Green.
\newblock Bayesian system identification of a nonlinear dynamical system using
  a novel variant of simulated annealing.
\newblock {\em Mechanical Systems and Signal Processing}, 52:133--146, 2015.

\bibitem{green-model2015}
PL~Green and K~Worden.
\newblock Bayesian and markov chain monte carlo methods for identifying
  nonlinear systems in the presence of uncertainty.
\newblock {\em Philosophical Transactions of the Royal Society A: Mathematical,
  Physical and Engineering Sciences}, 373(2051):20140405, 2015.

\bibitem{greydanus2019}
Samuel Greydanus, Misko Dzamba, and Jason Yosinski.
\newblock Hamiltonian neural networks.
\newblock {\em Advances in neural information processing systems}, 32, 2019.

\bibitem{haario2006}
Heikki Haario, Marko Laine, Antonietta Mira, and Eero Saksman.
\newblock Dram: efficient adaptive mcmc.
\newblock {\em Statistics and computing}, 16(4):339--354, 2006.

\bibitem{ho1966}
BL~Ho and Rudolf~E K{\'a}lm{\'a}n.
\newblock Effective construction of linear state-variable models from
  input/output functions.
\newblock {\em at-Automatisierungstechnik}, 14(1-12):545--548, 1966.

\bibitem{hossain2019}
Eklas Hossain, Imtiaj Khan, Fuad Un-Noor, Sarder~Shazali Sikander, and
  Md~Samiul~Haque Sunny.
\newblock Application of big data and machine learning in smart grid, and
  associated security concerns: A review.
\newblock {\em Ieee Access}, 7:13960--13988, 2019.

\bibitem{jia2019}
Xiaowei Jia, Jared Willard, Anuj Karpatne, Jordan Read, Jacob Zwart, Michael
  Steinbach, and Vipin Kumar.
\newblock Physics guided rnns for modeling dynamical systems: A case study in
  simulating lake temperature profiles.
\newblock In {\em Proceedings of the 2019 SIAM International Conference on Data
  Mining}, pages 558--566. SIAM, 2019.

\bibitem{jordan2007}
Dominic Jordan and Peter Smith.
\newblock {\em Nonlinear ordinary differential equations: an introduction for
  scientists and engineers}, volume~10.
\newblock Oxford University Press on Demand, 2007.

\bibitem{juang1985}
Jer-Nan Juang and Richard~S Pappa.
\newblock An eigensystem realization algorithm for modal parameter
  identification and model reduction.
\newblock {\em Journal of guidance, control, and dynamics}, 8(5):620--627,
  1985.

\bibitem{julier1997}
Simon~J Julier and Jeffrey~K Uhlmann.
\newblock New extension of the kalman filter to nonlinear systems.
\newblock In {\em Signal processing, sensor fusion, and target recognition VI},
  volume 3068, pages 182--193. International Society for Optics and Photonics,
  1997.

\bibitem{karevan2020}
Zahra Karevan and Johan~AK Suykens.
\newblock Transductive lstm for time-series prediction: An application to
  weather forecasting.
\newblock {\em Neural Networks}, 125:1--9, 2020.

\bibitem{khan2016}
Ahsan~Raza Khan, Anzar Mahmood, Awais Safdar, Zafar~A Khan, and Naveed~Ahmed
  Khan.
\newblock Load forecasting, dynamic pricing and dsm in smart grid: A review.
\newblock {\em Renewable and Sustainable Energy Reviews}, 54:1311--1322, 2016.

\bibitem{kirkpatrick1983}
Scott Kirkpatrick, C~Daniel Gelatt~Jr, and Mario~P Vecchi.
\newblock Optimization by simulated annealing.
\newblock {\em science}, 220(4598):671--680, 1983.

\bibitem{kramer2018}
Boris Kramer and Alex~A Gorodetsky.
\newblock System identification via cur-factored hankel approximation.
\newblock {\em SIAM Journal on Scientific Computing}, 40(2):A848--A866, 2018.

\bibitem{lale2020}
Sahin Lale, Kamyar Azizzadenesheli, Babak Hassibi, and Anima Anandkumar.
\newblock Logarithmic regret bound in partially observable linear dynamical
  systems.
\newblock {\em Advances in Neural Information Processing Systems},
  33:20876--20888, 2020.

\bibitem{li2019}
Kezhi Li, Chengyuan Liu, Taiyu Zhu, Pau Herrero, and Pantelis Georgiou.
\newblock Glunet: A deep learning framework for accurate glucose forecasting.
\newblock {\em IEEE journal of biomedical and health informatics},
  24(2):414--423, 2019.

\bibitem{ljung2010}
Lennart Ljung.
\newblock Perspectives on system identification.
\newblock {\em Annual Reviews in Control}, 34(1):1--12, 2010.

\bibitem{logg2012}
Anders Logg, Kent-Andre Mardal, and Garth Wells.
\newblock {\em Automated solution of differential equations by the finite
  element method: The FEniCS book}, volume~84.
\newblock Springer Science \& Business Media, 2012.

\bibitem{logg2010}
Anders Logg and Garth~N Wells.
\newblock Dolfin: Automated finite element computing.
\newblock {\em ACM Transactions on Mathematical Software (TOMS)}, 37(2):1--28,
  2010.

\bibitem{long2018}
Zichao Long, Yiping Lu, Xianzhong Ma, and Bin Dong.
\newblock Pde-net: Learning pdes from data.
\newblock In {\em International Conference on Machine Learning}, pages
  3208--3216. PMLR, 2018.

\bibitem{ma2011}
Zhanhua Ma, Sunil Ahuja, and Clarence~W Rowley.
\newblock Reduced-order models for control of fluids using the eigensystem
  realization algorithm.
\newblock {\em Theoretical and Computational Fluid Dynamics}, 25(1):233--247,
  2011.

\bibitem{mackay2003}
David~JC MacKay and David~JC Mac~Kay.
\newblock {\em Information theory, inference and learning algorithms}.
\newblock Cambridge university press, 2003.

\bibitem{masti2021}
Daniele Masti and Alberto Bemporad.
\newblock Learning nonlinear state--space models using autoencoders.
\newblock {\em Automatica}, 129:109666, 2021.

\bibitem{migliorati2013}
Giovanni Migliorati, Fabio Nobile, Erik von Schwerin, and Ra{\'u}l Tempone.
\newblock Approximation of quantities of interest in stochastic pdes by the
  random discrete l\^{}2 projection on polynomial spaces.
\newblock {\em SIAM Journal on Scientific Computing}, 35(3):A1440--A1460, 2013.

\bibitem{oymak2019}
Samet Oymak and Necmiye Ozay.
\newblock Non-asymptotic identification of lti systems from a single
  trajectory.
\newblock In {\em 2019 American Control Conference (ACC)}, pages 5655--5661.
  IEEE, 2019.

\bibitem{pappalardo2018}
Carmine~Maria Pappalardo and Domenico Guida.
\newblock System identification and experimental modal analysis of a frame
  structure.
\newblock {\em Engineering Letters}, 26(1), 2018.

\bibitem{pillonetto2016}
Gianluigi Pillonetto, Tianshi Chen, Alessandro Chiuso, Giuseppe De~Nicolao, and
  Lennart Ljung.
\newblock Regularized linear system identification using atomic, nuclear and
  kernel-based norms: The role of the stability constraint.
\newblock {\em Automatica}, 69:137--149, 2016.

\bibitem{piroddi2003}
Luigi Piroddi and William Spinelli.
\newblock An identification algorithm for polynomial narx models based on
  simulation error minimization.
\newblock {\em International Journal of Control}, 76(17):1767--1781, 2003.

\bibitem{plis2014}
Kevin Plis, Razvan Bunescu, Cindy Marling, Jay Shubrook, and Frank Schwartz.
\newblock A machine learning approach to predicting blood glucose levels for
  diabetes management.
\newblock In {\em Workshops at the Twenty-Eighth AAAI conference on artificial
  intelligence}, 2014.

\bibitem{rackauckas2020}
Christopher Rackauckas, Yingbo Ma, Julius Martensen, Collin Warner, Kirill
  Zubov, Rohit Supekar, Dominic Skinner, Ali Ramadhan, and Alan Edelman.
\newblock Universal differential equations for scientific machine learning.
\newblock {\em arXiv preprint arXiv:2001.04385}, 2020.

\bibitem{raissi2019}
Maziar Raissi, Paris Perdikaris, and George~E Karniadakis.
\newblock Physics-informed neural networks: A deep learning framework for
  solving forward and inverse problems involving nonlinear partial differential
  equations.
\newblock {\em Journal of Computational physics}, 378:686--707, 2019.

\bibitem{ribeiro2020}
Ant{\^o}nio~H Ribeiro, Koen Tiels, Jack Umenberger, Thomas~B Sch{\"o}n, and
  Luis~A Aguirre.
\newblock On the smoothness of nonlinear system identification.
\newblock {\em Automatica}, 121:109158, 2020.

\bibitem{sarkar2021}
Tuhin Sarkar, Alexander Rakhlin, and Munther~A Dahleh.
\newblock Finite time lti system identification.
\newblock {\em Journal of Machine Learning Research}, 22(26):1--61, 2021.

\bibitem{sarkka2013}
Simo S{\"a}rkk{\"a}.
\newblock {\em Bayesian filtering and smoothing}, volume~3.
\newblock Cambridge University Press, 2013.

\bibitem{schmid2010}
Peter~J Schmid.
\newblock Dynamic mode decomposition of numerical and experimental data.
\newblock {\em Journal of fluid mechanics}, 656:5--28, 2010.

\bibitem{schoukens2009}
J~Schoukens, Johan Suykens, and L~Ljung.
\newblock Wiener-hammerstein benchmark.
\newblock In {\em Proc. of the 15th IFAC symposium on System Identification
  (SYSID 2009)}, 2009.

\bibitem{schoukens2019}
Johan Schoukens and Lennart Ljung.
\newblock Nonlinear system identification: A user-oriented road map.
\newblock {\em IEEE Control Systems Magazine}, 39(6):28--99, 2019.

\bibitem{smith2018}
Samuel~L Smith and Quoc~V Le.
\newblock A bayesian perspective on generalization and stochastic gradient
  descent.
\newblock In {\em International Conference on Learning Representations}, 2018.

\bibitem{sun2020}
Yue Sun, Samet Oymak, and Maryam Fazel.
\newblock Finite sample system identification: Optimal rates and the role of
  regularization.
\newblock In {\em Learning for Dynamics and Control}, pages 16--25. PMLR, 2020.

\bibitem{tsiamis2019}
Anastasios Tsiamis and George~J Pappas.
\newblock Finite sample analysis of stochastic system identification.
\newblock In {\em 2019 IEEE 58th Conference on Decision and Control (CDC)},
  pages 3648--3654. IEEE, 2019.

\bibitem{tu2014}
Jonathan~H Tu, Clarence~W Rowley, Dirk~M Luchtenburg, Steven~L Brunton, and
  J~Nathan Kutz.
\newblock On dynamic mode decomposition: theory and applications.
\newblock {\em Journal of Computational Dynamics}, 1(2), 2014.

\bibitem{van2012}
Peter Van~Overschee and Bart De~Moor.
\newblock {\em Subspace identification for linear systems: Theory —
  Implementation — Applications}.
\newblock Springer Science \& Business Media, 2012.

\bibitem{viberg1995}
Mats Viberg.
\newblock Subspace-based methods for the identification of linear
  time-invariant systems.
\newblock {\em Automatica}, 31(12):1835--1851, 1995.

\bibitem{voss2004}
Henning~U Voss, Jens Timmer, and J{\"u}rgen Kurths.
\newblock Nonlinear dynamical system identification from uncertain and indirect
  measurements.
\newblock {\em International Journal of Bifurcation and Chaos},
  14(06):1905--1933, 2004.

\bibitem{zhang2021}
Chiyuan Zhang, Samy Bengio, Moritz Hardt, Benjamin Recht, and Oriol Vinyals.
\newblock Understanding deep learning (still) requires rethinking
  generalization.
\newblock {\em Communications of the ACM}, 64(3):107--115, 2021.

\bibitem{zhang2020}
Kunwu Zhang and Yang Shi.
\newblock Adaptive model predictive control for a class of constrained linear
  systems with parametric uncertainties.
\newblock {\em Automatica}, 117:108974, 2020.

\bibitem{zheng2020}
Yang Zheng and Na~Li.
\newblock Non-asymptotic identification of linear dynamical systems using
  multiple trajectories.
\newblock {\em IEEE Control Systems Letters}, 5(5):1693--1698, 2020.

\bibitem{zhong2019}
Yaofeng~Desmond Zhong, Biswadip Dey, and Amit Chakraborty.
\newblock Symplectic ode-net: Learning hamiltonian dynamics with control.
\newblock In {\em International Conference on Learning Representations}, 2019.

\end{thebibliography}
